\documentclass[12pt]{article}
\usepackage{epsf,epsfig,cite}

\graphicspath{{figs/}}

\input paperdef

\begin{document}
\begin{titlepage}
\pagestyle{empty}
\baselineskip=21pt
\begin{flushright}
CERN--PH--TH/2007-087\hfill
DCPT/07/50, IPPP/07/25\\
MPP--2007--64\hfill
UMN--TH--2606/07, FTPI--MINN--07/19 
\end{flushright}
\vskip 0.2in
\begin{center}
{\large{\bf The Supersymmetric Parameter Space in Light of \\
\boldmath{$B$}-physics Observables and Electroweak Precision Data}}
\end{center}
\begin{center}
\vskip 0.05in
{{\bf J.~Ellis}$^1$, 
{\bf S.~Heinemeyer}$^2$,
{\bf K.A.~Olive}$^{3}$,
{\bf A.M.~Weber}$^{4}$ 
and {\bf G.~Weiglein}$^{5}$}\\
\vskip 0.05in
{\it
$^1${TH Division, Physics Department, CERN, Geneva, Switzerland}\\
$^2$Instituto de Fisica de Cantabria (CSIC-UC), 
Santander,  Spain \\
$^3${William I.\ Fine Theoretical Physics Institute,\\
University of Minnesota, Minneapolis, MN~55455, USA}\\
$^4$Max-Planck-Institut f\"ur Physik, 
F\"ohringer Ring 6, D--80805 Munich, Germany \\
$^5${IPPP, University of Durham, Durham DH1~3LE, UK}\\
}
\vskip 0.1in
{\bf Abstract}
\end{center}
\baselineskip=18pt \noindent

{\small
Indirect information about the possible scale of
supersymmetry (SUSY) breaking is provided by $B$-physics observables (BPO) as
well as electroweak precision observables (EWPO). We combine the constraints
imposed by recent measurements of 
the  BPO
$\br(b \to s \ga)$, $\br(B_s \to \mu^+\mu^-)$, 
$\br(B_u \to \tau \nu_\tau)$ and $\De M_{B_s}$ with those obtained
from the experimental measurements of 
the EWPO $\MW$, $\sweff$, $\Ga_Z$, $(g-2)_\mu$ and $\Mh$, incorporating
the latest theoretical calculations of these observables within the
Standard Model and supersymmetric extensions. 
We perform a $\chi^2$~fit to the parameters of the constrained minimal
supersymmetric extension of the Standard Model (CMSSM), in which the
SUSY-breaking parameters are universal at the GUT scale, and the
non-universal Higgs model (NUHM), in which this constraint 
is relaxed for the soft SUSY-breaking contributions to the Higgs masses.
Assuming that the lightest supersymmetric particle (LSP) provides the
cold dark matter density preferred by WMAP and other cosmological data,
we scan over the remaining parameter space. 
Within the CMSSM, we confirm the preference found previously for a
relatively low SUSY-breaking scale, though there is some slight tension
between the EWPO and the BPO. 
In studies of some specific NUHM scenarios compatible with the cold dark
matter constraint we investigate \plane{\MA}{\tb}s and 
find preferred regions that have values of $\chi^2$ somewhat lower
than in the CMSSM. 
}


\vskip 0.15in
\leftline{\today}
\end{titlepage}
\baselineskip=18pt



\section {Introduction}

The dimensionality of the parameter space of the minimal supersymmetric
extension of the Standard Model (MSSM)~\cite{susy,susy2} is so high that
phenomenological analyses often make simplifying assumptions 
that reduce drastically the number of parameters. 
One assumption that is frequently employed is
that (at least some of) the soft SUSY-breaking parameters are universal
at some high input scale, before renormalization. 
One model based on this simplification is the 
constrained MSSM (CMSSM), in which all the soft SUSY-breaking scalar
masses $m_0$ are assumed to be universal at the GUT scale, as are the
soft SUSY-breaking gaugino masses $m_{1/2}$ and trilinear couplings
$A_0$. The assumption that squarks and sleptons with the same gauge
quantum numbers have the same masses is motivated by the absence of
identified supersymmetric contributions to flavour-changing neutral
interactions and rare decays (see \citere{hfag} and references therein).
Universality between squarks and 
sleptons with different gauge interactions may be motivated by some GUT
scenarios~\cite{GUTs}. 
However, the universality of the soft SUSY-breaking
contributions to the Higgs scalar masses is less motivated, and is
relaxed in the non-universal Higgs model (NUHM)~\cite{NUHM1,NUHM2,NUHMother}. 

There are different possible approaches to analyzing the reduced
parameter spaces of the CMSSM and the NUHM. One minimal approach would
be to approximate the various theoretical, phenomenological,
experimental, astrophysical and cosmological constraints naively by
$\theta$ functions, determine the domains of the SUSY parameters allowed
by their combination, and not attempt to estimate which values of the
parameters might be more or less likely. This approach would perhaps be
adequate if one were agnostic about the existence of low-energy SUSY. On
the other hand, if one were more positive about its existence, and keen
to find which SUSY parameter values were more `probable', one would make
a likelihood analysis and take seriously any possible hints that the
Standard Model (SM) might not fit perfectly the available data. This is the
approach taken in this paper.  
We perform a combined $\chi^2$~analysis of electroweak precision
observables (EWPO), going beyond previous such analyses~\cite{ehow3,ehow4} 
(see also \citere{other}), and of $B$-physics observables (BPO),
including some that have not been
included before in comprehensive analyses of the SUSY parameter
space (see, however, \citere{LSPlargeTB}). 
In the past, the set of EWPO included in such analyses have been
the $W$~boson mass $\MW$,  the effective leptonic weak mixing angle
$\sweff$, the anomalous magnetic moment of the  muon $(g-2)_\mu$, and
the mass of the lightest MSSM Higgs boson mass $\Mh$. Since our previous
study, the theoretical link between experimental observables and
$\sweff$ within the Standard Model has become more precise, changing the
$\chi^2$~distribution for the possible MSSM contribution. We also include
in this analysis a new EWPO, namely the total $Z$~boson width
$\Ga_Z$. In addition, we now include four BPO:  the branching ratios
$\br(b \to s \ga)$, $\br(B_s \to \mu^+ \mu^-)$ 
and $\br(B_u \to \tau \nu_\tau)$, and the $B_s$ mass mixing parameter
$\De M_{B_s}$. For each observable, we construct the $\chi^2$~function
including both theoretical and experimental systematic uncertainties, as
well as statistical errors. The largest theoretical systematic
uncertainty is that in $\br(b \to s \ga)$, mainly associated with the
renormalization-scale ambiguity. Since this is not a Gaussian error, we
do not add it in quadrature with the other errors. Instead, in order to
be conservative, we prefer to add it linearly.

For our CMSSM analysis, the fact that the cold dark matter density is
known from astrophysics and cosmology with an uncertainty smaller
than~$10~\%$ fixes with proportional precision one combination of the
SUSY parameters, enabling us to analyze the 
overall $\chi^2$~value as a function of
$m_{1/2}$ for fixed values of $\tb$ and $A_0$. The value of $|\mu|$ is
fixed by the electroweak vacuum conditions, the value of $m_0$ is fixed
with a small error by the dark matter density, and the Higgs mass
parameters are fixed by the universality assumption. As in previous
analyses, we consider various representative values of $A_0 \propto
m_{1/2}$  for the specific choices $\tb = 10, 50$.
Also as previously,
we find a marked preference for relatively small values of 
$m_{1/2} \sim 300, 600 \gev$ for $\tb = 10, 50$, respectively, driven
largely by $(g-2)_\mu$ with some assistance from $\MW$. This preference
would have been more marked if the BPO were not taken into
account. Indeed, there is a slight tension between the EWPO and the BPO,
with the latter disfavouring smaller $m_{1/2}$, particularly for large
$\tb$. As corollaries of this analysis, we present the 
$\chi^2$~distributions for the masses of various
MSSM particles, including the lightest Higgs boson mass $\Mh$. This shows
a strong preference for $\Mh \sim 115 \gev$, allowing 
$\Mh$ as high as $120 \gev$ with $\Delta \chi^2 \sim 4$.

In view of the slight tension between the EWPO and BPO
within the CMSSM, we have gone on to explore the NUHM, which effectively
has $\MA$ and $\mu$ as additional free parameters as compared to the CMSSM.
In particular, we have investigated whether the NUHM
reconciles more easily the EWPO and BPO, and specifically whether
there exist NUHM points with significantly lower $\chi^2$. As pointed
out previously, generic NUHM parameter planes in which the other
variables are held fixed do not satisfy the cold dark matter density
constraint imposed by WMAP et al. In this paper, we introduce
`WMAP surfaces', which are \plane{\MA}{\tb}s across in which the other
variables are adjusted continuously so as to maintain the LSP density
within the WMAP range. We then examine the $\chi^2$~values of the
EWPO and BPO in the NUHM as functions over these WMAP
surfaces~%
\footnote{A more complete characterization of these WMAP
  surfaces will be given elsewhere~\cite{EHHOW}, as well as a discussion
  of their 
  possible use as `benchmark scenarios' for evaluating the prospects for
  MSSM Higgs phenomenology at the Tevatron, the LHC and elsewhere.}%
.~In each of the WMAP surfaces we find localized regions preferred by
the EWPO and BPO and, in some cases, the minimum value of $\chi^2$ is
significantly lower than along the WMAP strips in the CMSSM, indicating
that the NUHM may help resolve the slight tension between the EWPO and
the BPO. We explore this possibility further by investigating lines that
explore further the NUHM parameter space in neighbourhoods of the
low-$\chi^2$ points in the WMAP surfaces.

In \refse{sec:expdata} we review the current status of the EWPO and BPO
that we use, our treatment of the available theoretical
calculations and their errors, as well as their present experimental values.
The analysis within the CMSSM can be found in \refse{sec:cmssm}, while
the NUHM investigation is presented in
\refse{sec:nuhm}. \refse{sec:conclusions} summarizes our principal
conclusions. 


\section{Current Experimental Data}
\label{sec:expdata}

The relevant data set includes five EWPO:
the mass of the $W$~boson, $\MW$, the effective leptonic
weak mixing angle, $\sweff$, the total $Z$~boson width, $\Ga_Z$,
the anomalous magnetic moment of the muon, $(g-2)_\mu$, and the mass of
the lightest MSSM Higgs boson, $\Mh$.
In addition, we include four BPO:
the branching ratios $\br(b \to s \ga)$, $\br(B_s \to \mu^+ \mu^-)$, 
and $\br(B_u \to \tau \nu_\tau)$ as well as the $B_s$~mass-mixing parameter
$\De M_{B_s}$. A detailed description of the EWPO and
$\br(b \to s \ga)$ can be found in~\citeres{ehow3,PomssmRep,ehow4,ZOweber}.

In this Section we start our analysis by recalling the current
precisions of the experimental results and the theoretical predictions
for all these observables. We also display the CMSSM predictions for
the EWPO (where new results are available), 
and also for the BPO. These predictions 
serve as examples of the expected ranges of the EWPO and BPO values once SUSY
corrections are taken into account.

In the following, we refer to the theoretical uncertainties from unknown
higher-order corrections as `intrinsic' theoretical uncertainties and
to the uncertainties induced by the experimental errors of the SM input
parameters as `parametric' theoretical uncertainties.
We do not discuss here the theoretical uncertainties in the
renormalization-group running between the high-scale input parameters
and the weak scale, see \citere{CDMRGEunc} for a recent discussion
in the context of calculations of the cold dark matter (CDM) density. At
present, these uncertainties are less important than the experimental and
theoretical uncertainties in the precision observables.

Assuming that the nine observables listed above are
uncorrelated, a $\chi^2$ fit 
has been performed with

\BE
\chi^2 \equiv \sum_{n=1}^{7} \KKL \KL
              \frac{R_n^{\rm exp} - R_n^{\rm theo}}{\si_n} \KR^2
              + 2 \log \KL \frac{\si_n}{\si_n^{\rm min}} \KR \KKR
                                               + \chi^2_{\Mh}
                                               + \chi^2_{B_s}.
\label{eq:chi2}
\EE
Here $R_n^{\rm exp}$ denotes the experimental central value of the
$n$th observable ($\MW$, $\sweff$, $\Ga_Z$, \mbox{$(g-2)_\mu$} and
$\br(b \to s \ga)$, $\br(B_u \to \tau \nu_\tau$), $\De M_{B_s}$),
$R_n^{\rm theo}$ is the corresponding MSSM prediction and $\si_n$
denotes the combined error, as specified below. 
Additionally,
$\si_n^{\rm min}$ is the minimum combined error over the parameter space of
each data set as explained below, and
$\chi^2_{\Mh}$ and $\chi^2_{B_s}$ denote the $\chi^2$ contribution
coming from the experimental limits 
on the lightest MSSM Higgs boson mass and on $\br(B_s \to \mu^+\mu^-)$,
respectively, which are also described below.

We also list below the parametric uncertainties in the predictions on the
observables induced by the experimental uncertainty in the top-
and bottom-quark masses. These errors neglect, however, the effects of
varying $\mt$ and $\mb$ on the SUSY spectrum that are induced via the
RGE running. 
In order to take the $\mt$ and $\mb$ parametric uncertainties correctly into
account, we evaluate the SUSY spectrum and the observables for
each data point first for the nominal values 
$\mt = 171.4 \gev$~\cite{mt1714}%
\footnote{Using the most recent experimental value, 
$\mt = 170.9 \gev$~\cite{mt1709} would have a minor impact on our
analysis.}%
~and $\mb(\mb) = 4.25 \gev$,
then for $\mt = (171.4 + 1.0) \gev$ and $\mb(\mb) = 4.25 \gev$,
and finally for $\mt = 171.4 \gev$ and $\mb(\mb) = (4.25 + 0.1) \gev$.
The latter two evaluations are used by appropriate rescaling to estimate
the full parametric uncertainties induced by the experimental uncertainties
$\de\mt^{\rm exp} = 2.1 \gev$~\cite{mt1714}%
\footnote{Using the most recent experimental $\mt$ error of 
$1.8 \gev$~\cite{mt1709} would also have a minor impact on our analysis.}%
~and $\de\mb(\mb)^{\rm exp} = 0.11 \gev$.
These parametric uncertainties are then added to the other errors
(intrinsic, parametric, and experimental) of the observables as described in
the text below.

\smallskip
We preface our discussion by describing our treatment of the cosmological
cold dark matter density, which guides our subsequent analysis of the
EWPO and BPO within the CMSSM and NUHM.


\subsection{Cold Dark Matter Density}

Throughout this analysis, we focus our attention on parameter points
that yield the correct value of the cold dark matter density inferred
from WMAP and other data, namely 
$0.094 < \Omega_{\rm CDM} h^2 < 0.129$~\cite{WMAP}. 
The fact that the density is relatively well known
restricts the SUSY parameter space to a thin, fuzzy `WMAP hypersurface',
effectively reducing its dimensionality by one. The variations in the
EWPO and BPO across this hypersurface may in general be neglected, so
that we may 
treat the cold dark matter constraint effectively as a $\delta$
function. For example, in the CMSSM we focus our attention on `WMAP
lines' in the \plane{m_{1/2}}{m_0}s for discrete values of the other
SUSY parameters $\tb$ and $A_0$~\cite{WMAPstrips,wmapothers}. 
Correspondingly, in the following, for 
each value of $m_{1/2}$, we present theoretical values for the EWPO and
BPO corresponding to the values of $m_0$ on WMAP strips. 

We note, however, that for any given value of $m_{1/2}$ there may be
more than one value of $m_0$ that yields a cold dark matter density
within the allowed range, implying that there may be more than one WMAP
line traversing the the $(m_{1/2}, m_0)$ plane.
 Specifically, in the CMSSM there is, in general, one WMAP line in the
coannihilation/rapid-annihilation funnel region and another in the
focus-point region, at higher $m_0$. Consequently, each EWPO and BPO may
have more than one value for any given value of $m_{1/2}$. In the
following, we restrict our study of the upper WMAP line to the part with
$m_0 < 2000 \gev$ for $\tb = 10$ and $m_0 < 3000 \gev$ for 
$\tb = 50$, restricting in turn the range of $m_{1/2}$. 

The NUHM, with $\MA$ and $\mu$, has two more parameters than the CMSSM,
which characterize the degrees of non-universality of the two Higgs
masses. The WMAP lines therefore should, in principle, be generalized to
three-volumes in the higher-dimensional NUHM parameter space where the
cold dark matter density remains within the WMAP range. We prefer here
to focus our attention on `WMAP surfaces' that
are slices through these
three-volumes with specific fixed values for (combinations of) the other
NUHM parameters. These WMAP surfaces are introduced in more detail in
the subsequent section describing our NUHM analysis, and will be
discussed in more detail in~\citere{EHHOW}. 

In regions that depend sensitively on the input values of $\mt$ and
$\mb(\mb)$, such as the focus-point region~\cite{focus} in the CMSSM, the
corresponding parametric uncertainty can become very large. In essence, the
`WMAP hypersurface' moves significantly as $\mt$ varies (and to a lesser extent
also $\mb(\mb)$), but remains thin. Incorporating this large
parametric uncertainty naively in \refeq{eq:chi2} would artificially
suppress the
overall $\chi^2$ value for such points. This artificial suppression is
avoided by adding the second term in \refeq{eq:chi2}, where 
$\si_n^{\rm min}$ is the value of the combined error evaluated for
parameter choices which minimize $\chi^2_n$ over the full data set.


\subsection{The $W$~Boson Mass}
\label{subsec:mw}

The $W$~boson mass can be evaluated from
\BE
\MW^2 \KL 1 - \frac{\MW^2}{\MZ^2}\right) = 
\frac{\pi \al}{\sqrt{2} \GF} \left(1 + \De r\KR ,
\label{eq:delr}
\EE
where $\al$ is the fine structure constant and $\GF$ the Fermi constant.
The radiative corrections are summarized 
in the quantity $\De r$~\cite{sirlin}.
The prediction for $\MW$ within the SM
or the MSSM is obtained by evaluating $\De r$ in these models and
solving \refeq{eq:delr} for $\MW$. 

We use the most precise available result for $\MW$ in the
MSSM~\cite{MWweber}. Besides the full SM result, for the MSSM it
includes the full set of one-loop
contributions~\cite{deltarMSSM1lA,deltarMSSM1lB,MWweber}   
as well as the corrections of \order{\al\als}~\cite{dr2lA} and of
\order{\al_{t,b}^2}~\cite{drMSSMal2B,drMSSMal2} to the quantity
$\De\rho$; see~\citere{MWweber} for details.

The remaining intrinsic theoretical uncertainty in the prediction for
$\MW$ within the MSSM is still significantly larger than in the SM. For
realistic parameters it has been estimated as~\cite{drMSSMal2}
\BE
\De\MW^{\rm intr,current} \lsim 10 \mev~,
\EE
depending on the mass scale of the supersymmetric particles.
The parametric uncertainties are dominated by the experimental error of
the top-quark mass
and the hadronic contribution to the shift in the
fine structure constant. Their current errors induce the following
parametric uncertainties~\cite{ZOweber}
\BEA
\de\mt^{\rm current} = 2.1 \gev &\Rightarrow&
\De\MW^{{\rm para},\mt, {\rm current}} \approx 13 \mev,  \\[.3em]
\de(\De\al_{\rm had}^{\rm current}) = 35 \times 10^{-5} &\Rightarrow&
\De\MW^{{\rm para},\De\al_{\rm had}, {\rm current}} \approx 6.3 \mev~.
\EEA
The present experimental value of $\MW$ 
is~\cite{lepewwg,LEPEWWG,TEVEWWG,MWcdf,MWworld}
\BE
\MW^{\rm exp,current} = 80.398 \pm 0.025 \gev.
\label{mwexp}
\EE
We add the experimental and theoretical errors for $\MW$ 
in quadrature in our analysis.

\begin{figure}[htb!]
\begin{center}
\includegraphics[width=.48\textwidth]{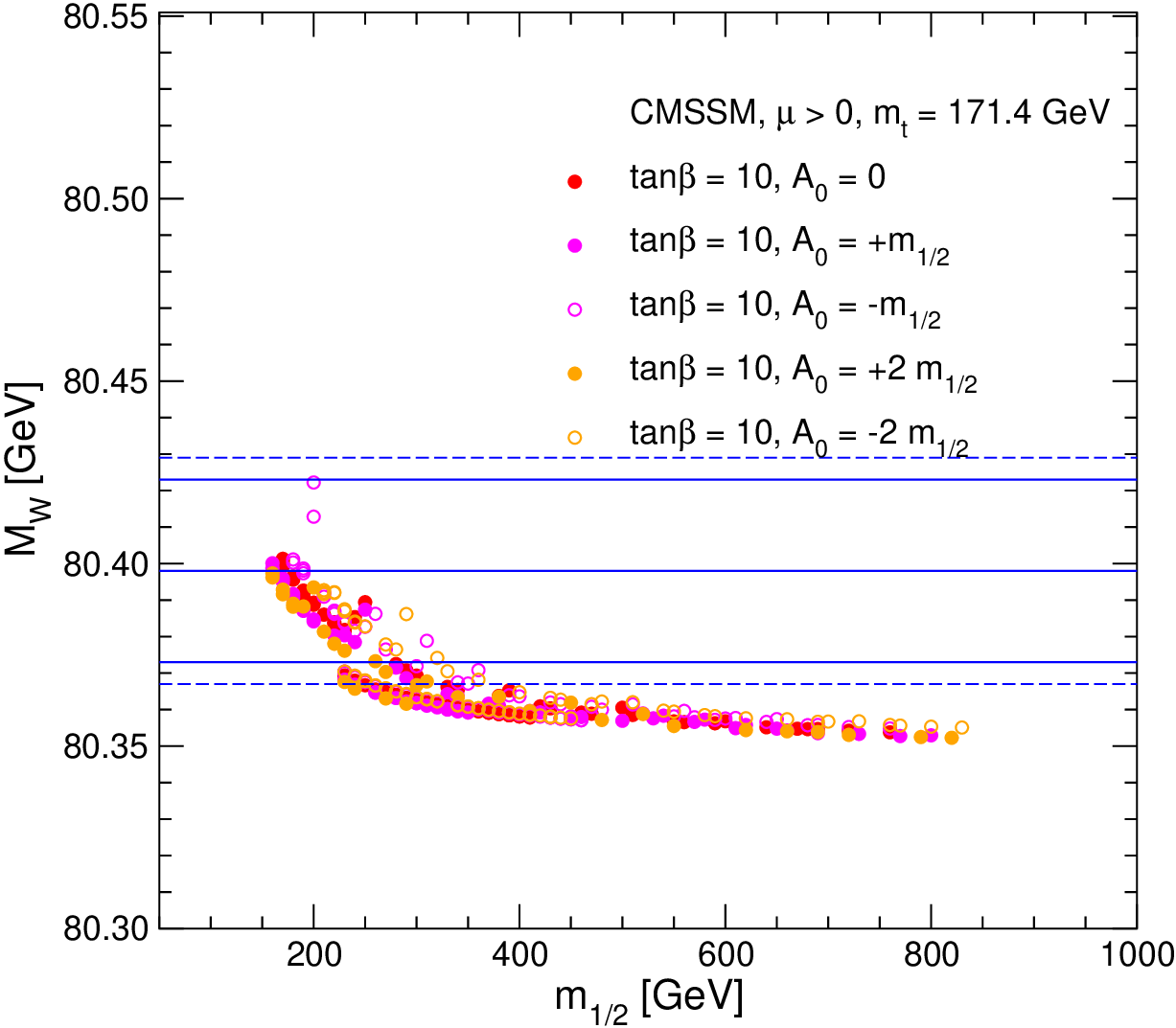}
\includegraphics[width=.48\textwidth]{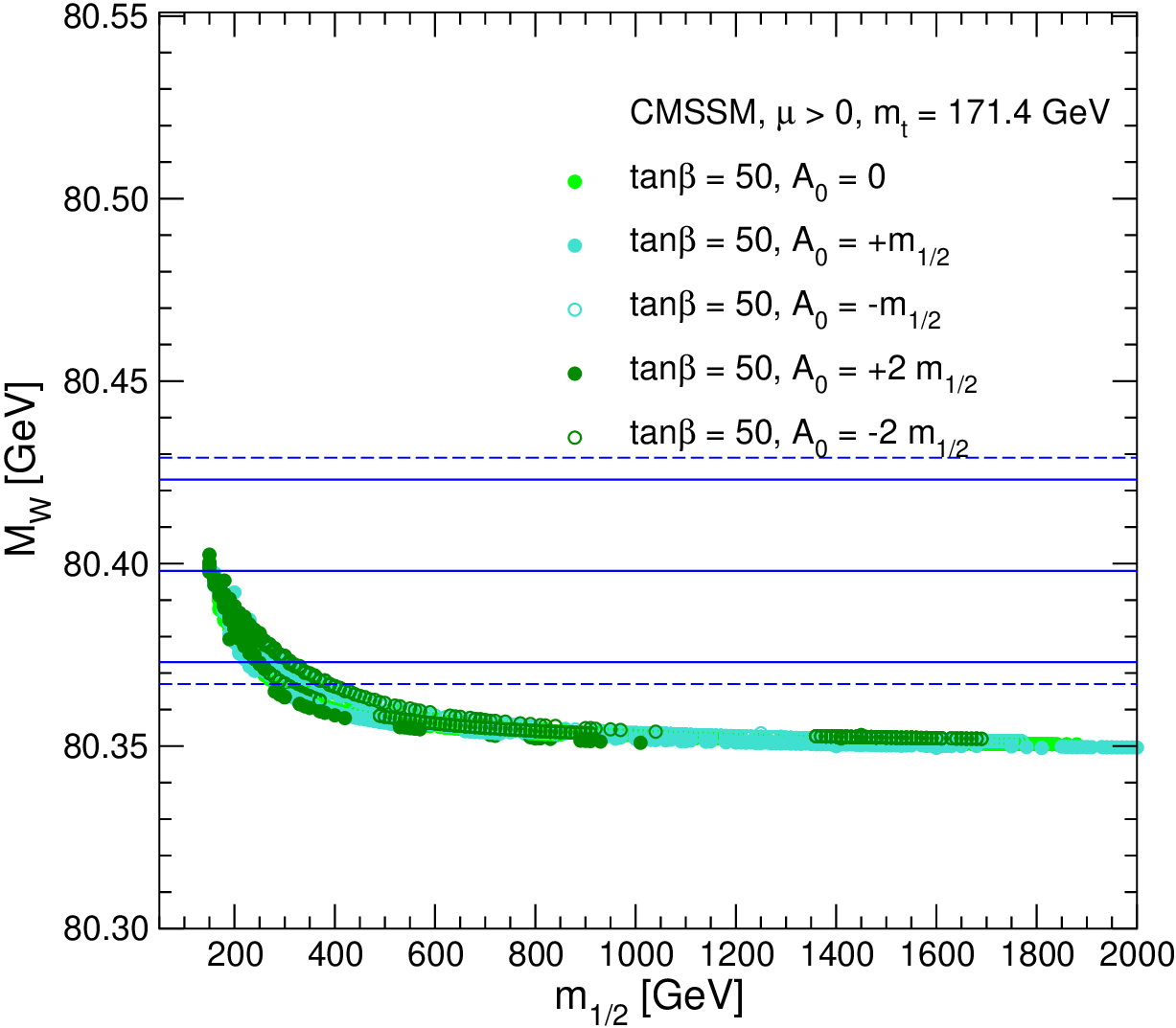}
\caption{%
The CMSSM predictions for $\MW$ are shown as functions of $m_{1/2}$ along the 
WMAP strips for (a) $\tb = 10$ and (b) $\tb = 50$ for various $A_0$
values. In each panel, the centre (solid) line is the present central
experimental value, and the (solid) outer lines show the current $\pm
1$-$\sigma$ range. 
The dashed lines correspond to the full error including also parametric
and intrinsic uncertainties.
}
\label{fig:MW}
\end{center}
\end{figure}

The current status of the MSSM prediction and the experimental
resolution is shown in \reffi{fig:MW}. We note that the CMSSM
predictions for $\MW$ in the coannihilation and focus-point regions
are quite similar, and depend little on $A_0$. We also see that
small values of $m_{1/2}$ are slightly preferred, reflecting the
familiar fact that the experimental value of $\MW$ is currently
somewhat higher than the SM prediction.

\subsection{The Effective Leptonic Weak Mixing Angle}

The effective leptonic weak mixing angle at the $Z$~boson peak
can be written as
\BE
 \sweff = \frac{1}{4} \, \left( 1 - \re \frac{v_{\rm eff}}{a_{\rm eff}}  
\right) \ ,
\EE
where $v_{\rm eff}$ and $a_{\rm eff}$ 
denote the effective vector and axial couplings
of the $Z$~boson to charged leptons.
We use the most precise available result for $\sweff$ in the
MSSM~\cite{ZOweber}. The prediction contains the same classes of
higher-order corrections as described in \refse{subsec:mw}.

In the MSSM with real parameters, the remaining intrinsic theoretical
uncertainty in the prediction  for $\sweff$ has been estimated
as~\cite{drMSSMal2} 
\BE
\De\sweff^{\rm intr,current} \lsim 7 \times 10^{-5}, 
\EE
depending on the SUSY mass scale.
The current experimental errors of $\mt$ and $\De\al_{\rm had}$
induce the following parametric uncertainties~\cite{ZOweber}
\BEA
\de\mt^{\rm current} = 2.1 \gev &\Rightarrow&
\De\sweff^{{\rm para},\mt, {\rm current}} \approx 6.3 \times 10^{-5}, \\[.3em]
\de(\De\al_{\rm had}^{\rm current}) = 35 \times 10^{-5} &\Rightarrow&
\De\sweff^{{\rm para},\De\al_{\rm had}, {\rm current}} \approx 
12 \times 10^{-5} .
\EEA
The experimental value is~\cite{lepewwg,LEPEWWG}
\BE
\sweff^{\rm exp,current} = 0.23153 \pm 0.00016~.
\label{swfit}
\EE
We add the experimental and theoretical errors for $\sweff$ 
in quadrature in our analysis.

\begin{figure}[htb!]
\begin{center}
\includegraphics[width=.48\textwidth]{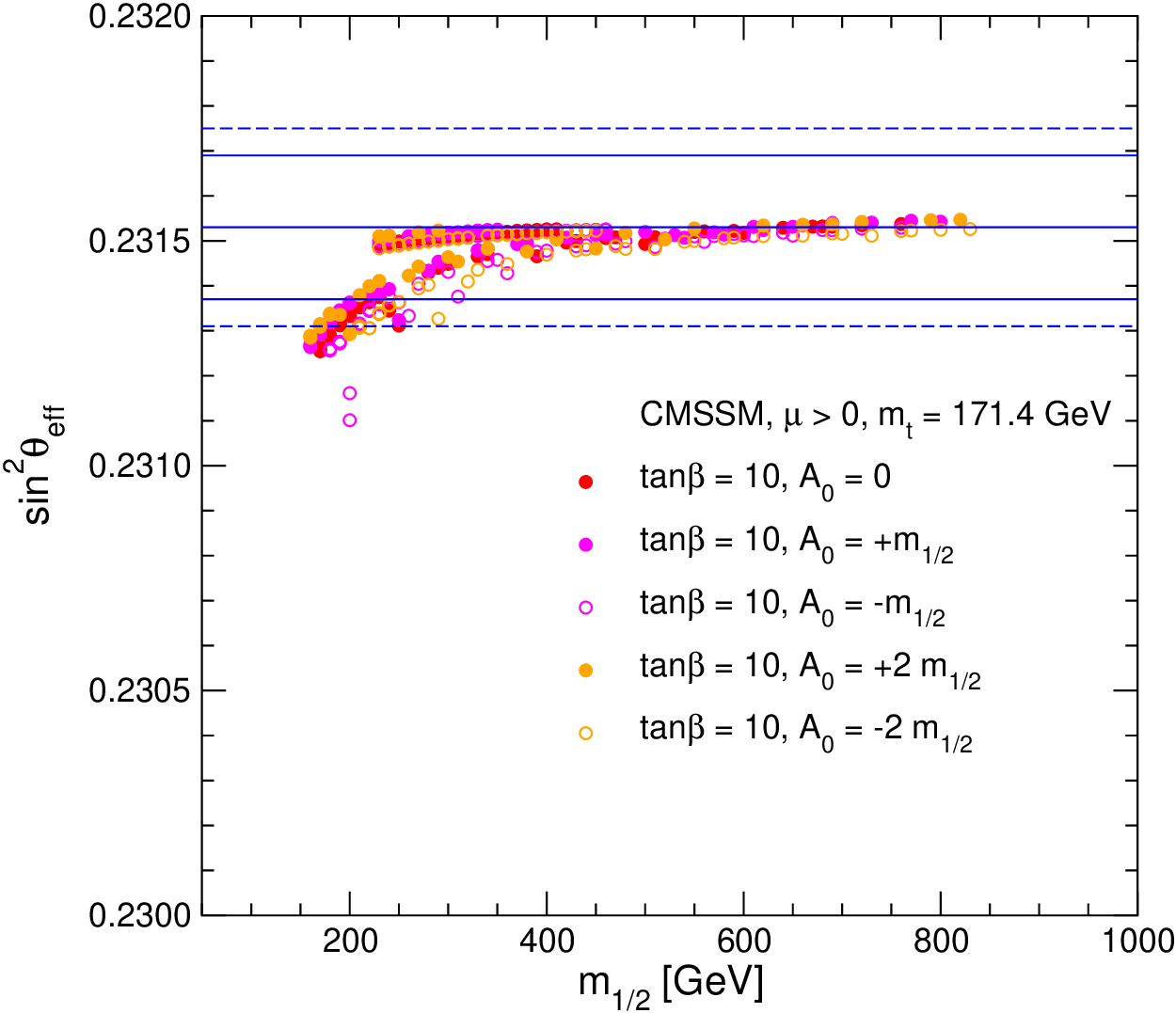}
\includegraphics[width=.48\textwidth]{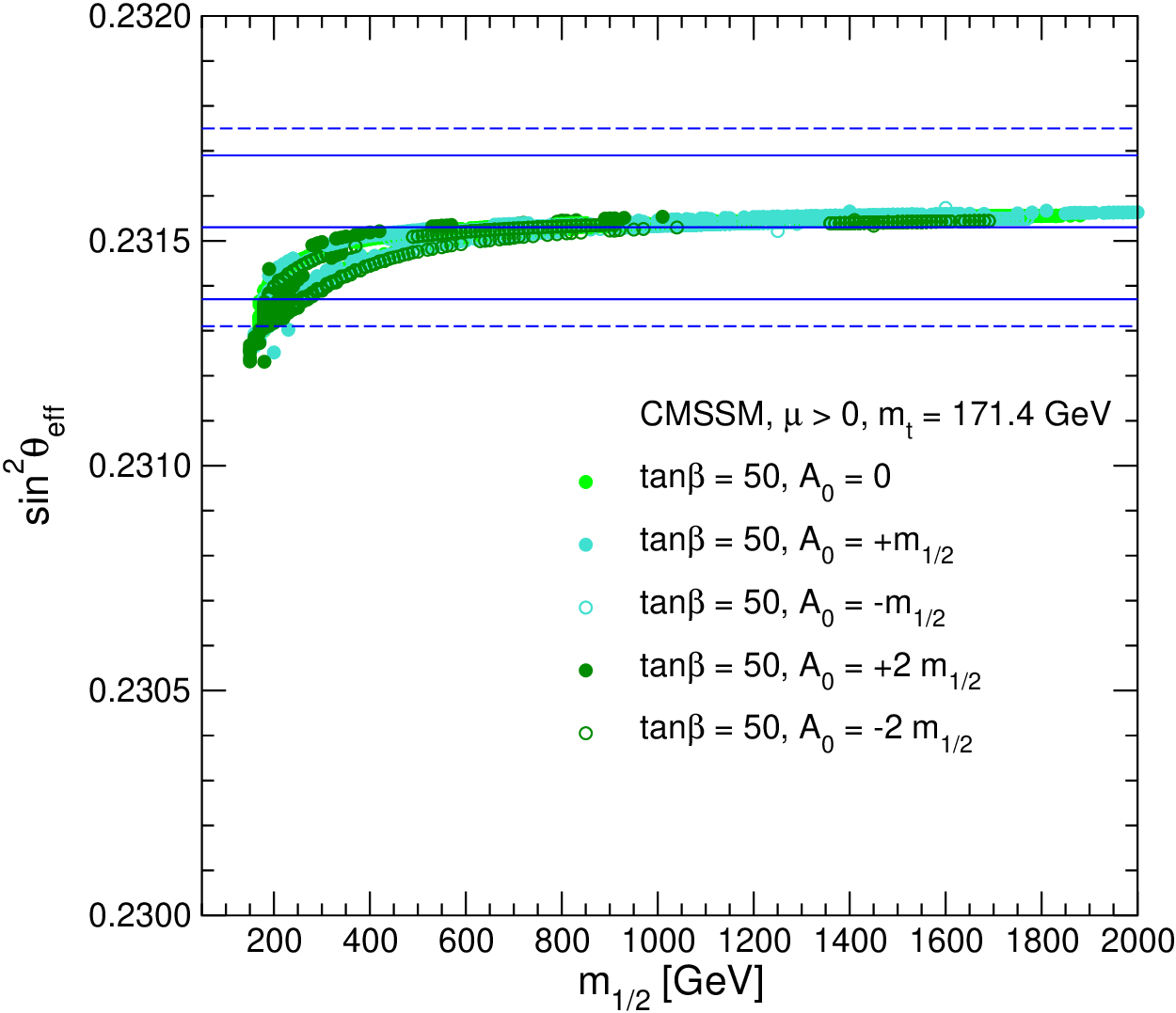}
\caption{%
The CMSSM predictions for $\sweff$ as functions of $m_{1/2}$ along the 
WMAP strips for (a) $\tb = 10$ and (b) $\tb = 50$ for various $A_0$
values. In each panel, the  centre (solid) line is the present central
experimental value, and the (solid) outer lines show the current $\pm
1$-$\sigma$ range. 
The dashed lines correspond to the full error including also parametric
and intrinsic uncertainties.
}
\label{fig:SW}
\end{center}
\end{figure}

As compared with our older analyses~\cite{ehow3,ehow4} we now use a
new result for $\sweff$, obtained recently, that differs non-negligibly
from that used previously, due to the inclusion of more higher-order
corrections (which also result in a smaller intrinsic error). The
corresponding new results in the CMSSM are shown in \reffi{fig:SW} for
$\tb = 10$ (left) and $\tb = 50$ (right) as functions of
$m_{1/2}$. Whereas previously the agreement with the experimental result
was best for $m_{1/2} \approx 300 \gev$, we now find best agreement for
large $m_{1/2}$ values. However, taking all uncertainties into account, the
deviation for $m_{1/2}$ generally stays below the level of one sigma.
We note that the predictions for $\sweff$ in the
coannihilation and focus-point regions are somewhat different.\\

\subsection{The Total \boldmath{$Z$}~Boson Decay Width}

The total $Z$~boson decay width, $\Ga_Z$, is given by 
\BE
\Ga_Z = \Ga_l + \Ga_h + \Ga_{\neu{1}}~,
\EE
where $\Ga_{l,h}$ are the rates for decays into SM leptons and quarks,
respectively, and
$\Ga_{\neu{1}}$ denotes the decay width to the lightest neutralino.
We have checked that, for the parameters analyzed in this paper, always
$\Ga_{\neu{1}} = 0$. However, SUSY particles enter via virtual
corrections to $\Ga_l$ and $\Ga_h$. 
We use the most precise available result for $\Ga_Z$ in the
MSSM~\cite{ZOweber}. The prediction contains the same classes of
MSSM 
higher-order corrections as described in \refse{subsec:mw}.

So far no estimate has been made of the intrinsic uncertainty in the
prediction for $\Ga_Z$ in the MSSM. Following the numerical analysis
in~\citere{ZOweber}, we use a conservative value of
\BE
\De\Ga_Z^{\rm intr,current} \lsim 1.0 \mev
\EE
The current experimental errors of $\mt$ and
$\De\al_{\rm had}$ induce the following parametric uncertainties~\cite{ZOweber}
\BEA
\de\mt^{\rm current} = 2.1 \gev &\Rightarrow&
\De\Ga_Z^{{\rm para},\mt, {\rm current}} \approx 0.51 \mev,  \\[.3em]
\de(\De\al_{\rm had}^{\rm current}) = 35 \times 10^{-5} &\Rightarrow&
\De\Ga_Z^{{\rm para},\De\al_{\rm had}, {\rm current}} \approx 
0.32 \mev .
\EEA
The experimental value is~\cite{lepewwg,LEPEWWG}
\BE
\Ga_Z^{\rm exp,current} = 2495.2 \pm 2.3 \mev~.
\EE
We add the experimental and theoretical errors for $\Ga_Z$  in
quadrature in our analysis. 

\begin{figure}[htb!]
\begin{center}
\includegraphics[width=.48\textwidth]{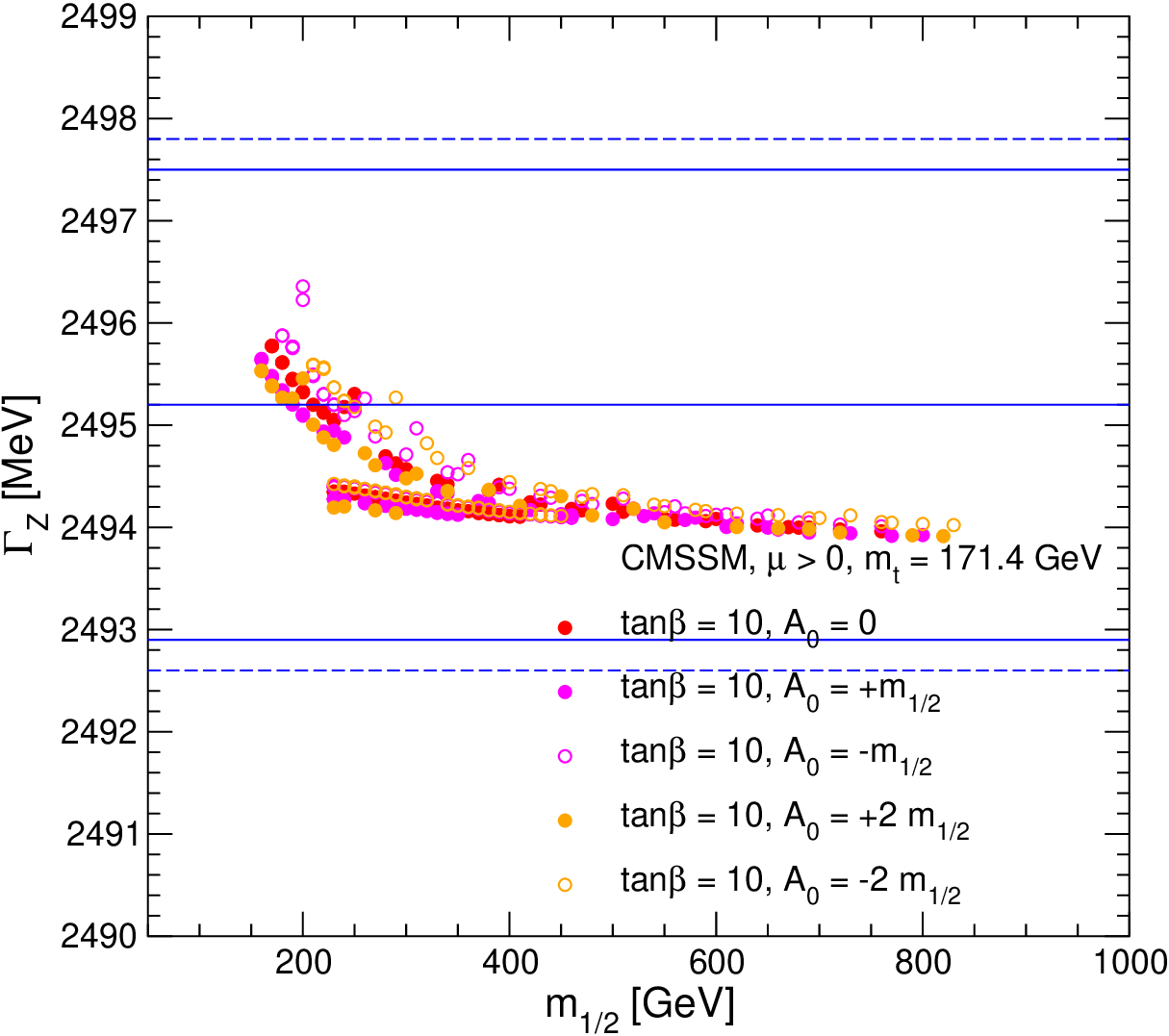}
\includegraphics[width=.48\textwidth]{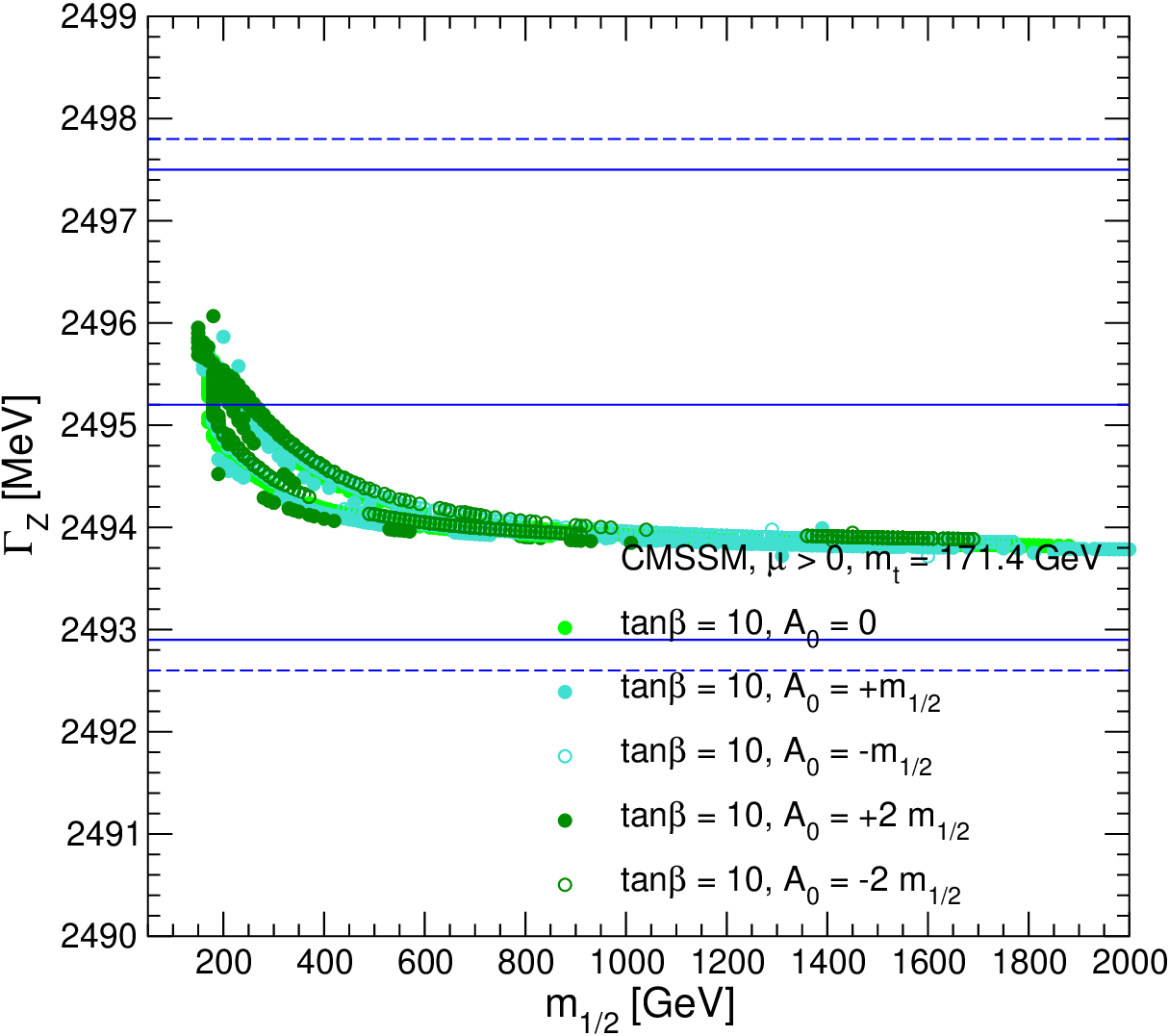}
\caption{%
The CMSSM predictions for $\Ga_Z$ as functions of $m_{1/2}$ along the 
WMAP strips for (a) $\tb = 10$ and (b) $\tb = 50$ for various $A_0$
values. In each panel, the  centre (solid) line is the present central
experimental value, and the (solid) outer lines show the current $\pm
1$-$\sigma$ range. 
The dashed lines correspond to the full error including also parametric
and intrinsic uncertainties.
}
\label{fig:GZ}
\end{center}
\end{figure}

A comparison of the MSSM prediction with the experimental value is shown
in \reffi{fig:GZ}. We see that the experimental value is within 
$\sim 1/2$ a standard deviation of the CMSSM value at large $m_{1/2}$,
which corresponds to the SM value with the same Higgs boson mass.
The marginal improvement in the CMSSM
prediction at small $m_{1/2}$ is not significant. We note that the
predictions for $\Ga_Z$ in the coannihilation and focus-point regions are
somewhat different.

\subsection{The Anomalous Magnetic Moment of the Muon}

The SM prediction for the anomalous magnetic moment of 
the muon 
(see~\citeres{g-2review,g-2review2,g-2reviewDS,g-2reviewMRR,g-2reviewFJ}
for reviews) depends on the evaluation of QED contributions (see
\citeres{Kinoshita,g-2QEDmassdep} for recent updates), the
hadronic vacuum polarization and light-by-light (LBL) contributions. The
former  have been evaluated
in~\citeres{DEHZ,g-2HMNT,g-2HMNT2,Jegerlehner,g-2reviewFJ,Yndurain,DDDD} 
and the latter in~\citeres{LBLwrongsign1,LBLwrongsign2,LBL,LBLnew,LBLnew2}. 
The evaluations of the 
hadronic vacuum polarization contributions using $e^+ e^-$ and $\tau$ 
decay data give somewhat different results. In view of the fact that
recent $e^+ e^-$ measurements tend to confirm earlier results, whereas
the correspondence between previous $\tau$ data and preliminary data
from BELLE is not so clear, and also in view of the additional
uncertainties associated with  
the isospin transformation from $\tau$ decay, we use here the latest
estimate based on $e^+e^-$ data~\cite{DDDD}:
\BE
\amutheo = 
(11\, 659\, 180.5 \pm 4.4_{\rm had} \pm 3.5_{\rm LBL} \pm 0.2_{\rm QED+EW})
 \times 10^{-10},
\label{eq:amutheo}
\EE
where the source of each error is labeled. We note that the new $e^+e^-$
data sets that have recently been published in~\citeres{KLOE,CMD2,SND} have
been partially included in the updated estimate of $(g - 2)_\mu$. 

The SM prediction is to be compared with
the final result of the Brookhaven $(g-2)_\mu$ experiment 
E821~\cite{g-2exp,g-2exp2}, namely:
\BE
\amuexp = (11\, 659\, 208.0 \pm 6.3) \times 10^{-10},
\label{eq:amuexp}
\EE
leading to an estimated discrepancy~\cite{DDDD,g-2SEtalk}
\BE
\amuexp-\amutheo = (27.5 \pm 8.4) \times 10^{-10},
\label{delamu}
\EE
equivalent to a 3.3-$\sigma$ effect%
\footnote{Three other recent evaluations yield slightly different
  numbers~\cite{g-2reviewFJ,g-2HMNT2,g-2reviewMRR}, 
  but similar discrepancies with the SM prediction.}%
.~While it would be premature to regard this deviation as a firm
evidence for new physics, within the context of SUSY, it does indicate a
preference for a non-zero contribution. 

Concerning the MSSM contribution, the complete one-loop
result was evaluated a decade ago~\cite{g-2MSSMf1l}. 
In view of the correlation between the signs of $(g - 2)_\mu$ and of
$\mu$~\cite{correlation}, variants of the MSSM with
$\mu < 0$ are already severely challenged by the
present data on $\amu$, whether one uses either the $e^+ e^-$ or 
$\tau$ decay data, so we restrict our attention in this paper to
models with $\mu > 0$. 
In addition to the full one-loop contributions, the leading QED
two-loop corrections have also been
evaluated~\cite{g-2MSSMlog2l}. Further corrections at the two-loop
level have been obtained recently~\cite{g-2FSf,g-2CNH}, 
leading to corrections to the one-loop result that are $\lsim 10\%$. These
corrections are taken into account in our analysis according to the
approximate formulae given in~\citeres{g-2FSf,g-2CNH}.

\begin{figure}[htb!]
\begin{center}
\includegraphics[width=.48\textwidth]{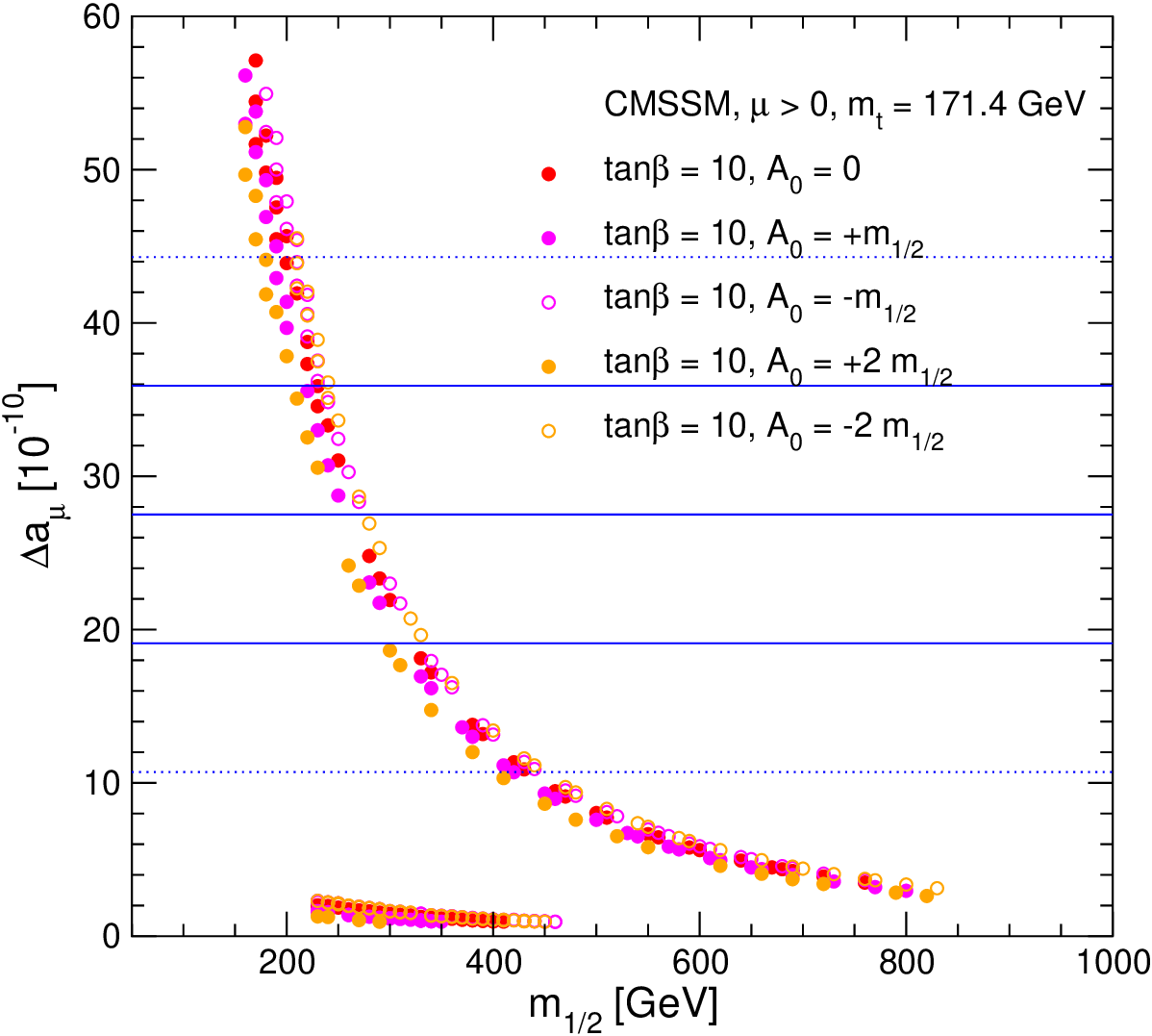}
\includegraphics[width=.48\textwidth]{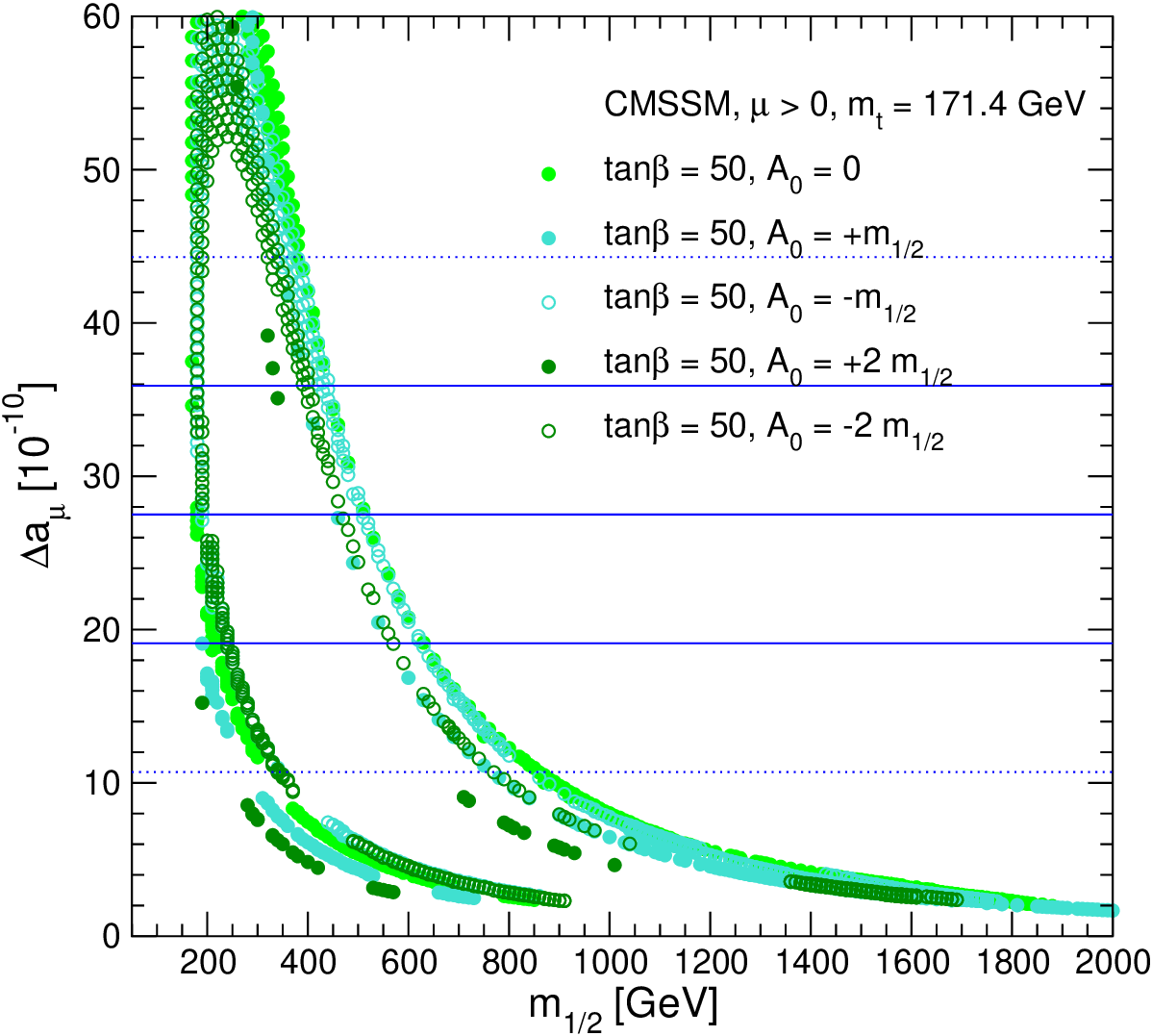}
\caption{%
The CMSSM predictions for $(g-2)_\mu$, $\De\amu$, as functions of
$m_{1/2}$ along the  WMAP strips for (a) $\tb = 10$ and (b) $\tb = 50$
for various $A_0$ values. In each panel, the  centre (solid) line is the
present central experimental value, and the solid (dotted) outer lines
show the current $\pm 1 (2)$-$\sigma$ ranges.
}
\label{fig:AMU}
\end{center}
\end{figure}

The current status of the CMSSM prediction and the experimental
resolution is shown in \reffi{fig:AMU}, where the 1- and 2-$\si$ bands
are shown. We note that the coannihilation and focus-point region predictions
for $\amu$ are quite different. For $\tb = 10$, the focus-point
prediction agrees less well with the data, whereas for $\tb = 50$
the focus-point prediction does agree well in a limited range of
$m_{1/2} \sim 200 \gev$. 

\subsection{The Mass of the Lightest MSSM Higgs Boson}

The mass of the lightest $\cp$-even MSSM Higgs boson can be predicted in 
terms of the other MSSM parameters. At the tree level, the two
$\cp$-even Higgs  boson masses are obtained as functions of $\MZ$, the
$\cp$-odd Higgs boson mass $\MA$, and $\tb$, whereas other parameters enter
into the loop corrections. 
We employ the Feynman-diagrammatic method for the theoretical prediction
of $\Mh$, using the code 
{\tt FeynHiggs}~\cite{feynhiggs,mhiggslong,mhcMSSMlong},
which includes all numerically relevant known higher-order corrections.
The status of these results 
can be summarized as follows. For the
one-loop part, the complete result within the MSSM is 
known~\cite{ERZ,mhiggsf1lB,mhiggsf1lC}. 
Computation of the two-loop
effects is quite advanced: see~\citere{mhiggsAEC} and
references therein. These include the strong corrections
at \order{\al_t\als} and Yukawa corrections at \order{\al_t^2}
to the dominant one-loop \order{\al_t} term, and the strong
corrections from the bottom/sbottom sector at \order{\al_b\als}. 
In the case of the $b/\Sbot$~sector
corrections, an all-order resummation of the $\Tb$-enhanced terms,
\order{\al_b(\als\tb)^n}, is also known~\cite{deltamb,deltamb1}.
Most recently, the \order{\al_t \al_b} and \order{\al_b^2} corrections
have been derived~\cite{mhiggsEP5}~%
\footnote{
A two-loop effective potential calculation has been presented 
in~\citere{fullEP2l}, including now even the leading three-loop
corrections~\cite{mhiggs3l}, but no public code based on this result
is currently available.
}%
. The current intrinsic error of $\Mh$ due to
unknown higher-order corrections has been estimated to 
be~\cite{mhiggsAEC,mhiggsFDalbals,PomssmRep,mhiggsWN}
\BE
\De\Mh^{\rm intr,current} = 3 \gev~,
\EE
which we interpret effectively as a $\sim 95~\%$ confidence level limit:
see below.

It should be noted that, for the unconstrained MSSM with small values
of $\MA$ and values of $\tb$ 
which 
are not too small, a significant
suppression of the $hZZ$ coupling can occur compared to the SM value, in
which case the experimental lower bound on $\Mh$ may be more than 20~GeV
below the 
SM value~\cite{LEPHiggsMSSM}. However, we have checked that within the 
CMSSM and the other models studied in this paper, the $hZZ$ coupling is
always very close to the SM value. Accordingly, the
bounds from the SM Higgs search at LEP~\cite{LEPHiggsSM}
can be taken over directly (see e.g.\ \citeres{asbs1,ehow1}).

Concerning the $\chi^2$ analysis, we 
use the complete likelihood information available from LEP.
Accordingly, we evaluate as follows the $\Mh$ contribution to the 
overall $\chi^2$
function~%
\footnote{
We thank P.~Bechtle and K.~Desch for detailed discussions and
explanations.
}%
. Our starting points are the $CL_s(\Mh)$ values provided by the 
final LEP results on the SM Higgs boson search, see Fig.~9 
in~\citere{LEPHiggsSM}~%
\footnote{
We thank A.~Read for providing us with the $CL_s$ values.
}%
. We obtain by inversion from  $CL_s(\Mh)$
the corresponding value of ${\tilde \chi}^2(\Mh)$ determined from~\citere{PDG}
\BE
\edz {\rm erfc}(\sqrt{\edz\tilde\chi^2(\Mh)}) \equiv CL_s(\Mh)~,
\EE
and note the fact that $CL_s(\Mh = 116.4 \gev) = 0.5$ implies that
$\tilde\chi^2(116.4 \gev) = 0$ as is appropriate for a one-sided
limit. Correspondingly we set $\tilde\chi^2(\Mh > 116.4 \gev) = 0$.
The theoretical uncertainty is included by convolving the likelihood function
associated with
$\tilde\chi^2(\Mh)$ and a Gaussian function, $\tilde\Phi_{1.5}(x)$,
normalized to unity and centered around $\Mh$, whose width is $1.5 \gev$:
\BE
\chi^2(\Mh) = -2 \log\KL
\int_{-\infty}^{\infty} 
    e^{-\tilde\chi^2(x)/2} \; \tilde\Phi_{1.5}(\Mh - x) \, {\rm d}x \KR~.
\EE
In this way, a theoretical uncertainty of up to $3 \gev$ is assigned for 
$\sim 95\%$ of
all $\Mh$ values corresponding to one parameter point. 
The final $\chi^2_{\Mh}$ is then obtained as 
\BEA
\chi^2_{\Mh} = \chi^2(\Mh) - \chi^2(116.4 \gev) &{\rm ~for~}& 
                                                \Mh \le 116.4 \gev~, \\
\chi^2_{\Mh} = 0 &{\rm ~for~}& \Mh > 116.4 \gev ~,
\EEA
and is then combined with the corresponding quantities for the other 
observables we consider, see \refeq{eq:chi2}. 

\begin{figure}[htb!]
\begin{center}
\includegraphics[width=.48\textwidth]{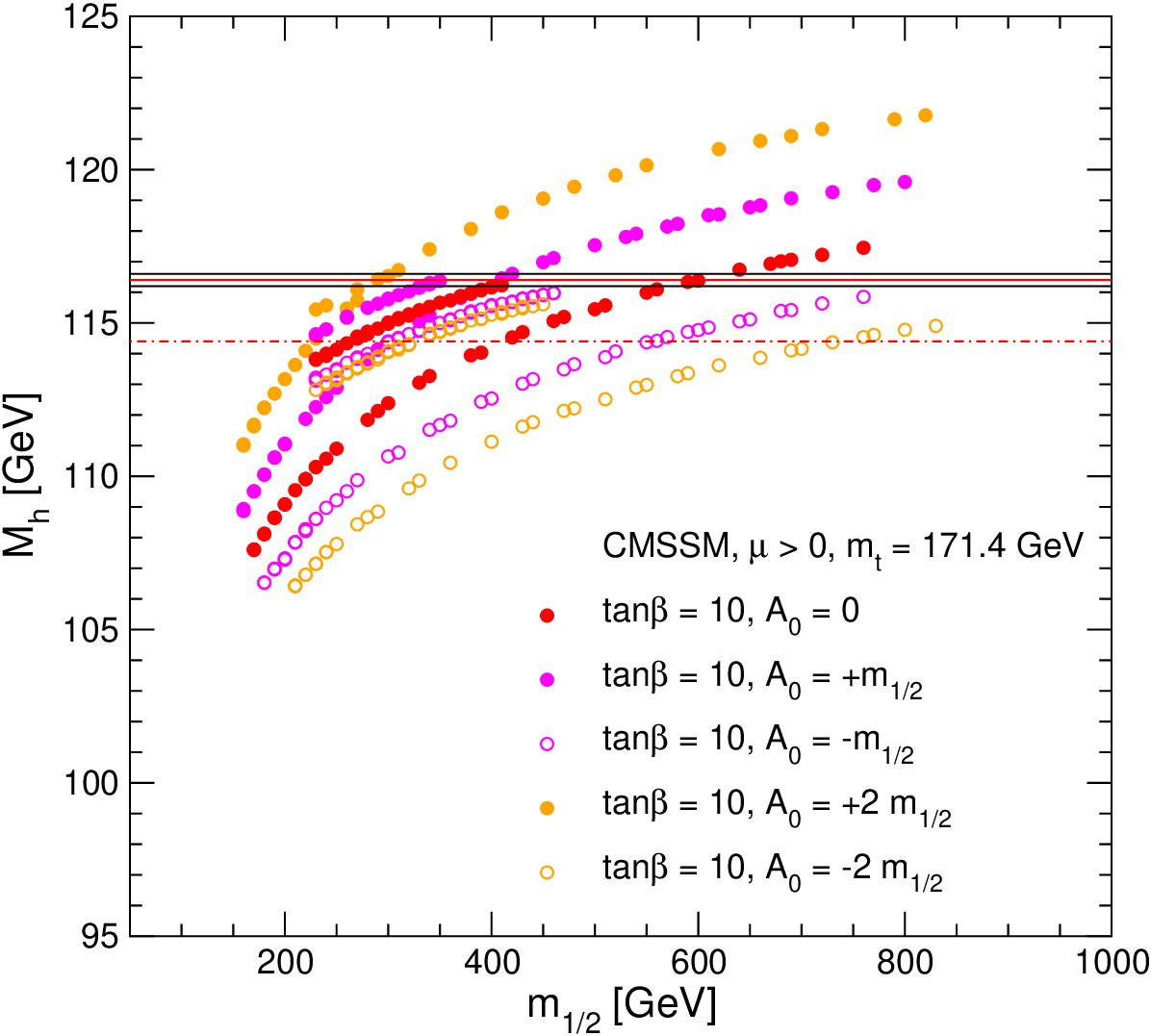}
\includegraphics[width=.48\textwidth]{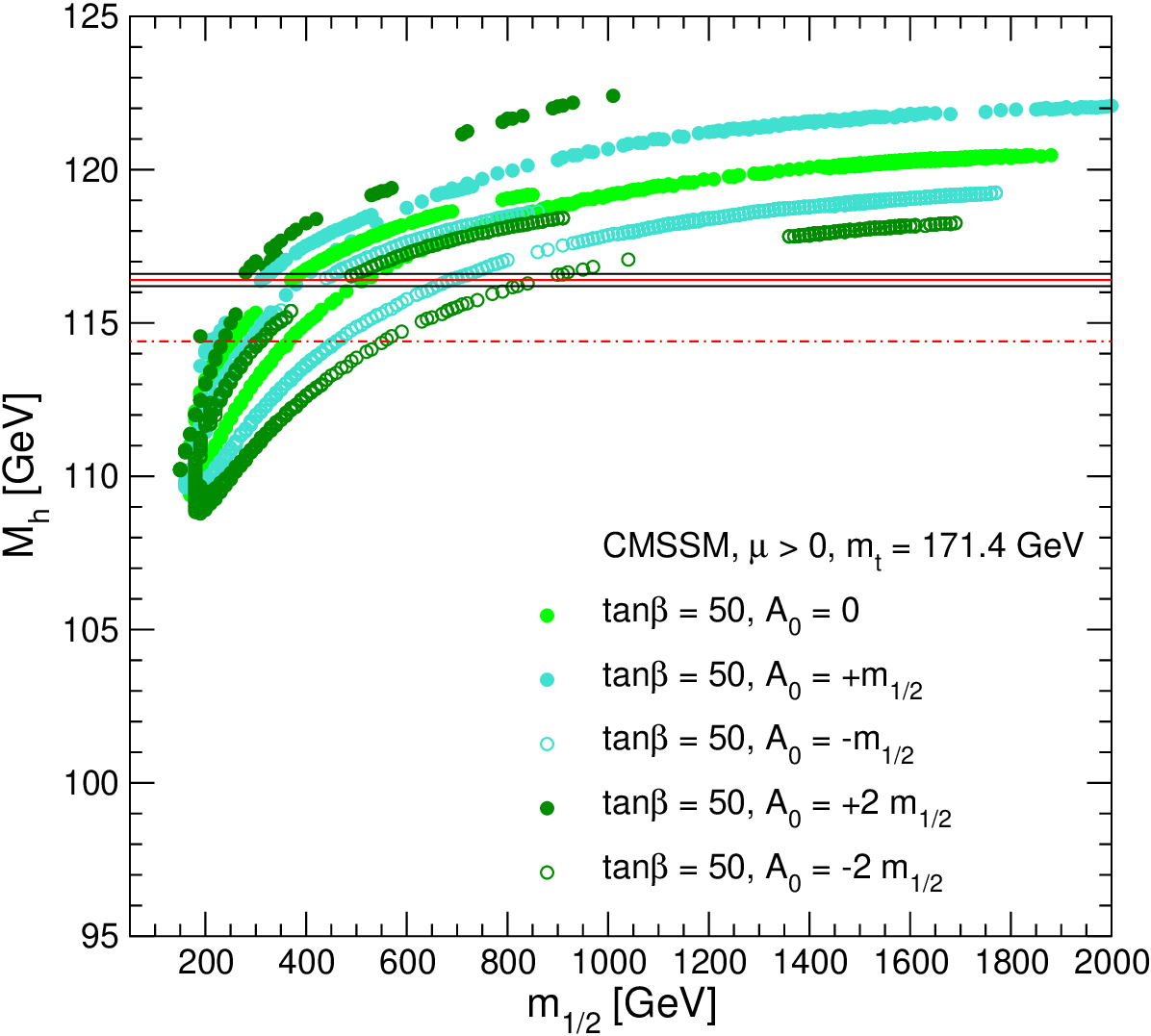}
\caption{
The CMSSM predictions for $\Mh$ as functions of $m_{1/2}$ with 
(a) $\tb = 10$ and (b) $\tb = 50$ for various $A_0$. 
We also show the present 95\%~C.L.\ exclusion limit of $114.4 \gev$ and
a hypothetical LHC measurement of $\Mh = 116.4 \pm 0.2 \gev$.
}
\label{fig:Mh}
\end{center}
\vspace{-1em}
\end{figure}

We show in \reffi{fig:Mh} the predictions for $\Mh$ in the CMSSM for  
$\tb = 10$ (left) and $\tb = 50$ (right).
The predicted values of $\Mh$ are similar in the coannihilation
and focus-point regions. They depend significantly on
$A_0$, particularly in the coannihilation region, where
negative values of $A_0$ tend to predict very low values of $\Mh$ that are 
disfavoured by the LEP direct search. 
Also shown in \reffi{fig:Mh} is the present nominal 95~\%~C.L.\ exclusion
limit for a SM-like Higgs boson, namely $114.4 \gev$~\cite{LEPHiggsSM}, and a 
hypothetical LHC measurement of $\Mh = 116.4 \pm 0.2\gev$. We recall that we
use the numerical value of the LEP Higgs likelihood function in our combined
analysis.

\subsection{The decay $b \to s \ga$}

Since this decay occurs at the loop level in the SM, the MSSM 
contribution might {\it a priori} be of similar magnitude. A
recent
theoretical estimate of the SM contribution to the branching ratio at
the NNLO QCD level is~\cite{bsgtheonew}
\BE
\br( b \to s \ga ) = (3.15 \pm 0.23) \times 10^{-4}~.
\label{bsga}
\EE
We record that the error estimate for $\br(b \to s \ga)$ is still under
debate, and that other SM contributions to $b \to s \ga$ have been
calculated~\citeres{hulupo,bsgneubert}, but these corrections are 
small compared with the theoretical uncertainty quoted in (\ref{bsga}).

For comparison, the present experimental 
value estimated by the Heavy Flavour Averaging Group (HFAG)
is~\cite{bsgexp,hfag}
\BE
\br( b \to s \ga ) = (3.55 \pm 0.24 {}^{+0.09}_{-0.10} \pm 0.03) \times 10^{-4},
\label{bsgaexp}
\EE
where the first error is the combined statistical and uncorrelated systematic 
uncertainty, the latter two errors are correlated systematic theoretical uncertainties
 and corrections respectively. 

Our numerical results have been derived with the 
$\br(b \to s \ga)$ evaluation provided in \citeres{bsgGH,ali,ali2},
incorporating also the latest SM corrections provided in~\citere{bsgtheonew}. 
The calculation has been checked against other
approaches~\cite{bsgMicro,bsgKO1,bsgKO2}.
For the current theoretical intrinsic uncertainty of the MSSM prediction for 
$\br(b \to s \ga)$ we use the SM uncertainty given in \refeq{bsga} and
add linearly the intrinsic MSSM corrections 
$0.15 \times 10^{-4}$~\cite{bsgKO1,bsgKO2} and the last two errors
given by HFAG of $\simeq 0.13 \times 10^{-4}$~\cite{hfag}. The full intrinsic
error is then added  linearly to the sum in quadrature of the
experimental error given by HFAG as $0.24$
and the parametric error. 

\begin{figure}[htb!]
\begin{center}
\includegraphics[width=.48\textwidth]{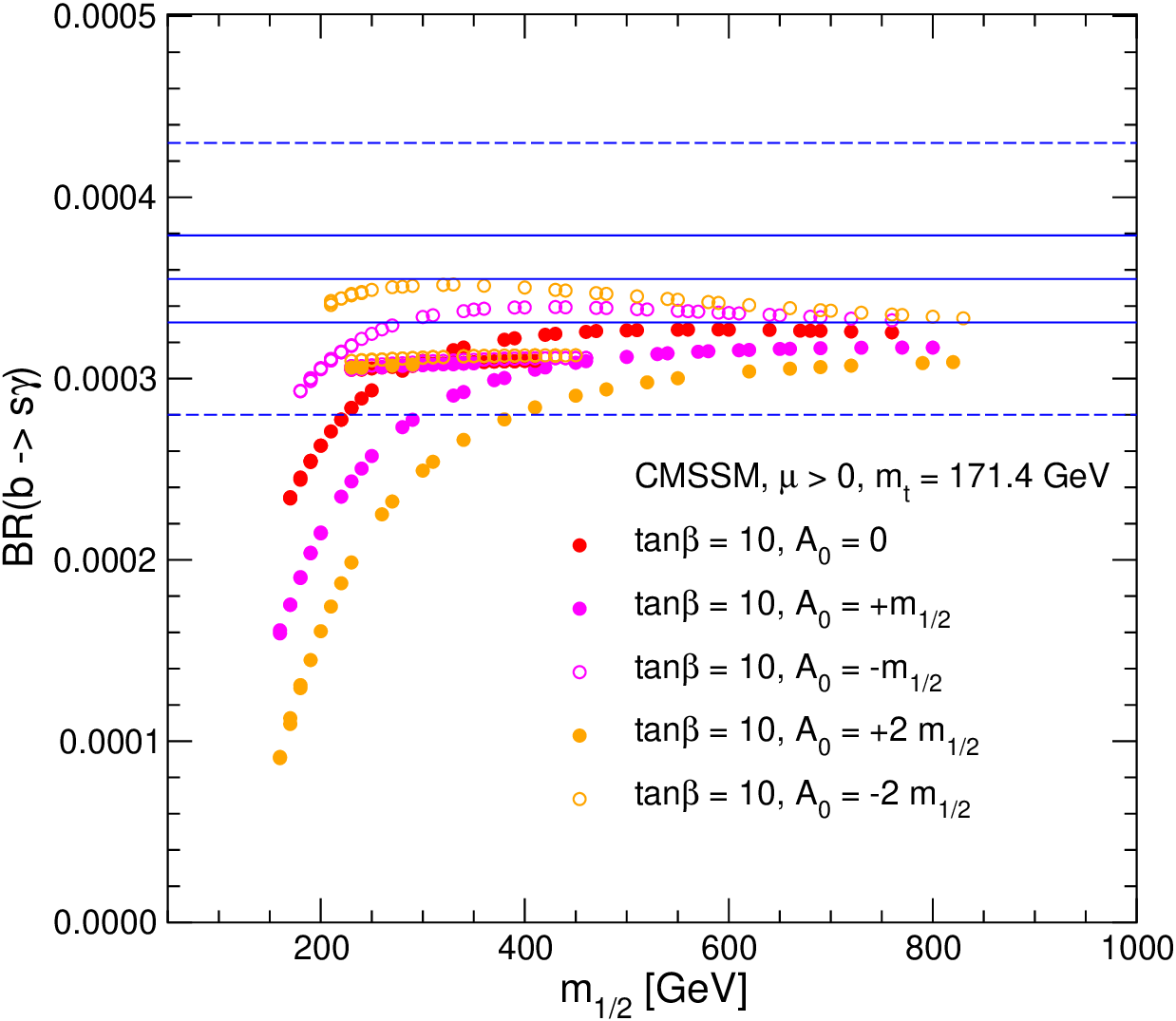}
\includegraphics[width=.48\textwidth]{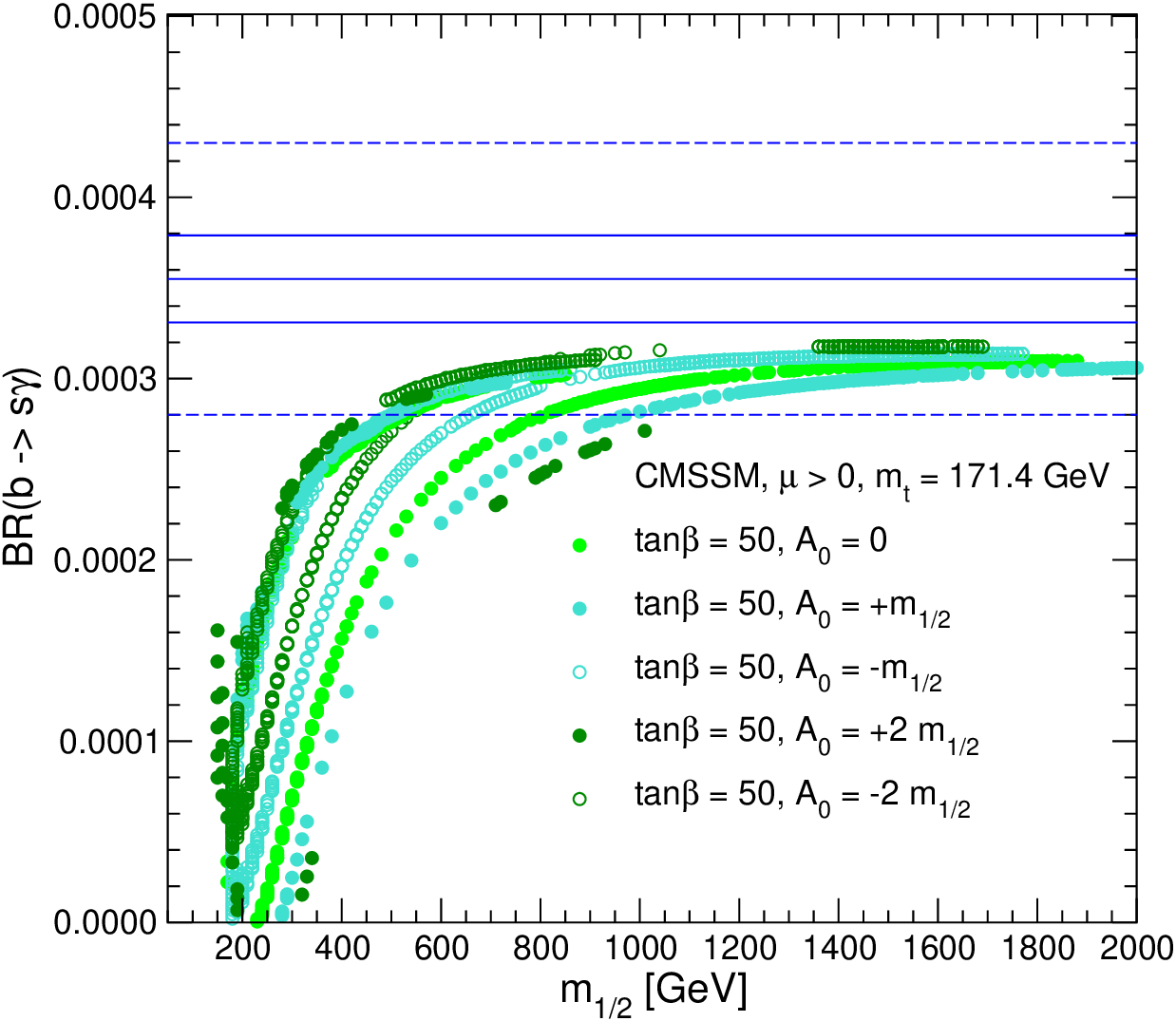}
\caption{%
The CMSSM predictions for $\br(b \to s \ga)$ as functions of $m_{1/2}$ 
along the WMAP strips for 
$\tb = 10$ (left) and $\tb = 50$ (right) and various choices of $A_0$. 
The central (solid) line indicates the current experimental
central value, and the other solid lines show the current 
$\pm 1$-$\si$ experimental range. The dashed line is the $\pm 1$-$\si$ error
including also the full intrinsic error (see text).
}
\label{fig:BSG}
\end{center}
\vspace{-1em}
\end{figure}

In \reffi{fig:BSG} we show the predictions in the CMSSM for 
$\br(b \to s \ga)$ for $\tb = 10, 50$ as functions of $m_{1/2}$,
compared with the 1-$\si$ experimental error (full line) and the
full error (dashed line, but assuming a negligible parametric error).
For $\tb = 10$, we see that positive values of $A_0$ are disfavoured
at small $m_{1/2}$, and that small values of $m_{1/2}$ are
disfavoured for all the studied values of $A_0$ if $\tb = 50$.

\subsection{The Branching Ratio for $B_s \to \mu^+\mu^-$}
\label{subsec:bsmm}

The SM prediction for this branching ratio is 
$(3.4 \pm 0.5) \times 10^{-9}$~\cite{bsmmtheosm}, and 
the present experimental upper limit from the Fermilab Tevatron collider
is $1.0 \times 10^{-7}$ at the $95\%$ C.L.~\cite{bsmmexp}, providing ample 
room for the MSSM to dominate the SM contribution. The current Tevatron
sensitivity is based on an integrated luminosity of about 780~\ipb\
collected at CDF. The exclusion bounds can be translated into a
$\chi^2$~function for each value of $\br(B_s \to \mu^+ \mu^-)$~%
\footnote{
We thank C.-J.~Stephen and M.~Herndon for providing the
$\chi^2$~numbers. A slightly more stringent upper limit of
$0.93 \times 10^{-7}$ at the $95\%$ C.L. has been announced more
recently by the D0~Collaboration~\cite{bsmmD0}. However, the
corresponding $\chi^2$~function is not available to us. Since the
difference to the result employed here is small, we expect only a minor
impact on our analysis.
}:%
\BE
\tilde\chi^2(B_s) \equiv \chi^2(\br(B_s \to \mu^+ \mu^-))~,
\EE
with $\tilde\chi^2(\br(B_s \to \mu+\mu-) < 0.266 \times 10^{-7}) = 0$.
The theory uncertainty is included by convolving the likelihood function
associated with
$\tilde\chi^2(B_s)$ and a Gaussian function, $\tilde\Phi_{\rm th}(x)$,
normalized to unity and centered around $\br(B_s \to \mu^+\mu^-)$, 
whose width is given by the theory uncertainty, see below. Consequently,
\BE
\label{eq:chibmm}
\chi^2(B_s) = -2 \log\KL
\int_{-\infty}^{\infty} 
    e^{-\tilde\chi^2(x)/2} \; 
    \tilde\Phi_{\rm th}(\br(B_s \to \mu^+\mu^-) - x) \, {\rm d}x \KR~.
\EE
The final $\chi^2_{B_s}$ is then obtained as 
\BEA
\chi^2_{B_s} = \chi^2(B_s) - \chi^2(0.266 \times 10^{-7}) &{\rm ~for~}& 
               \br(B_s \to \mu^+\mu^-) \ge 0.266 \times 10^{-7}~, \\
\chi^2_{B_s} = 0 &{\rm ~for~}& 
               \br(B_s \to \mu^+\mu^-) < 0.266 \times 10^{-7}~.
\EEA
The Tevatron sensitivity is expected to improve significantly in the
future. The limit that could be reached at the end of Run~II is 
$\sim 2 \times 10^{-8}$ assuming 8~\ifb\ collected with each
detector~\cite{bsmmexpfuture}. 
A sensitivity even down to the SM value can be
expected at the LHC. Assuming the SM value, i.e.\
$\br(B_s \to \mu^+ \mu^-) \approx 3.4 \times 10^{-9}$, it has been
estimated~\cite{lhcb} that LHCb can observe 33~signal events
over 10~background events within 3~years of low-luminosity
running. Therefore this process offers good prospects for probing the MSSM.

For the theoretical prediction we use results from~\citere{bsmumu}, 
which are in good agreement with \citere{ourBmumu}. This calculation
includes the full one-loop evaluation and the leading two-loop 
QCD corrections. 

The theory error is estimated as follows.
We take into account the parametric uncertainty induced by~\cite{fbstheo}
\BE
f_{B_s} = 230 \pm 30 \mev~.
\EE
The most important SUSY contribution to $\br(B_s \to \mu^+\mu^-)$ scales as
\BE
\label{eq:bsmmtheo}
\br(B_s \to \mu^+\mu^-) \sim \frac{f_{B_s}^2}{\MA^4}~.
\EE
In the models that predict the value of $\MA$ at the low-energy scale,
i.e.\ in our case the CMSSM, we additionally include the
parametric uncertainty due to the shift in $\MA$ in
\refeq{eq:bsmmtheo} that is induced by the experimental errors of
$\mt$ and $\mb$ in the RGE running~\cite{ourBmumu}.
These errors are added in quadrature. 
The intrinsic error is estimated to be negligible as compared to the
parametric error. Thus the parametric error constitutes our theory
error entering in \refeq{eq:chibmm}.

In \reffi{fig:BMM} the CMSSM predictions for $\br(B_s \to \mu^+\mu^-)$ 
for $\tb = 10, 50$ as functions of $m_{1/2}$ are compared with the
present Tevatron limit. For $\tb = 10$ (left plot) the CMSSM prediction
is significantly below the present and future Tevatron 
sensitivity. However, already with the current sensitivity, the Tevatron
starts to probe the CMSSM coannihilation region for $\tb = 50$ and 
$A_0 \ge 0$, whereas the CMSSM prediction in the focus-point region is
significantly below the current 
sensitivity.

\begin{figure}[htb!]
\begin{center}
\includegraphics[width=.48\textwidth]{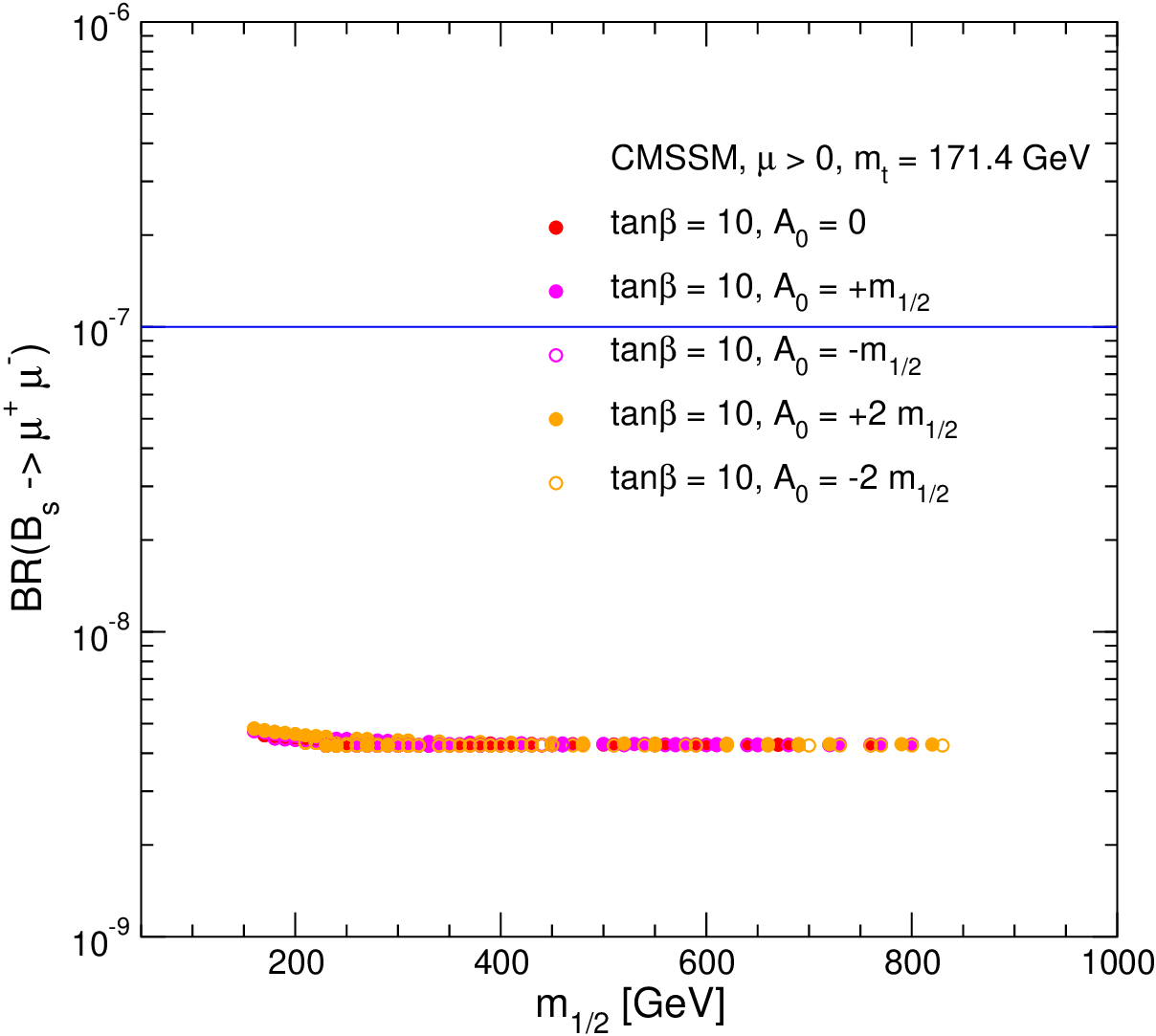}
\includegraphics[width=.48\textwidth]{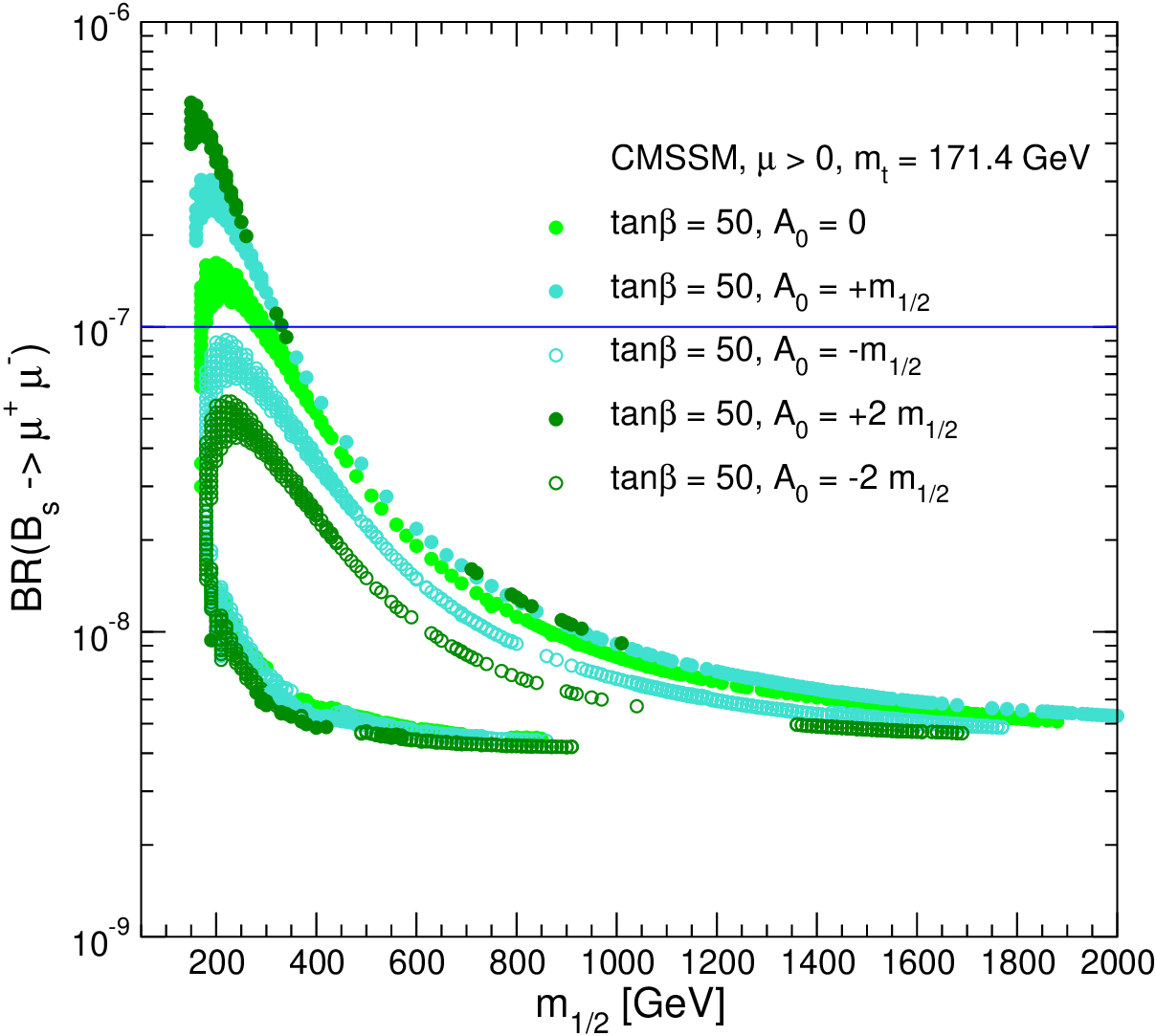}
\caption{%
The CMSSM predictions for $\br(B_s \to \mu^+\mu^-)$ as functions of 
$m_{1/2}$  along the WMAP strips for 
$\tb = 10$ (left) and $\tb = 50$ (right) and various choices of $A_0$. 
The solid line indicates the current experimental 95\%~C.L. exclusion bound.
}
\label{fig:BMM}
\end{center}
\end{figure}


\subsection{The Branching Ratio for $B_u \to \tau \nu_\tau$}
\label{subsec:btn}

The decay $B_u \to \tau \nu_\tau$ has recently been
observed by BELLE~\cite{btnexp}, and the experimental world average is given
by~\cite{btnexp,btnexp2,LSPlargeTB}
\BE
\br(B_u \to \tau \nu_\tau)_{\rm exp} = (1.31 \pm  0.49) \; \times 10^{-4}~.
\EE
We follow \citere{BPOtheo} for the theoretical evaluation of this
decay. The main new contribution within the MSSM comes from the
direct-exchange of a virtual charged 
Higgs boson decaying into $\tau \nu_\tau$. Taking
into account the resummation of the leading $\tb$~enhanced
corrections, within scenarios with minimal flavor violation such as the
CMSSM and the NUHM, the ratio of the MSSM
result over the SM result can be written as
\BE
\frac{\br(B_u \to \tau \nu_\tau)_{\rm MSSM}}
     {\br(B_u \to \tau \nu_\tau)_{\rm SM}}   =
\KKL 1 - \KL \frac{m_{B_u}^2}{\MHp^2} \KR
     \frac{\TQb}{1 + \eps_0 \tb} \KKR^2~.
\label{eq:btntheo}
\EE
Here $\eps_0$ denotes the effective coupling of the charged Higgs
boson to up- and down-type quarks, see \citere{BPOtheo} for details. 
The deviation of the experimental result from the SM prediction can be
expressed as 
\BE
\frac{\br(B_u \to \tau \nu_\tau)_{\rm epx}}
     {\br(B_u \to \tau \nu_\tau)_{\rm SM}}   = 0.93 \pm 0.41~,
\label{eq:btnexp}
\EE
where the error includes the experimental error as well as the parametric
errors from the various SM inputs. 
We use \refeq{eq:btntheo} for our theory evaluation, which can then
be compared with \refeq{eq:btnexp}, provided that the value of 
$\De M_{B_d}$ agrees sufficiently well in 
the SM and in the MSSM (which we assume here). As an error
estimate we use the combined experimental and parametric error from 
\refeq{eq:btnexp}, an estimated intrinsic error of $\sim 2\%$, and
in the CMSSM, as for $\br(B_s \to \mu^+\mu^-)$, an additional parametric
error from $\MHp$, evaluated from RGE running. These errors
have been added in quadrature.

We show in \reffi{fig:BTN} the theoretical results for the ratio of
CMSSM/SM for $\br(B_u \to \tau \nu_\tau)$ as functions of $m_{1/2}$ for
$\tb = 10, 50$. These results are also compared with the present
experimental result. The central (solid) line indicates the current
experimental central value, and the other solid (dotted)
lines show the current $\pm 1(2)$-$\sigma$ ranges from \refeq{eq:btnexp}.
For $\tb = 10$ the SM result is reproduced over most of the parameter
space. Only very small $m_{1/2}$ values give a ratio visibly smaller
than 1. For $\tb = 50$ the result varies strongly between~0 and~1, and the
CMSSM could easily account for the small deviation of the central value
of the experimental result from the SM prediction, should that become
necessary. The prediction in the focus-point region is somewhat closer
to the SM value. 

\begin{figure}[htb!]
\begin{center}
\includegraphics[width=.48\textwidth]{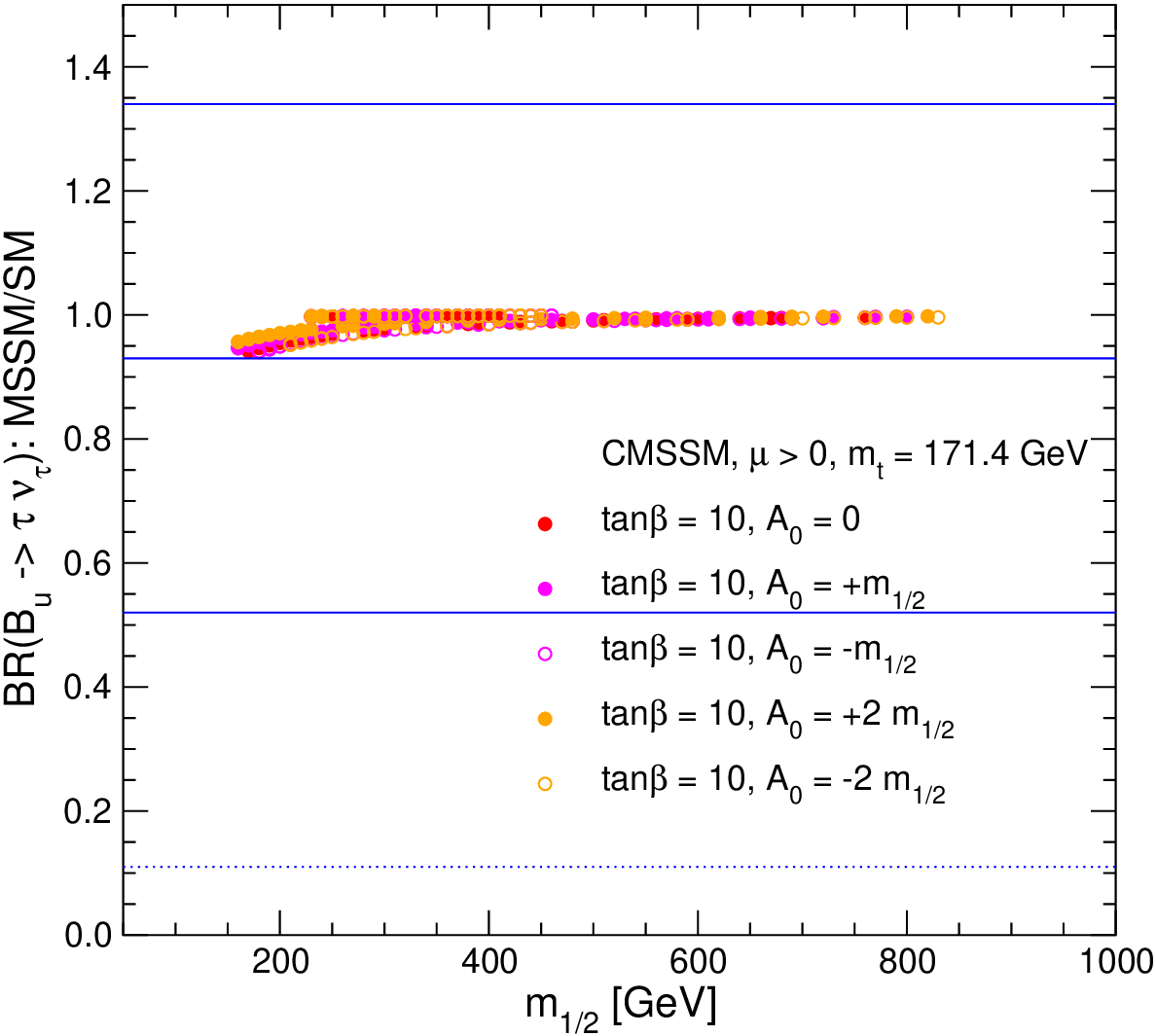}
\includegraphics[width=.48\textwidth]{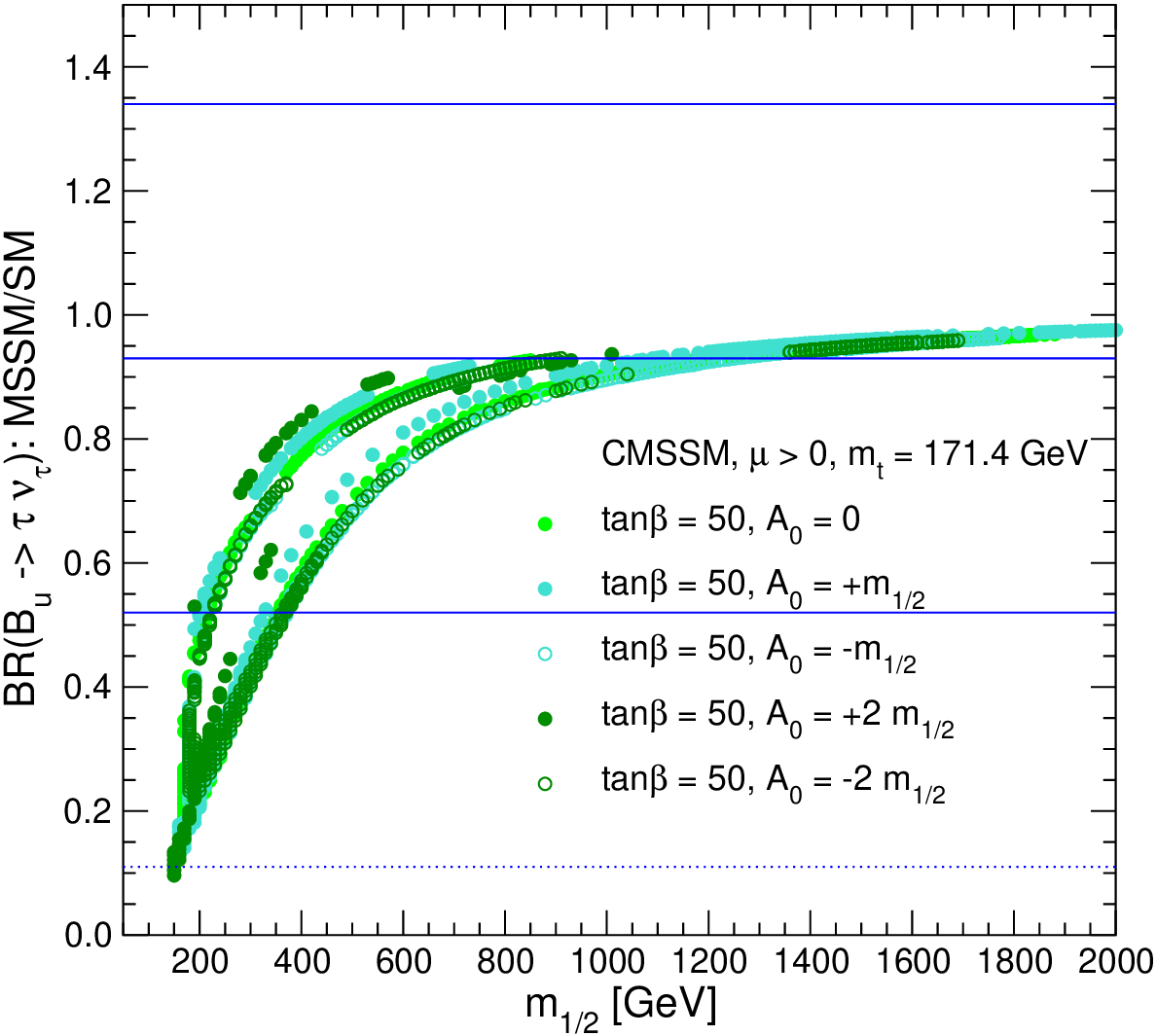}
\caption{%
The predictions for the ratio CMSSM/SM for $\br(B_u \to \tau \nu_\tau)$ 
as functions of  $m_{1/2}$  along the WMAP strips for 
$\tb = 10$ (left) and $\tb = 50$ (right) and various choices of $A_0$. 
The central (solid) line indicates the current experimental
central value, and the other solid (dotted)
lines show the current $\pm 1(2)$-$\sigma$ ranges.
}
\label{fig:BTN}
\end{center}
\end{figure}


\subsection{The $B_s$--$\bar B_s$ Mass Difference $\De M_{B_s}$}
\label{subsec:dms}

The $B_s$--$\bar B_s$ oscillation frequency and consequently the the
$B_s$--$\bar B_s$ mass difference has recently been measured by the
CDF Collaboration~\cite{dmsCDF}, 
\BE
(\De M_{B_s})_{\rm exp} = 17.77 \pm 0.12 {\rm ~ps}^{-1}~,
\EE
which is compatible with the broader range of the result from
D0~\cite{dmsD0}. 

We follow \citere{BPOtheo} for the theory evaluation. The main
MSSM contribution to the $B_s$--$\bar B_s$ oscillation comes from the
exchange of neutral Higgs bosons, but we use here the full result given
in \citere{BPOtheo} (taken from \citere{dmstheo}), where the leading
dependence is given as
\BE
1 - \frac{(\De M_{B_s})_{\rm MSSM}}{(\De M_{B_s})_{\rm SM}} \sim
  \frac{\mb(\mb) \, m_s(\mb)}{\MA^2}~.
\label{eq:dmstheo}
\EE
The SM value, obtained from a global fit, is given by~\cite{dmsUTfit}
\BE
(\De M_{B_s})_{\rm SM} = 19.0 \pm 2.4 {\rm ~ps}^{-1}~,
\EE
resulting in 
\BE
\frac{(\De M_{B_s})_{\rm exp}}{(\De M_{B_s})_{\rm SM}} = 0.93 \pm 0.13~.
\label{eq:dmsexp}
\EE
The error in \refeq{eq:dmsexp} is supplemented by the parametric
errors in \refeq{eq:dmstheo} from $m_s(\mb) = 93 \pm 17 \mev$ and, 
in the case of the CMSSM, as for $\br(B_s \to \mu^+\mu^-)$, an
additional parametric error from $\MA$. These errors are added in
quadrature. The intrinsic error, in comparison, is assumed to be negligible.

\begin{figure}[tbh!]
\begin{center}
\includegraphics[width=.48\textwidth]{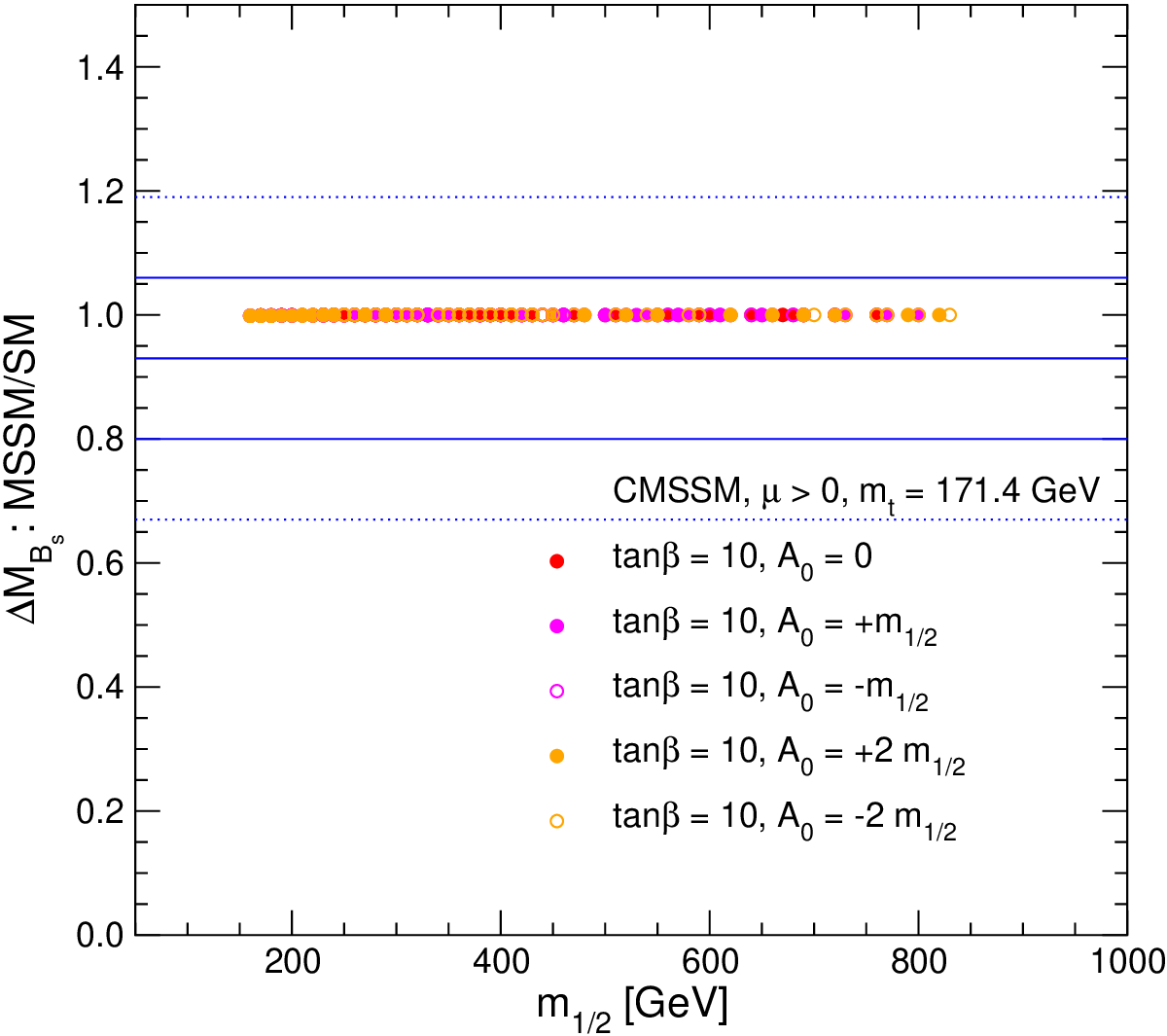}
\includegraphics[width=.48\textwidth]{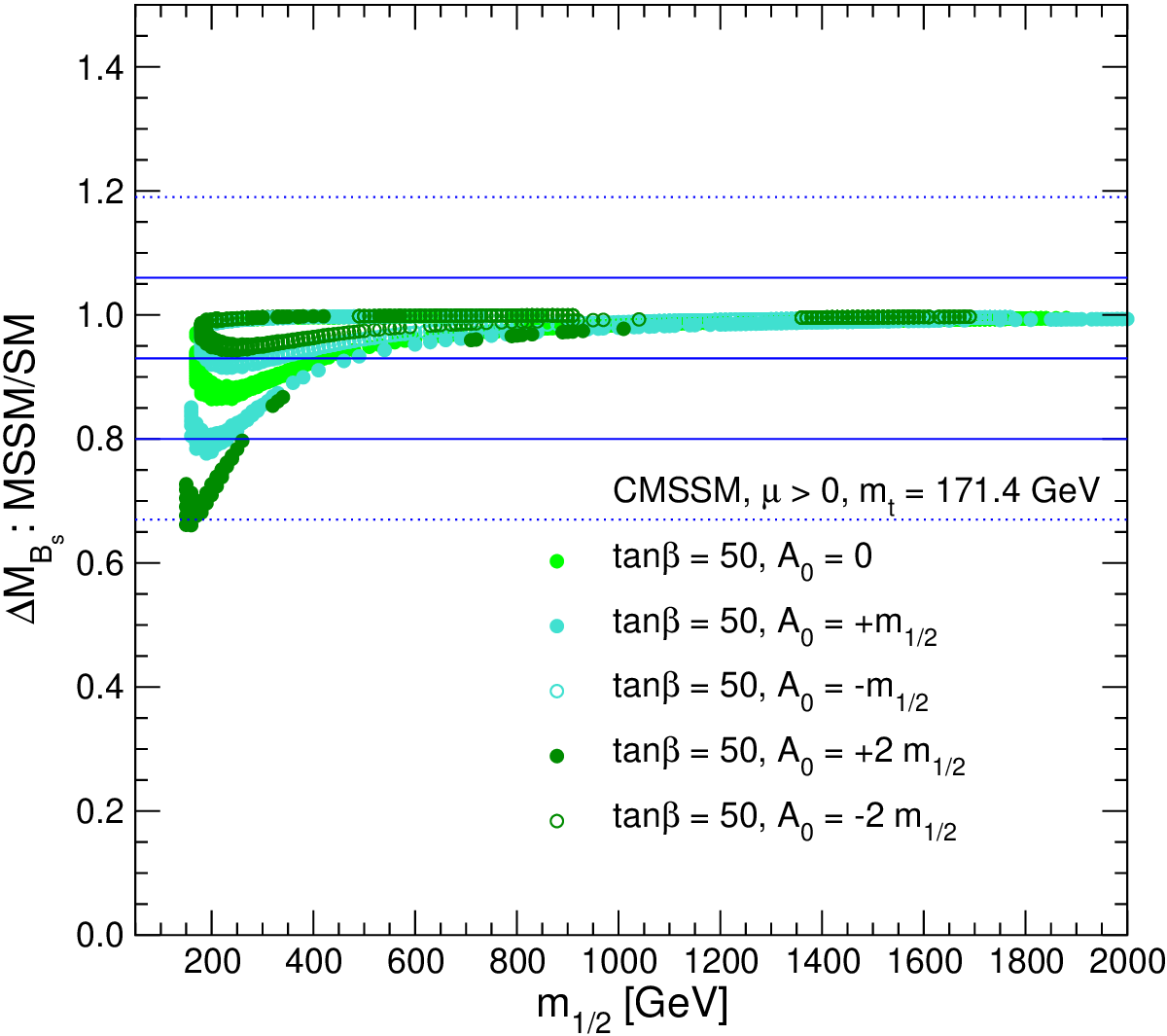}
\caption{%
The predictions for the ratio CMSSM/SM for $\De M_{B_s}$ as
functions of  $m_{1/2}$  along the WMAP strips for 
$\tb = 10$ (left) and $\tb = 50$ (right) and various choices of $A_0$. 
The central (solid) line indicates the current experimental
central value, and the other solid (dotted)
lines show the current $\pm 1(2)$-$\sigma$ ranges.
}
\label{fig:DMS}
\end{center}
\vspace{-1em}
\end{figure}

In \reffi{fig:DMS} we show the results for the ratio of CMSSM/SM for 
$\De M_{B_s}$ as functions of $m_{1/2}$ for $\tb = 10, 50$.
These are also compared with the present experimental result. 
The central (solid) line indicates the current experimental
central value, and the other solid (dotted)
lines show the current $\pm 1(2)$-$\sigma$ ranges from \refeq{eq:dmsexp}.
For $\tb = 10$ the SM result is reproduced over the whole parameter
space. Only for $\tb = 50$ and $m_{1/2} \lsim 500 \gev$ in the
coannihilation region can the CMSSM
prediction be significantly lower than~1.
Here the CMSSM could account for the small deviation of the experimental
result from the central value SM prediction, should that be necessary.


\section{CMSSM Analysis Including EWPO and BPO}
\label{sec:cmssm}

We now use the analyses of the previous
Section to estimate the combined $\chi^2$~function for the CMSSM as
a function of $m_{1/2}$, using the master formula (\ref{eq:chi2}). As
a first step, Fig.~\ref{fig:chi_ewpo} displays the $\chi^2$~distribution
for the EWPO alone. 

In the case $\tb = 10$ (left panel of \reffi{fig:chi_ewpo}), we see a
well-defined minimum of 
$\chi^2$ for $m_{1/2} \sim 300 \gev$ when $A_0 > 0$, which disappears
for large negative $A_0$ and is not present in the focus-point
region. The rise at small $m_{1/2}$ is due both to the lower limit on
$\Mh$ coming from the direct search at LEP and to $(g - 2)_\mu$, whilst
the rise at large $m_{1/2}$ is mainly due to $(g - 2)_\mu$ (see
\reffi{fig:AMU}). 
The measurement of $\MW$ (see \reffi{fig:MW})
leads to a slightly lower minimal value of~$\chi^2$,
but there are no substantial contributions from
any of the other EWPO. The preference for $A_0 > 0$ in the
coannihilation region is due to $\Mh$ (see \reffi{fig:Mh}), and the
relative disfavour for the focus-point regions is due to its mismatch
with $(g - 2)_\mu$ (see \reffi{fig:AMU}). 

In the case $\tb = 50$ (right panel of \reffi{fig:chi_ewpo}), we again
see a well-defined minimum of $\chi^2$, this time for $m_{1/2} \sim 400$
to 500~GeV, which is similar for all the studied values of $A_0$. In
this case, there is also a similar minimum of $\chi^2$ for the
focus-point region at $m_{1/2} \sim 200 \gev$. The increase in $\chi^2$
at small $m_{1/2}$ is due to $(g - 2)_\mu$ as well as $\Mh$, whereas the
increase at large $m_{1/2}$ is essentially due to $(g - 2)_\mu$. We note
that the overall minimum of $\chi^2 \sim 2$ is similar for both values
of $\tb$, and represents an excellent fit in each case.

\reffi{fig:chi_bpo} shows the corresponding combined $\chi^2$ for the
BPO alone. For both values of $\tb$, these prefer large values of
$m_{1/2}$, reflecting the fact that there is no hint of any deviation
from the SM, and the overall quality of the fit is good. Small values
of $m_{1/2}$ are disfavoured, particularly in the coannihilation region
with $A_0 > 0$, mainly due to $b \to s \ga$. The focus-point region is
generally in very good 
agreement with the BPO data, except at very low $m_{1/2} \lsim 400 \gev$
for $\tb = 50$.

\begin{figure}[tbh!]
\begin{center}
\includegraphics[width=.48\textwidth]{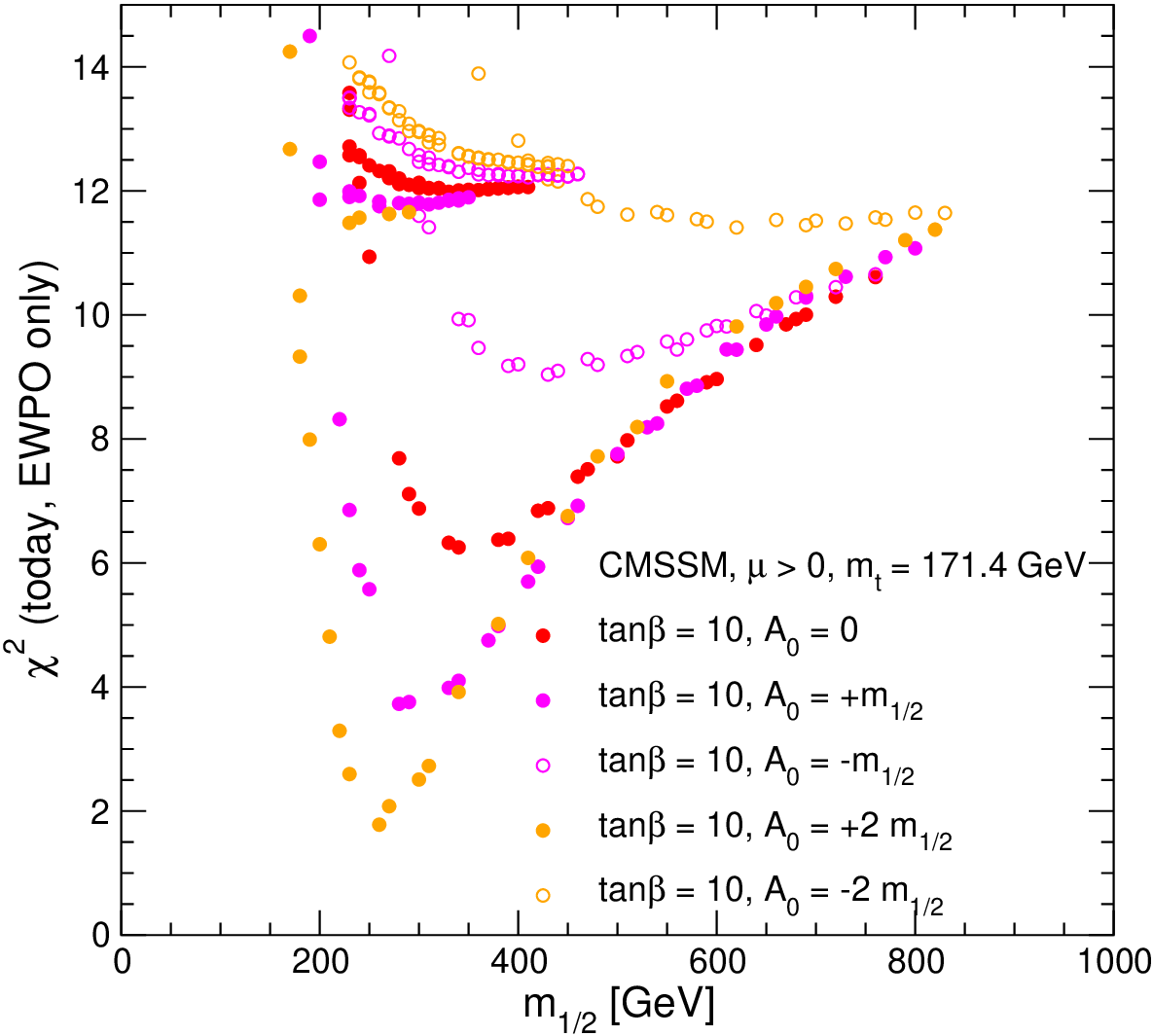}
\includegraphics[width=.48\textwidth]{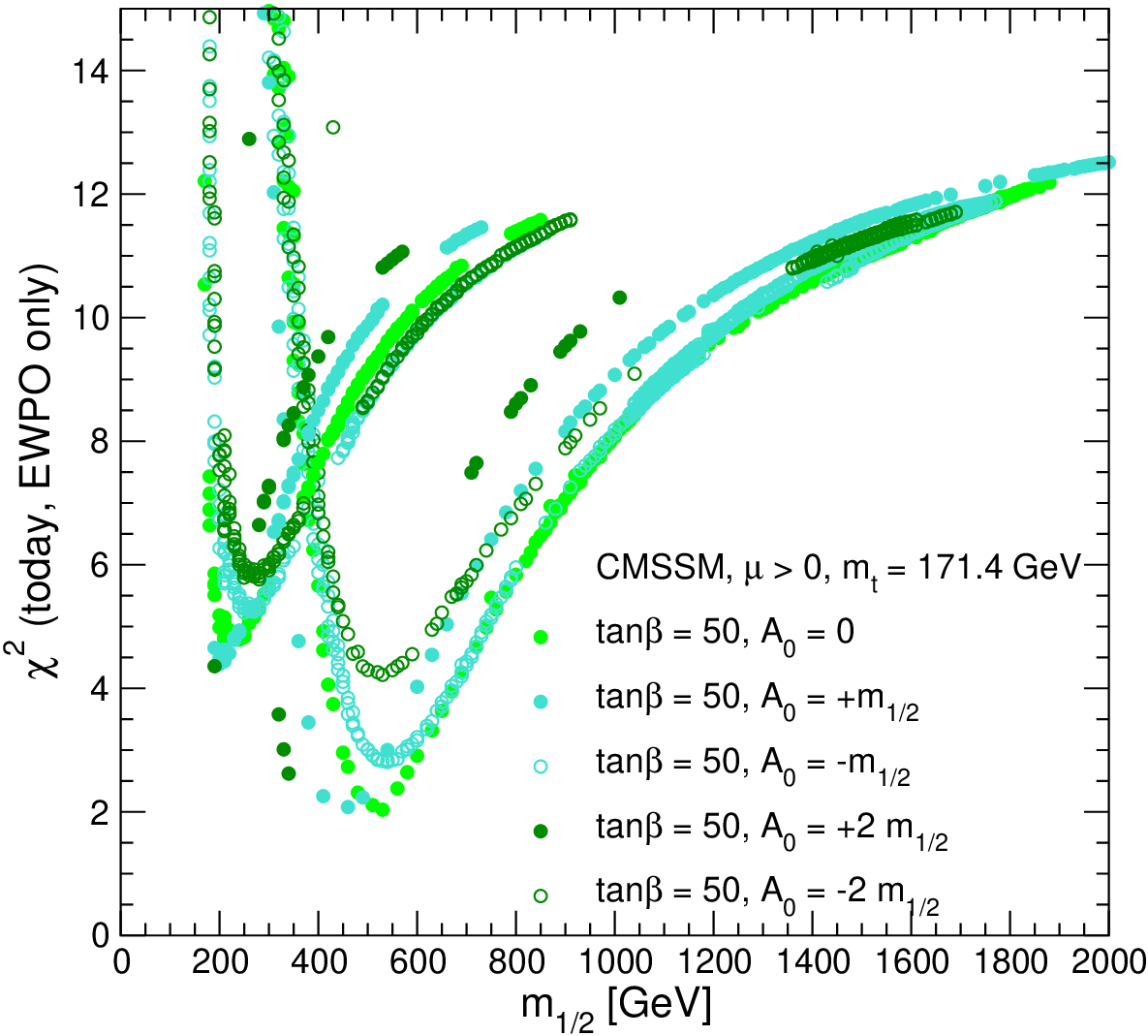}
\caption{%
The combined $\chi^2$~function for the electroweak
observables $\MW$, $\sweff$, $\Ga_Z$, $(g - 2)_\mu$ and $\Mh$, 
evaluated in the CMSSM for $\tb = 10$ (left) and
$\tb = 50$ (right) for various discrete values of $A_0$.
We use $\mt = 171.4 \pm 2.1 \gev$ and $\mb(\mb) = 4.25 \pm 0.11 \gev$,
and $m_0$ is chosen to yield the central value of the cold dark matter
density indicated by WMAP and other observations for the central values
of $\mt$ and $\mb(\mb)$.}
\label{fig:chi_ewpo}
\end{center}
\vspace{-1em}
\end{figure}

\begin{figure}[tbh!]
\begin{center}
\includegraphics[width=.48\textwidth]{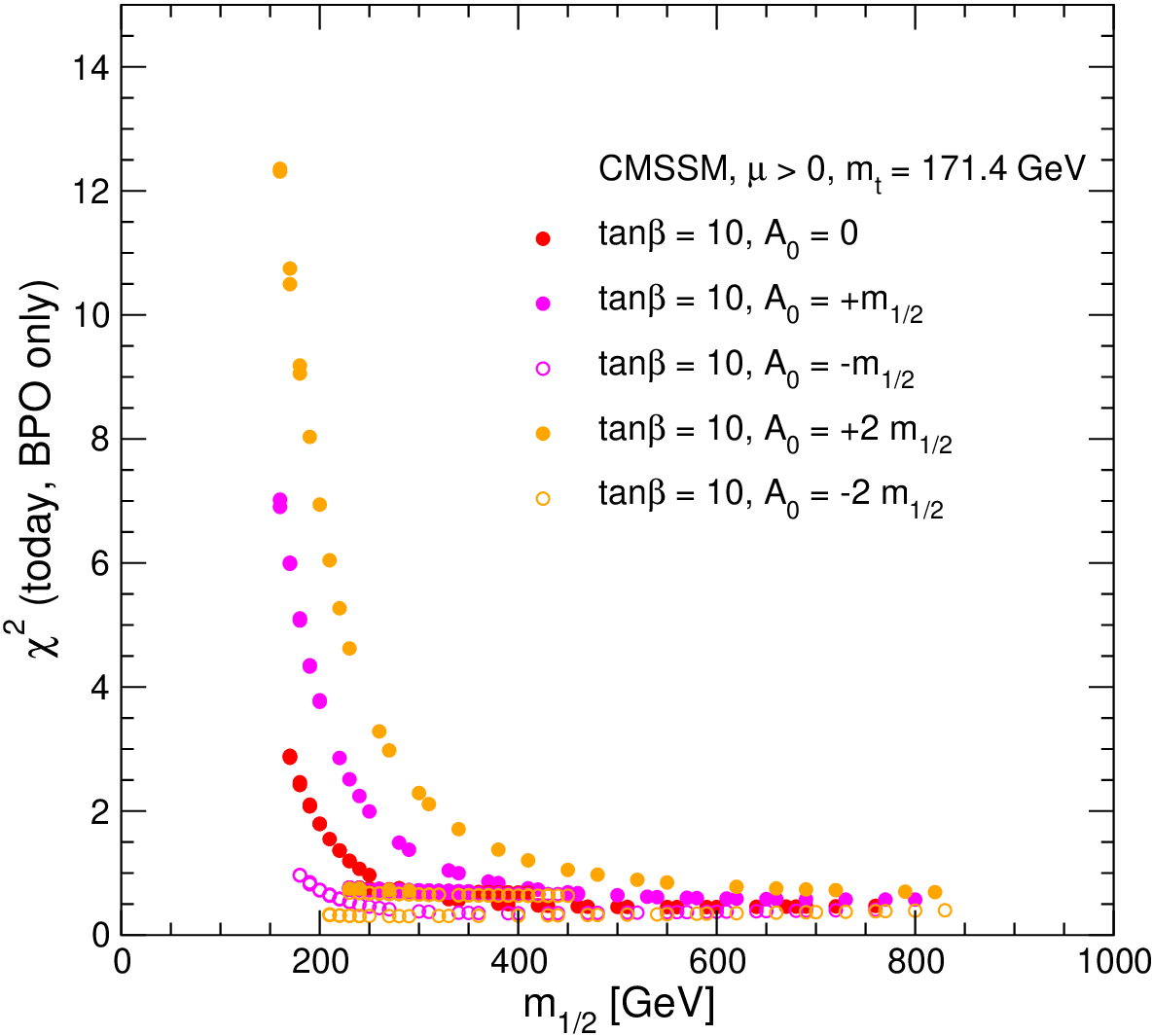}
\includegraphics[width=.48\textwidth]{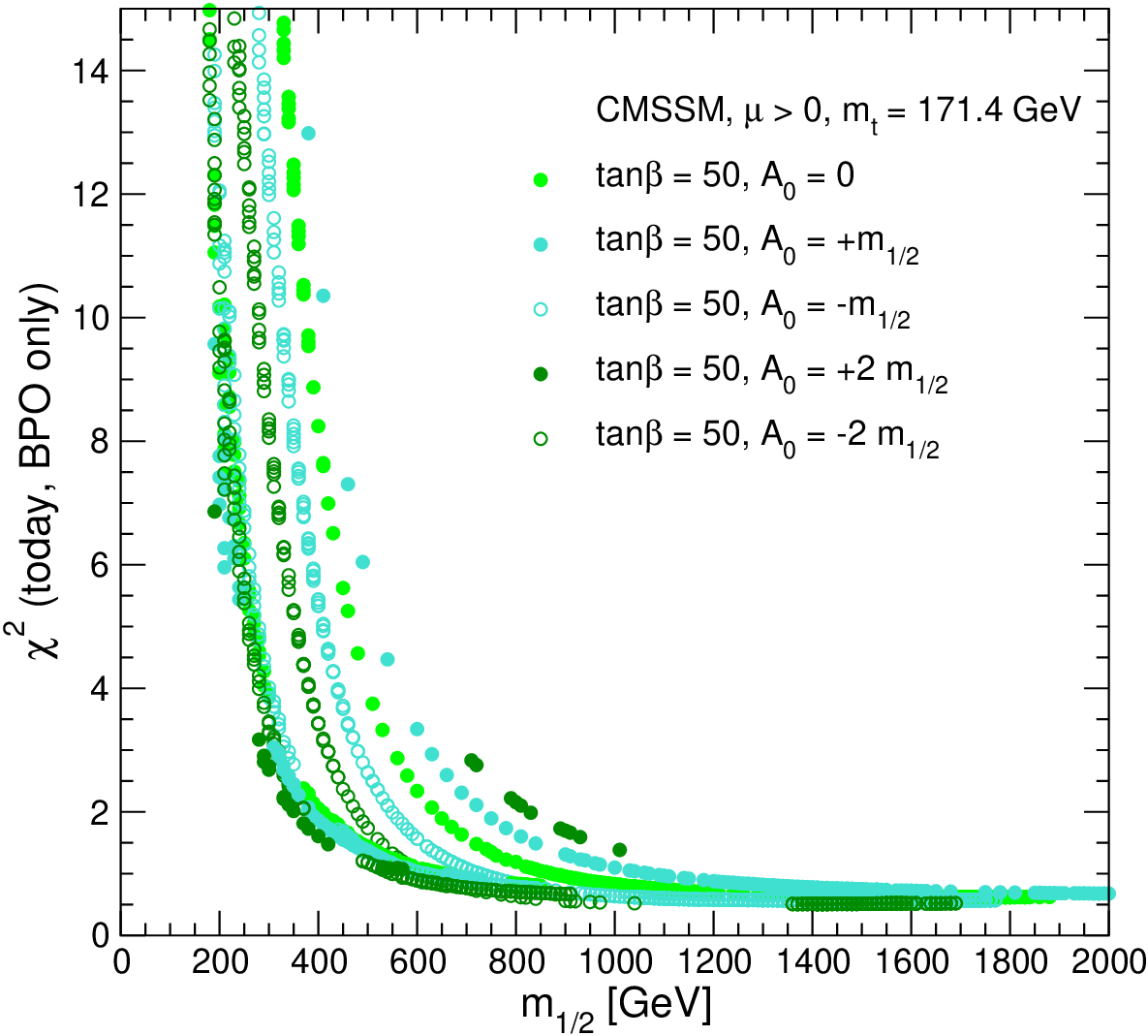}
\caption{%
The combined $\chi^2$~function for the $b$~physics observables
$\br(b \to s \ga)$, $\br(B_s \to \mu^+\mu^-)$, $\br(B_u \to \tau \nu_\tau)$
and $\De M_{B_s}$, evaluated in the CMSSM for $\tb = 10$ (left) and
$\tb = 50$ (right) for various discrete values of $A_0$.
We use
$\mt = 171.4 \pm 2.1 \gev$ and $\mb(\mb) = 4.25 \pm 0.11 \gev$, and 
$m_0$ is chosen to yield the central value of the cold dark matter
density indicated by WMAP and other observations for the central values
of $\mt$ and $\mb(\mb)$.}
\label{fig:chi_bpo}
\end{center}
\vspace{-1em}
\end{figure}

Finally, we show in \reffi{fig:chi} the combined $\chi^2$~values
for the EWPO and BPO, computed in accordance with \refeq{eq:chi2}.
We see that the global minimum of $\chi^2 \sim 4.5$
for both values of $\tb$. This is quite a good fit for the number of
experimental observables being fitted, and the $\chi^2/{\rm d.o.f.}$ is
similar to the one for the EWPO alone. This increase in the total
$\chi^2$ reflects the fact that the BPO
exhibit no tendency to reinforce the preference of the EWPO for small
$m_{1/2}$: rather the reverse, in fact. 
For both values of $\tb$, the focus-point region is disfavoured by comparison
with the coannihilation region, though this effect is less important for
$\tb = 50$. 
For $\tb = 10$, $m_{1/2} \sim 300 \gev$ and $A_0 > 0$ are preferred, whereas,
for $\tb = 50$, $m_{1/2} \sim 600 \gev$ and $A_0 < 0$ are preferred. This
change-over is largely due to the impact of the LEP $\Mh$ constraint for
$\tb = 10$ and the $b \to s \ga$ constraint for $\tb = 50$.

\begin{figure}[tbh!]
\begin{center}
\includegraphics[width=.48\textwidth]{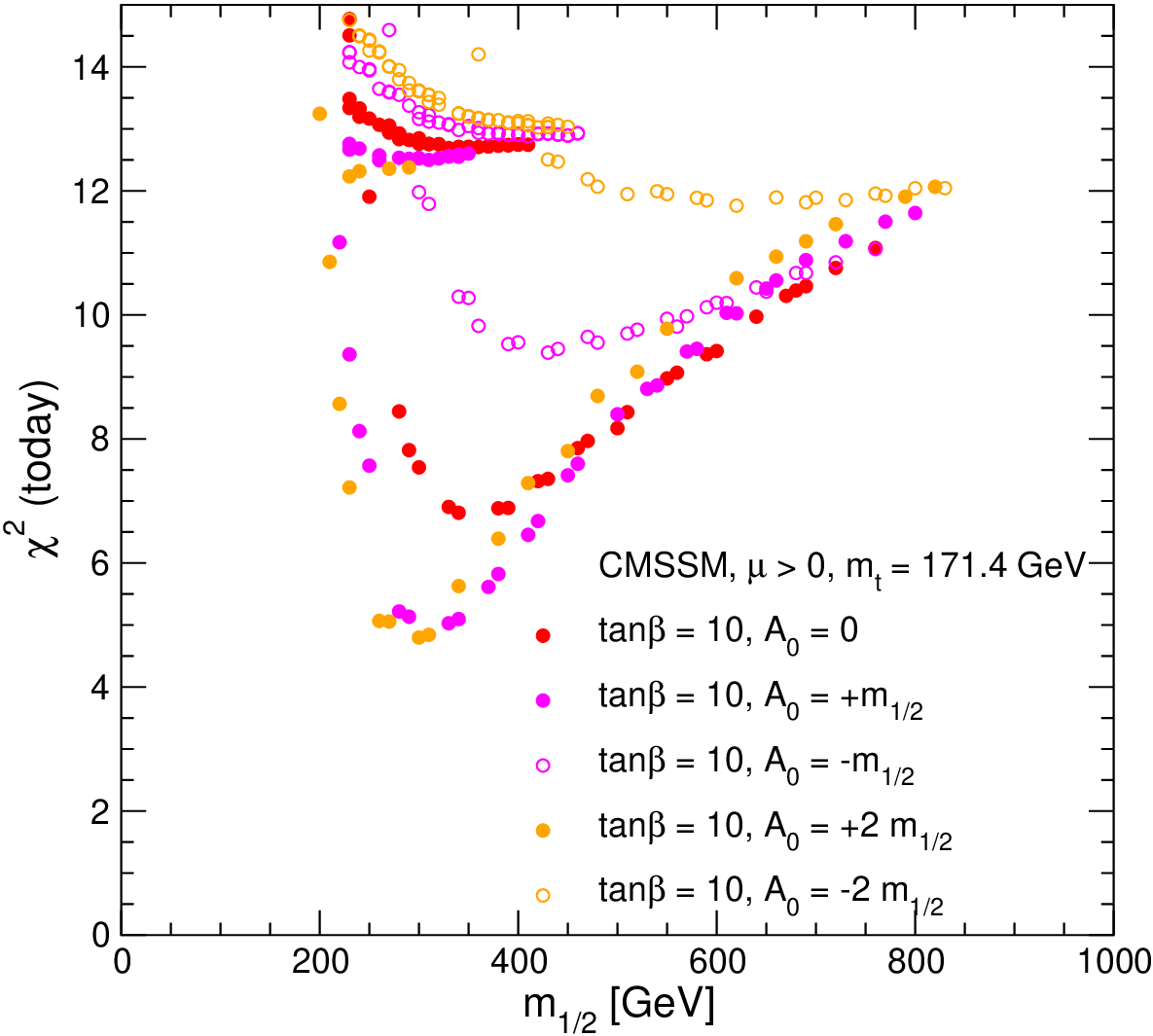}
\includegraphics[width=.48\textwidth]{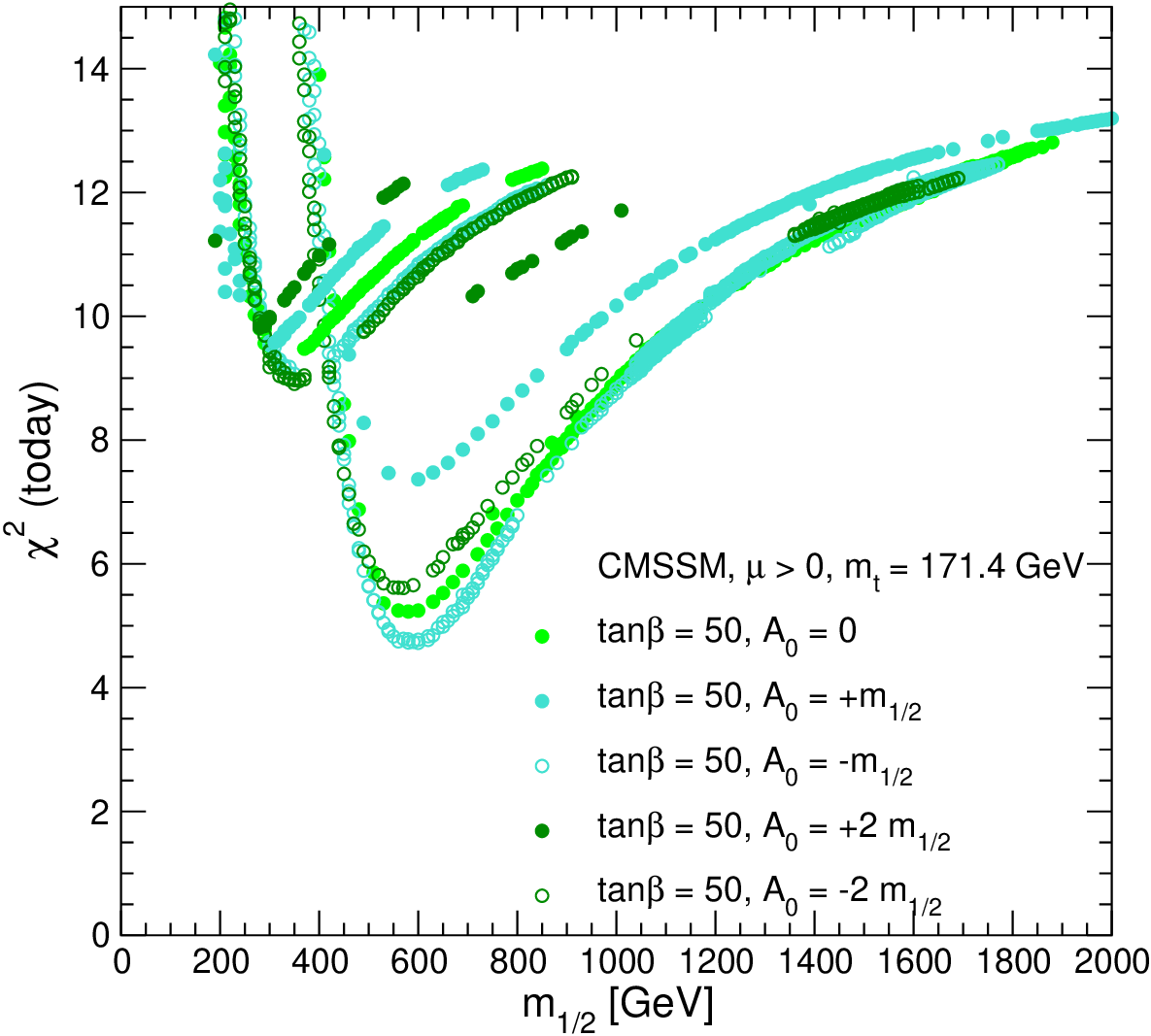}
\caption{%
The combined $\chi^2$~function for the electroweak
observables $\MW$, $\sweff$, $\Ga_Z$, $(g - 2)_\mu$, $\Mh$, 
and the $b$~physics observables
$\br(b \to s \ga)$, $\br(B_s \to \mu^+\mu^-)$, $\br(B_u \to \tau \nu_\tau)$
and $\De M_{B_s}$, evaluated in the CMSSM for $\tb = 10$ (left) and
$\tb = 50$ (right) for various discrete values of $A_0$. We use
$\mt = 171.4 \pm 2.1 \gev$ and $\mb(\mb) = 4.25 \pm 0.11 \gev$, and 
$m_0$ is chosen to yield the central value of the cold dark matter
density indicated by WMAP and other observations for the central values
of $\mt$ and $\mb(\mb)$.}
\label{fig:chi}
\end{center}
\vspace{-1em}
\end{figure}

\begin{figure}[tbh!]
\begin{center}
\includegraphics[width=.45\textwidth,height=0.36\textwidth]{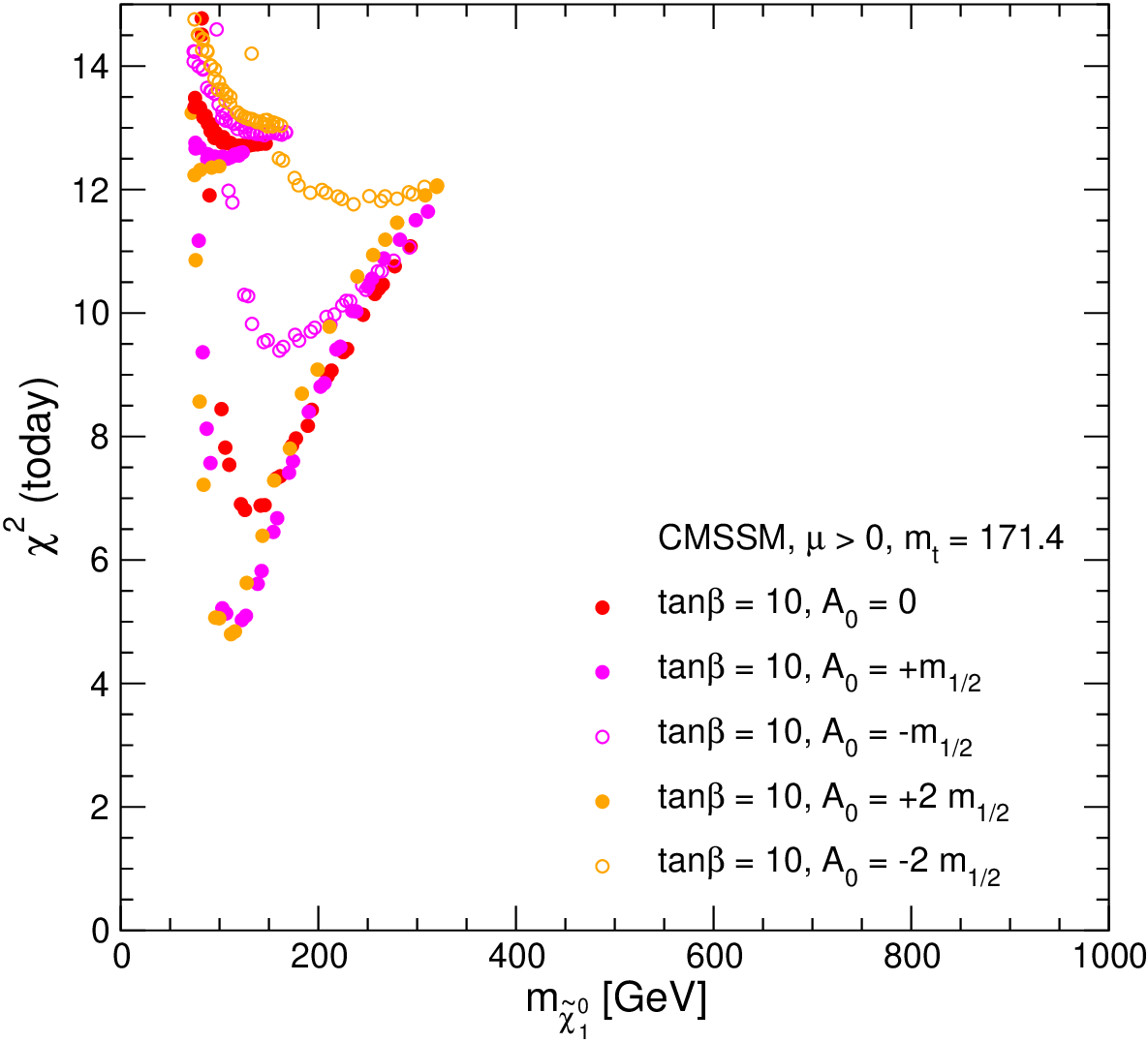}
\includegraphics[width=.45\textwidth,height=0.36\textwidth]{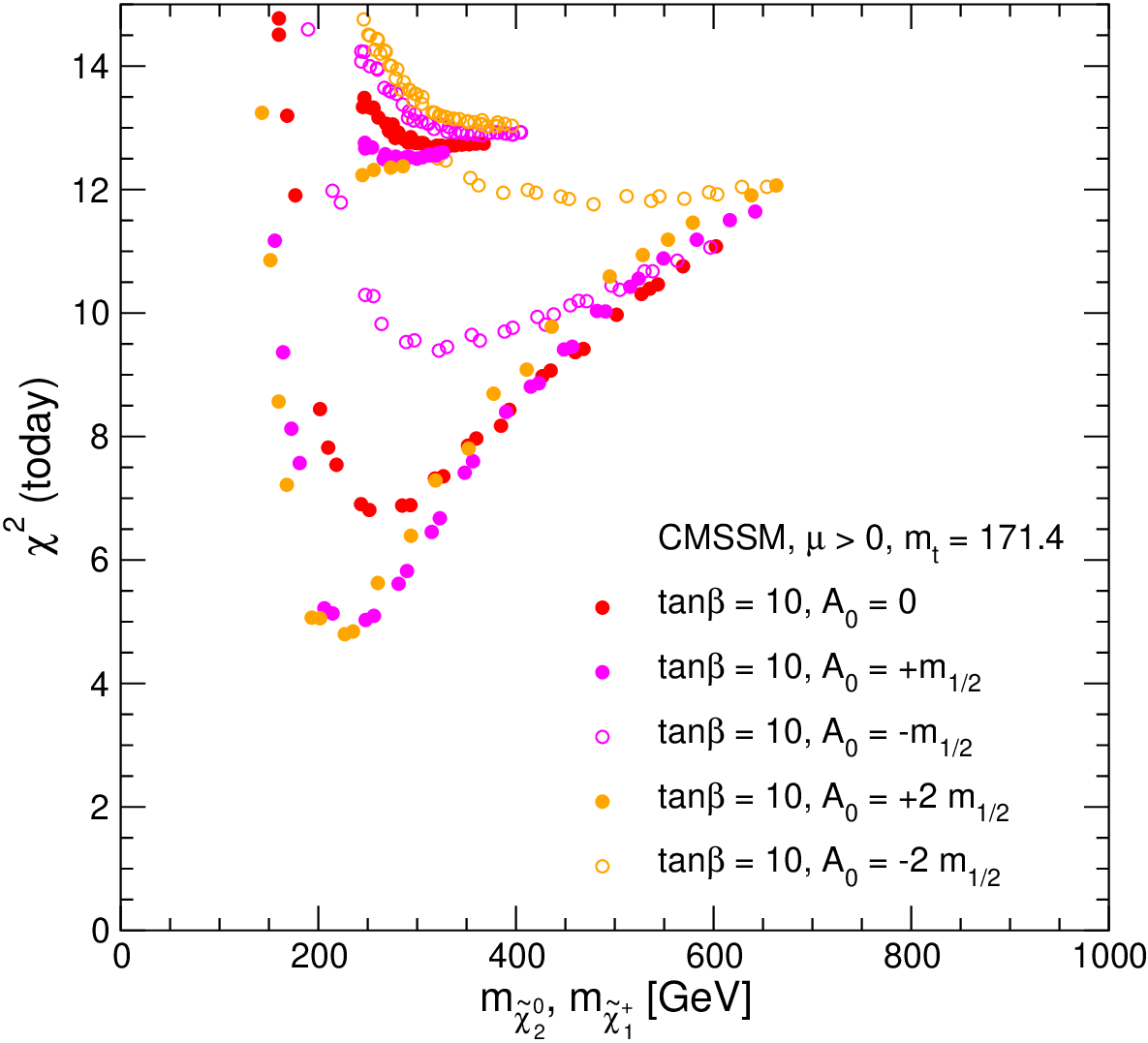}
\end{center}
\begin{center}
\includegraphics[width=.45\textwidth,height=0.36\textwidth]{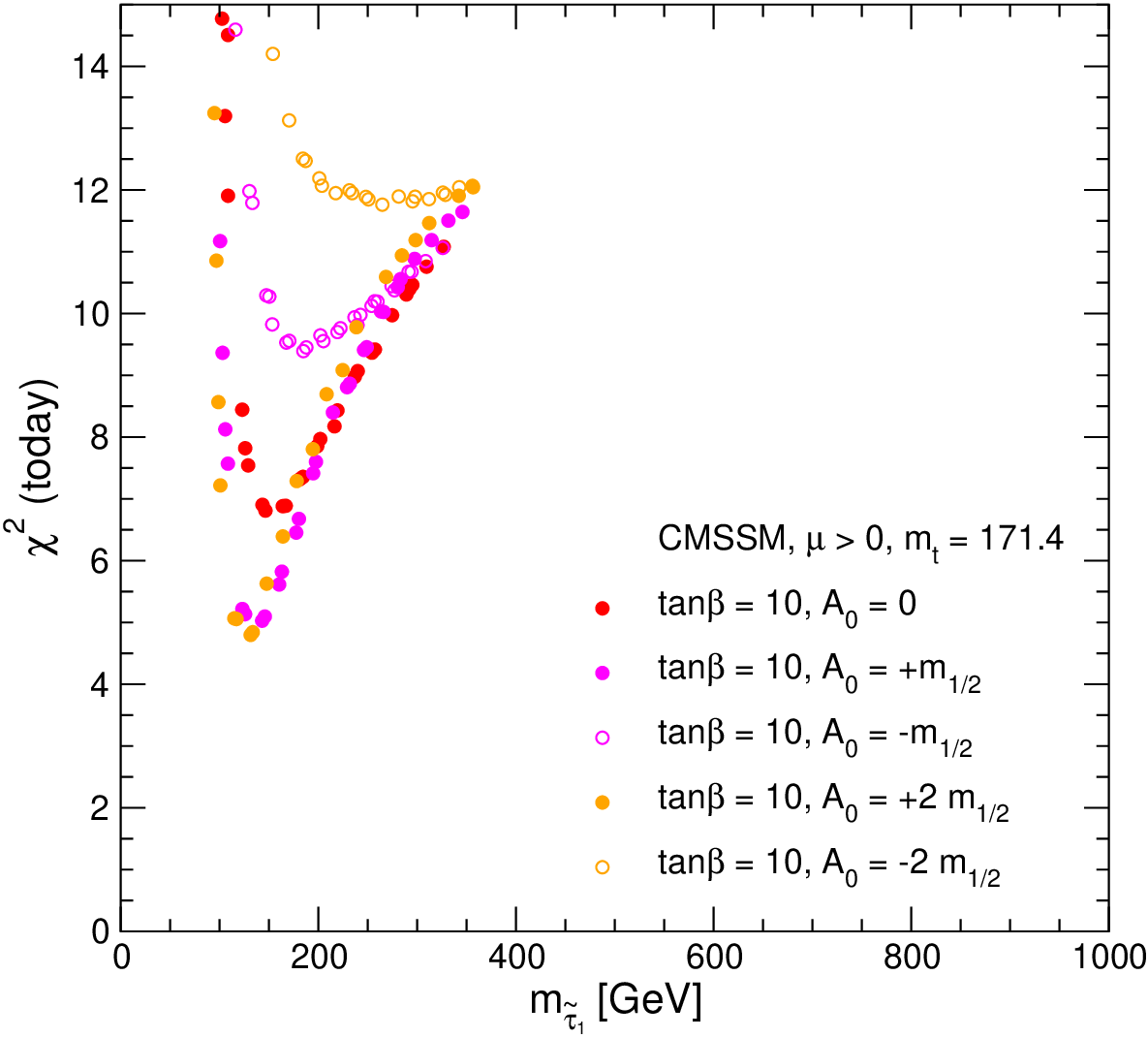}
\includegraphics[width=.45\textwidth,height=0.36\textwidth]{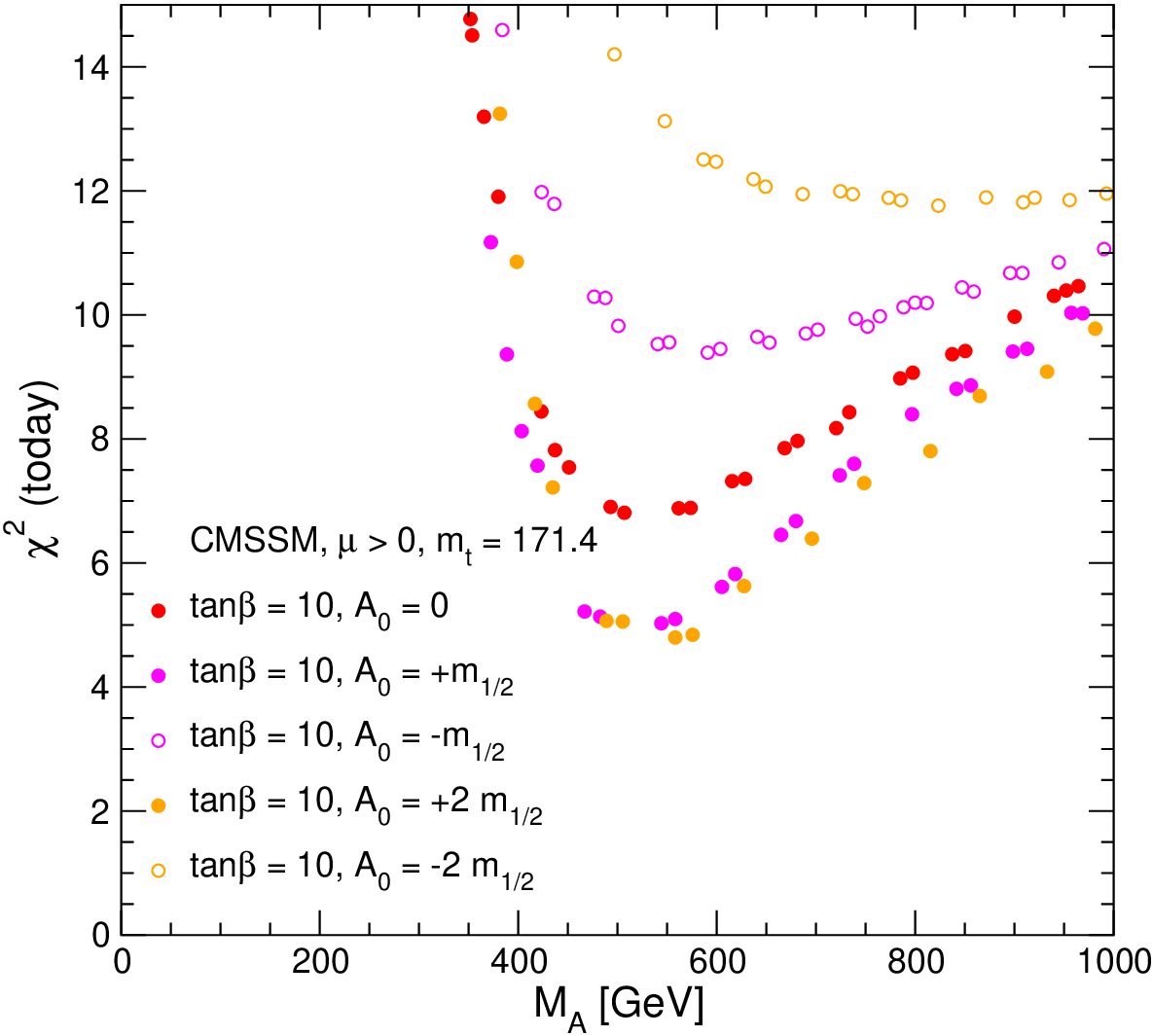}
\end{center}
\begin{center}
\includegraphics[width=.45\textwidth,height=0.36\textwidth]{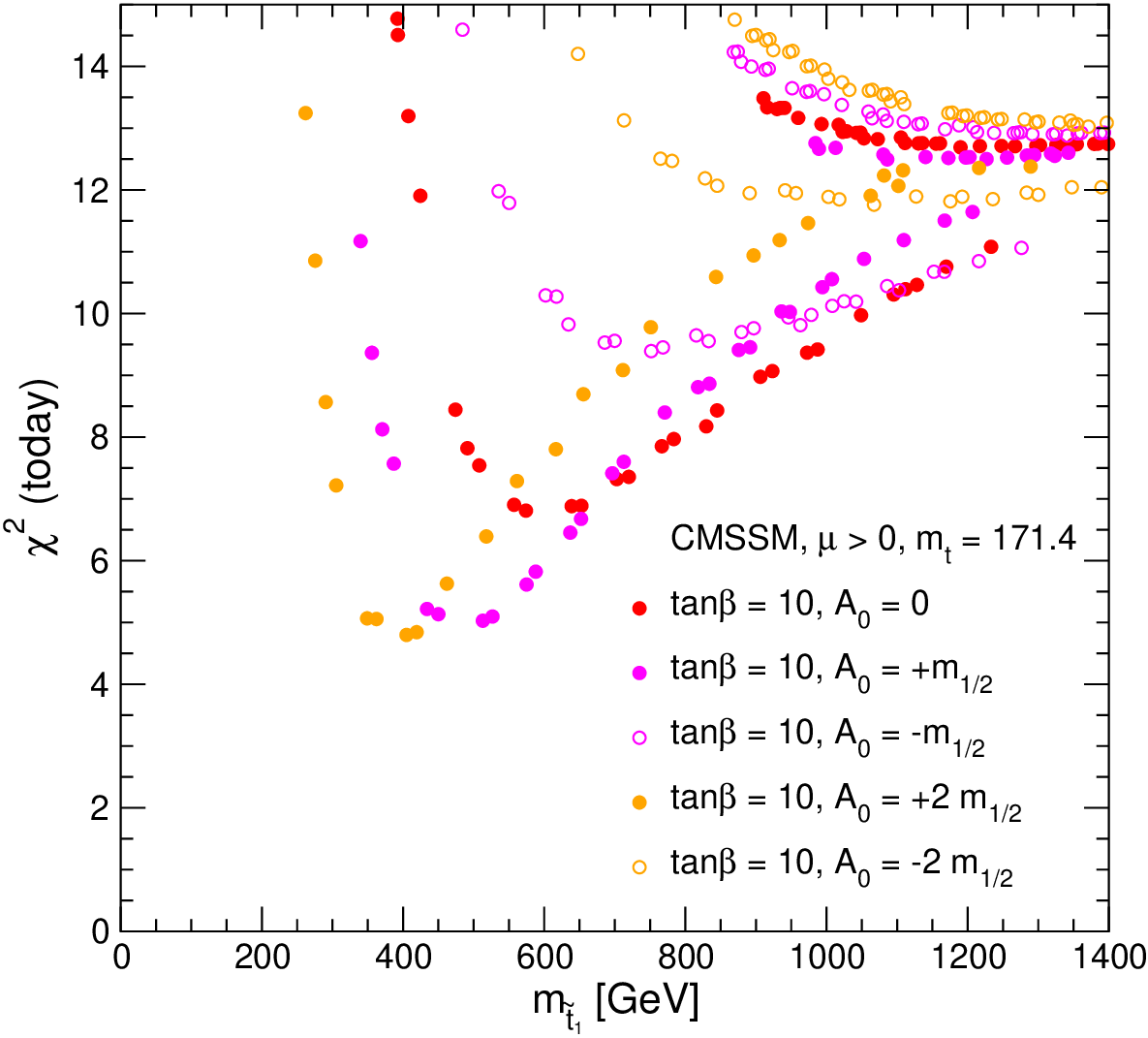}
\includegraphics[width=.45\textwidth,height=0.36\textwidth]{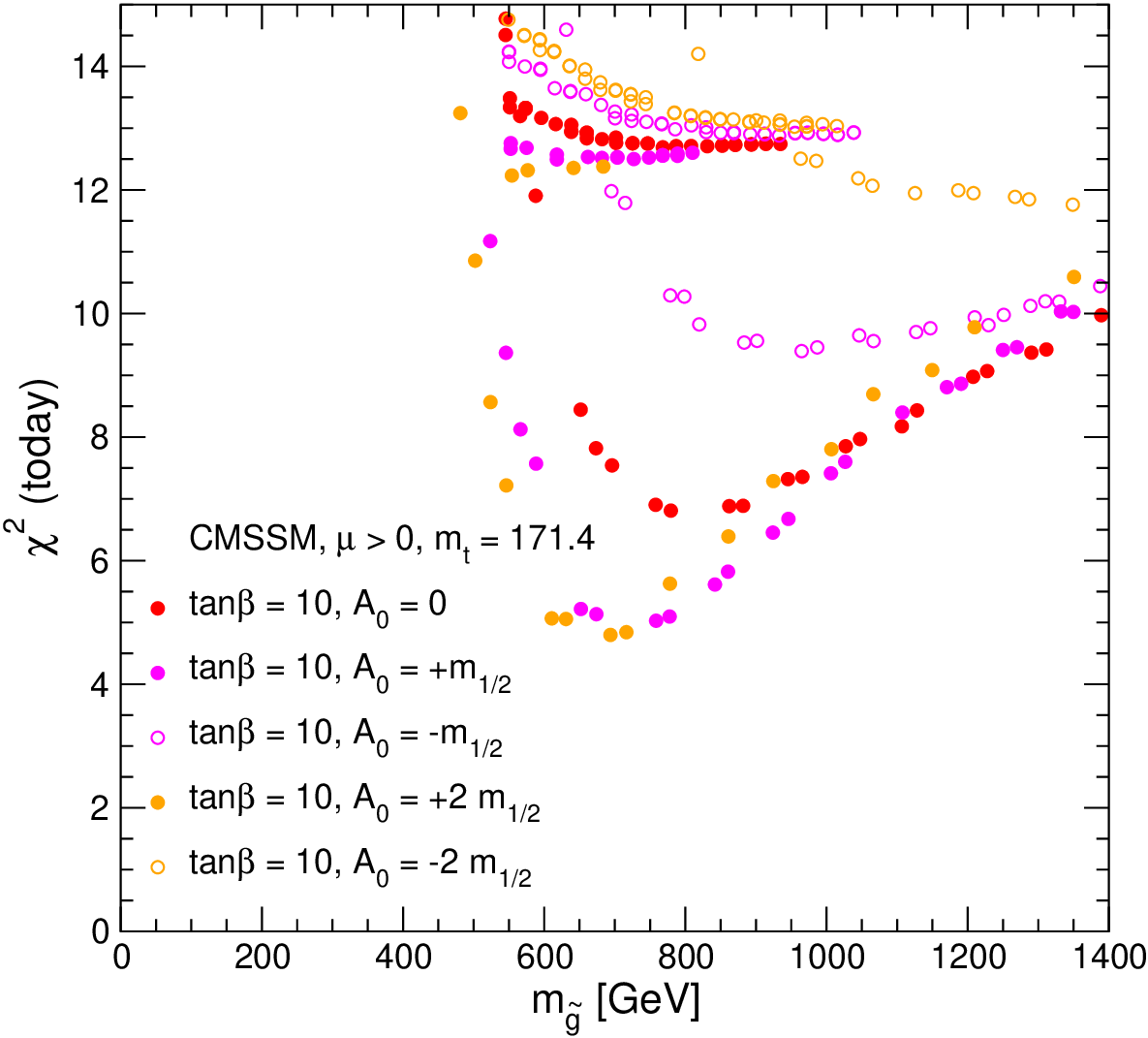}
\caption{%
Various SUSY masses are presented with their respective $\chi^2$ value in
the CMSSM for $\tb = 10$. The panels show (a)~$\mneu{1}$, (b)~$\mneu{2}$
and $\mcha{1}$ (which are very similar), (c)~$\mstaue$, (d)~$\MA$, 
(e)~$\mste$ and (f)~$\mgl$.
}
\label{fig:10masses}
\end{center}
\vspace{-1em}
\end{figure}

We display in \reffi{fig:10masses} the $\chi^2$~functions
for various SUSY masses 
in the CMSSM for $\tb = 10$, including (a)~$\mneu{1}$, (b)~$\mneu{2}$
and $\mcha{1}$ (which are very similar), (c)~$\mstaue$, (d)~$\MA$, 
(e)~$\mste$ and (f)~$\mgl$. We see two distinct populations of points,
corresponding to the $\chi - {\tilde \tau_1}$ coannihilation (which is
favoured) and focus-point regions (which is disfavoured). In the latter region,
very low values of $m_{1/2}$ are preferred, as can be seen in panels (a)
and (f), relatively small values of $\mu$, as can be seen in panel (b),
large values of $m_0$, as can be seen in panels (c) and (e), and large
values of $\MA$, as can (not) be seen in panel (d). Compared to the
analysis in \citere{ehow4}, where $\br(b \to s \ga)$ was the only BPO
included, and where a top quark mass of $172.7 \gev$ was used, there is
no significant shift of the values of the masses where $\chi^2$ has its
minimum, which is in the coannihilation region. As before, the present
analysis gives hope for seeing squarks and gluinos in the early days of
the LHC (panels (e) and (f)), and also hope for seeing charginos,
neutralinos and staus at the ILC (panels (a), (b) and (c)), whereas
observing the heavier Higgs bosons would be more challenging (panel\,(d)). 

In \reffi{fig:50masses} we show the analogous $\chi^2$~functions for
various SUSY masses  in the CMSSM for $\tb = 50$: (a)~$\mneu{1}$,
(b)~$\mneu{2}$ and $\mcha{1}$ (which are very similar), (c)~$\mstaue$,
(d)~$\MA$,  (e)~$\mste$ and (f)~$\mgl$. We again see the clear
separation between the focus-point and coannihilation regions, 
interpolated by a light-Higgs pole strip, and that
the coannihilation region is somewhat preferred. As for lower $\tb$, small
values of $m_{1/2}$ and larger values of $m_0$ are preferred, and also
small values of $\mu$ and larger values of $\MA$. Again as for 
$\tb = 10$, compared to the analysis in \citere{ehow4}, where 
$\br(b \to s \ga)$ was  the only BPO included and where a top quark mass
of $172.7 \gev$ was used, we do not find a significant shift in the
values of the masses with lowest $\chi^2$. The sparticle masses are
generally higher 
than for $\tb = 10$: finding squarks and gluinos should still be
`easy' at the LHC, but seeing charginos, neutralinos and staus at the
ILC would be more challenging, depending on its center-of-mass energy.

\begin{figure}[tbh!]
\begin{center}
\includegraphics[width=.45\textwidth,height=0.36\textwidth]{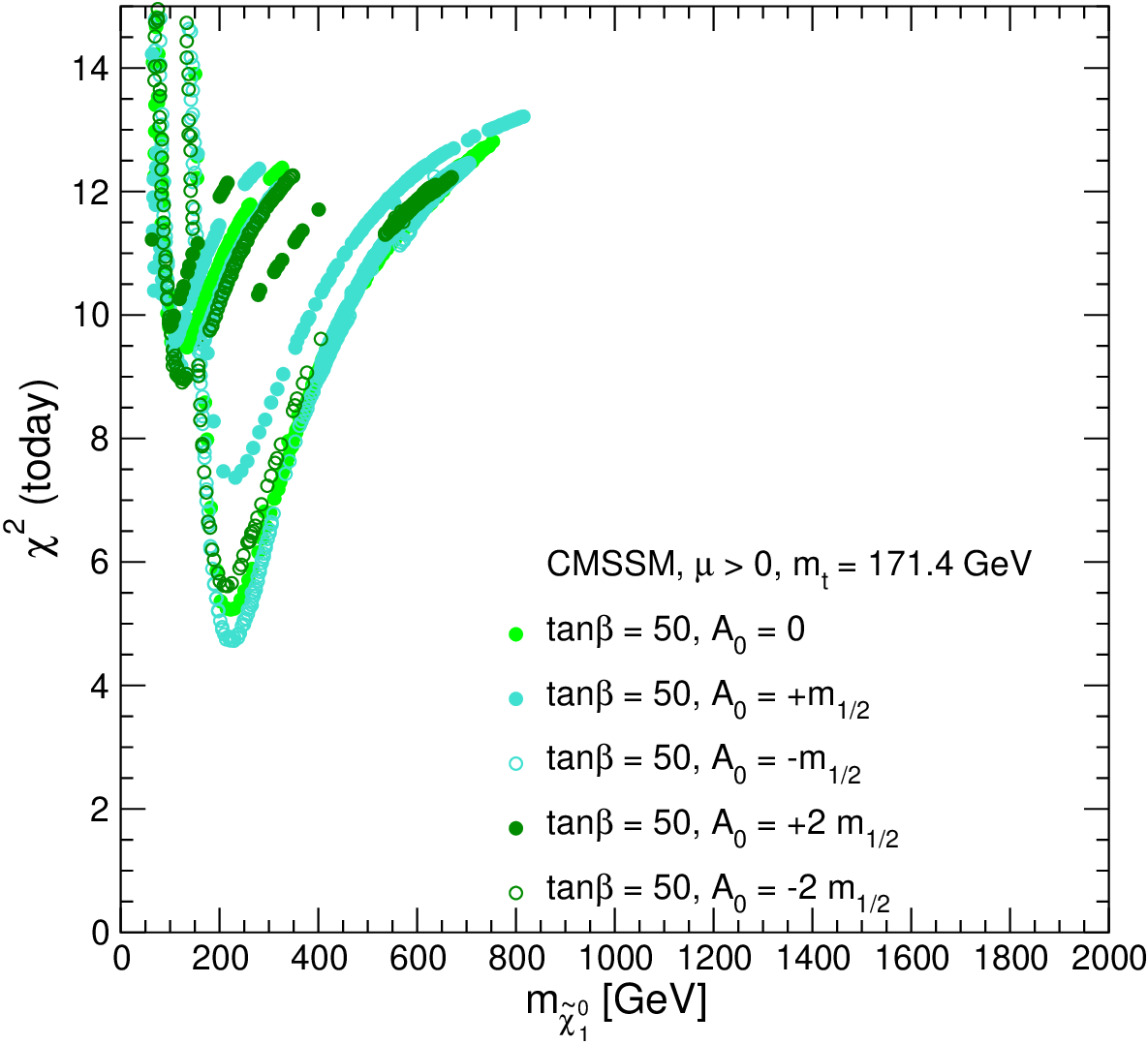}
\includegraphics[width=.45\textwidth,height=0.36\textwidth]{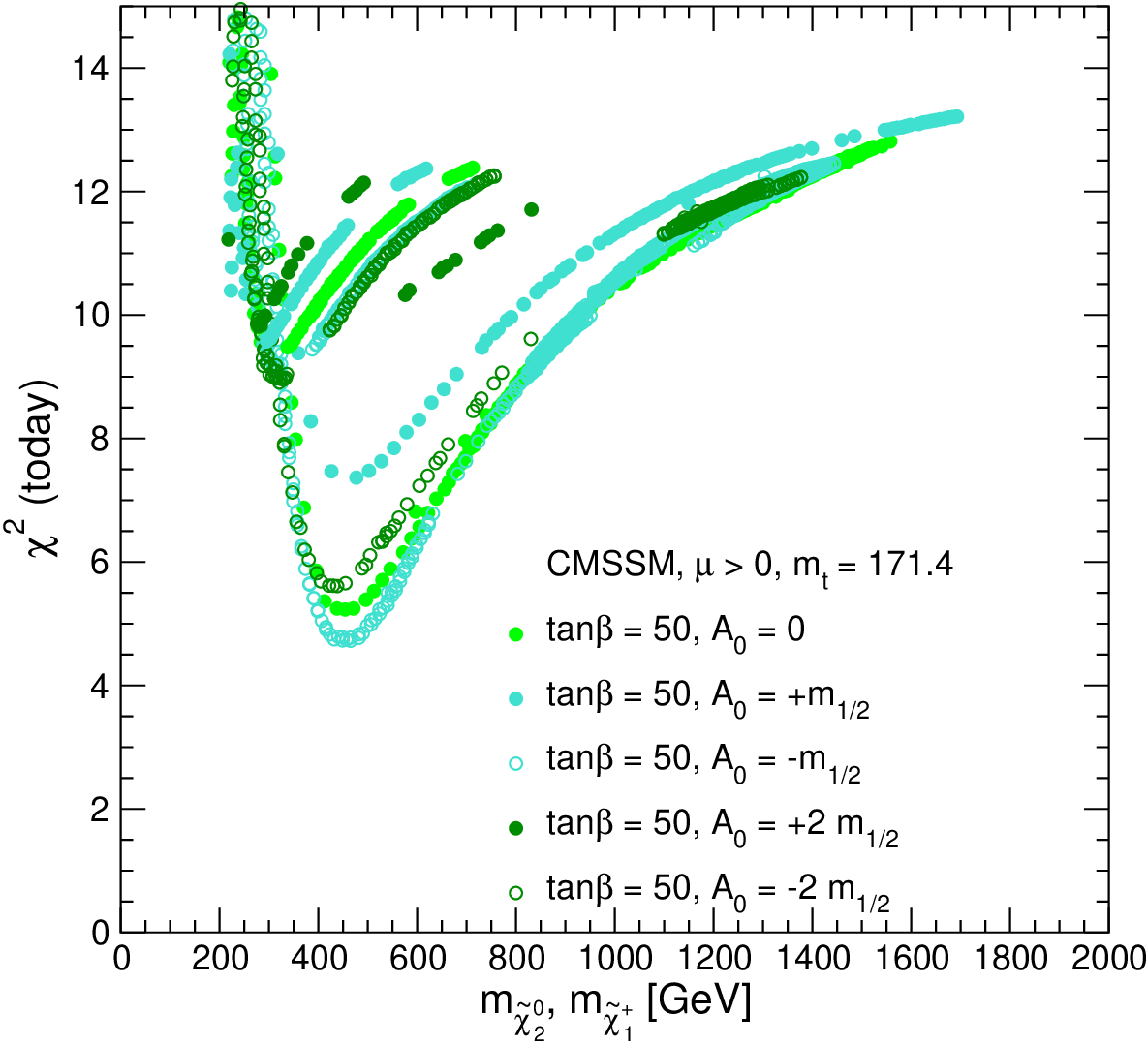}
\end{center}
\begin{center}
\includegraphics[width=.45\textwidth,height=0.36\textwidth]{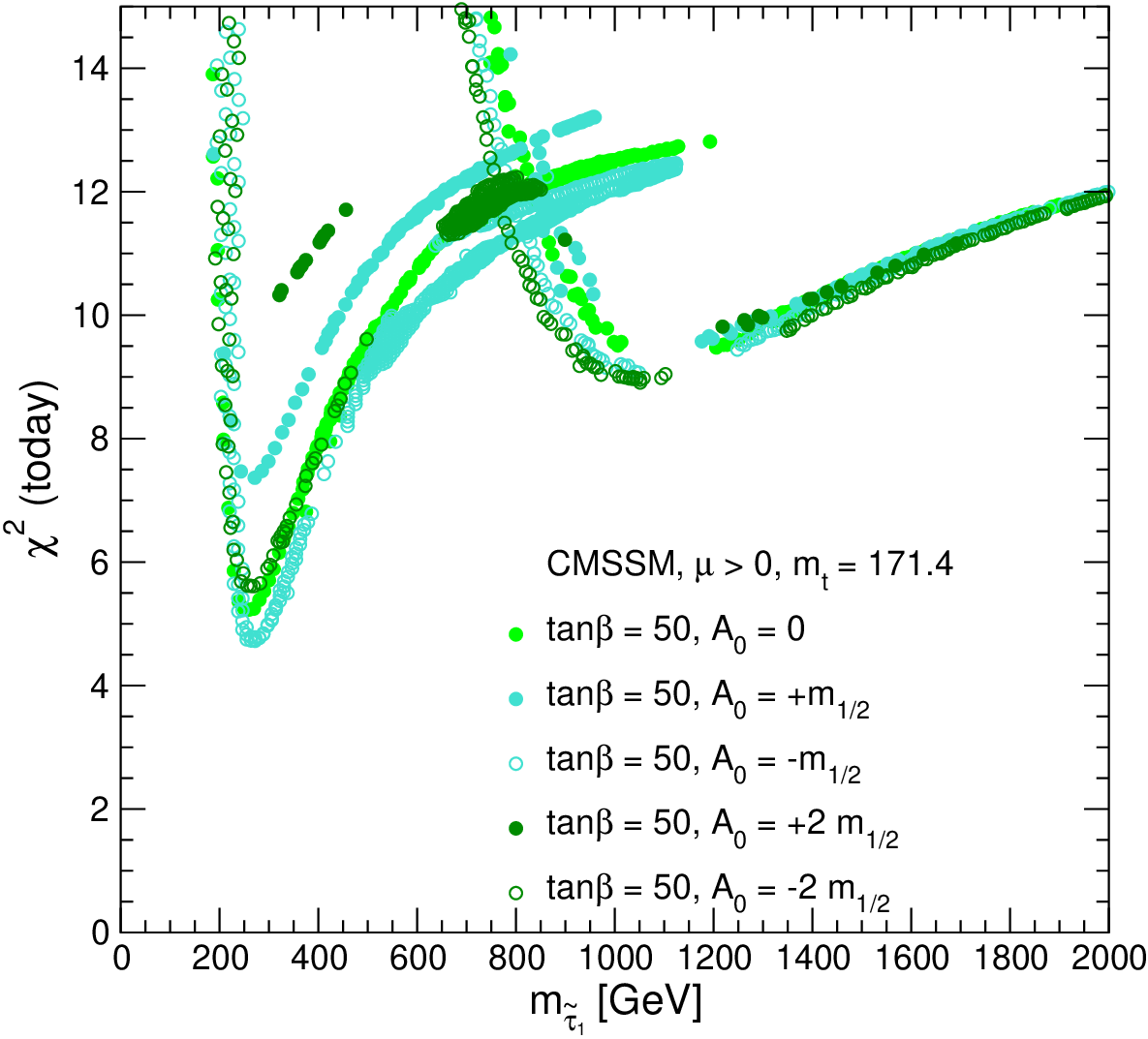}
\includegraphics[width=.45\textwidth,height=0.36\textwidth]{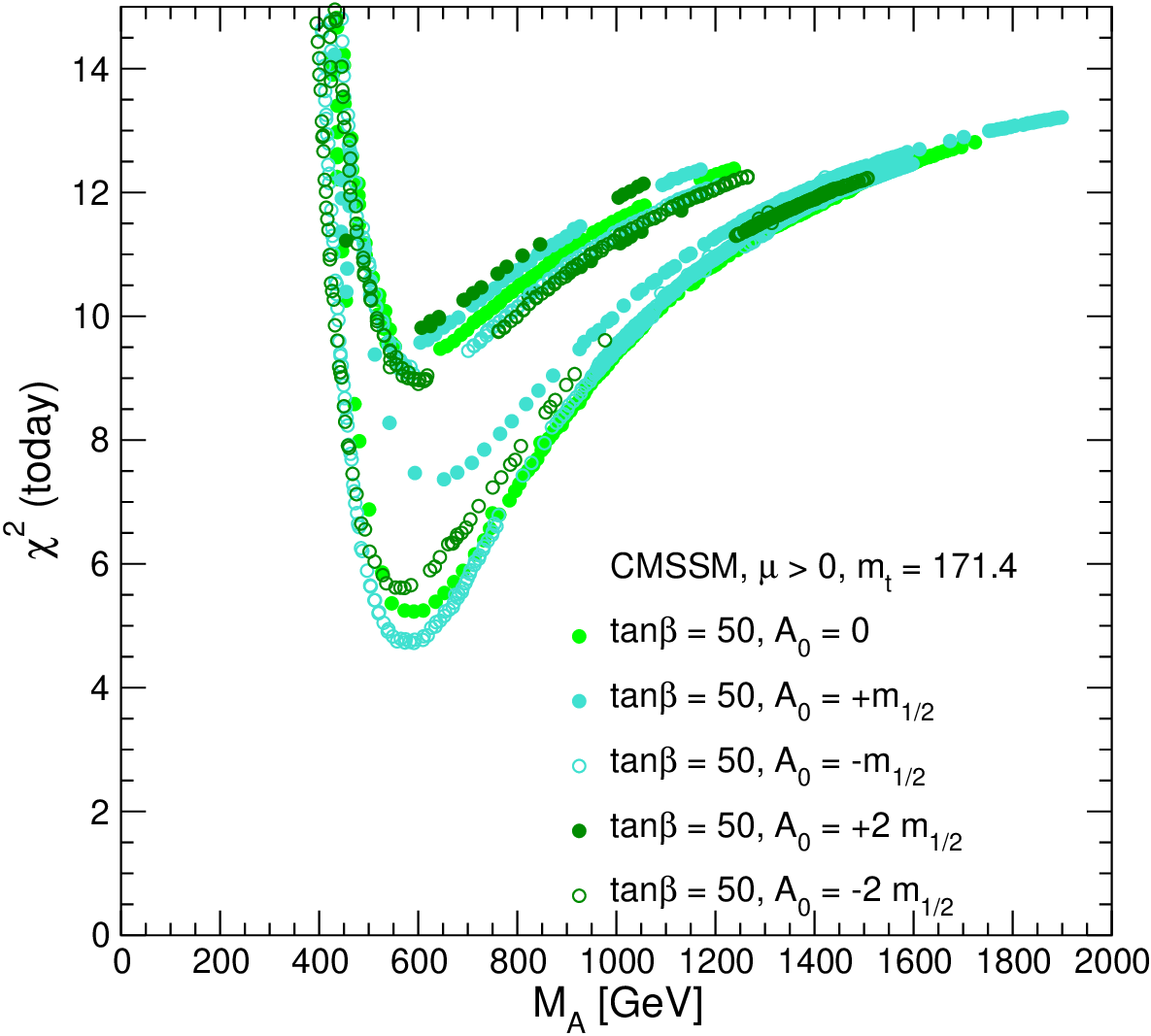}
\end{center}
\begin{center}
\includegraphics[width=.45\textwidth,height=0.36\textwidth]{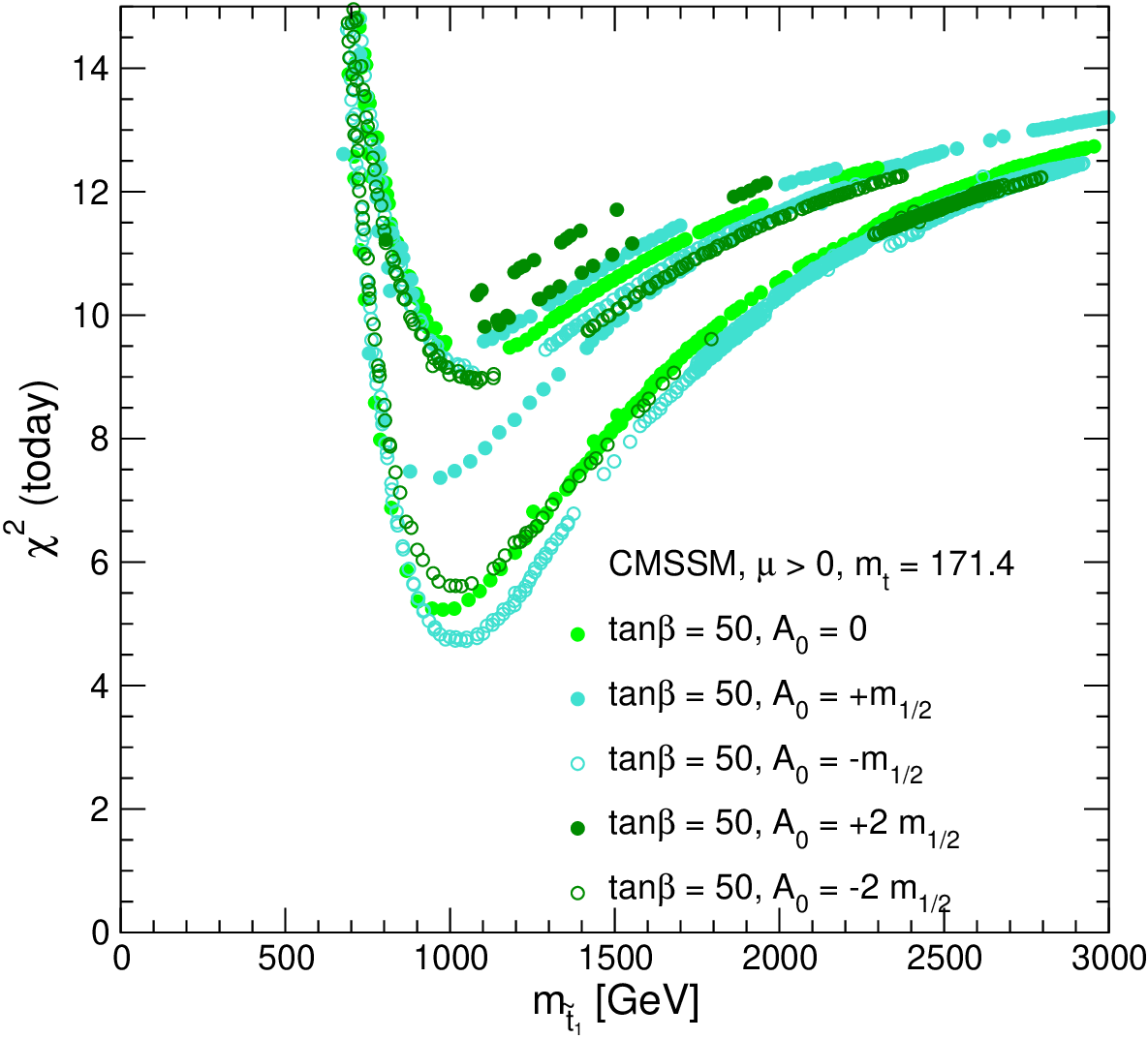}
\includegraphics[width=.45\textwidth,height=0.36\textwidth]{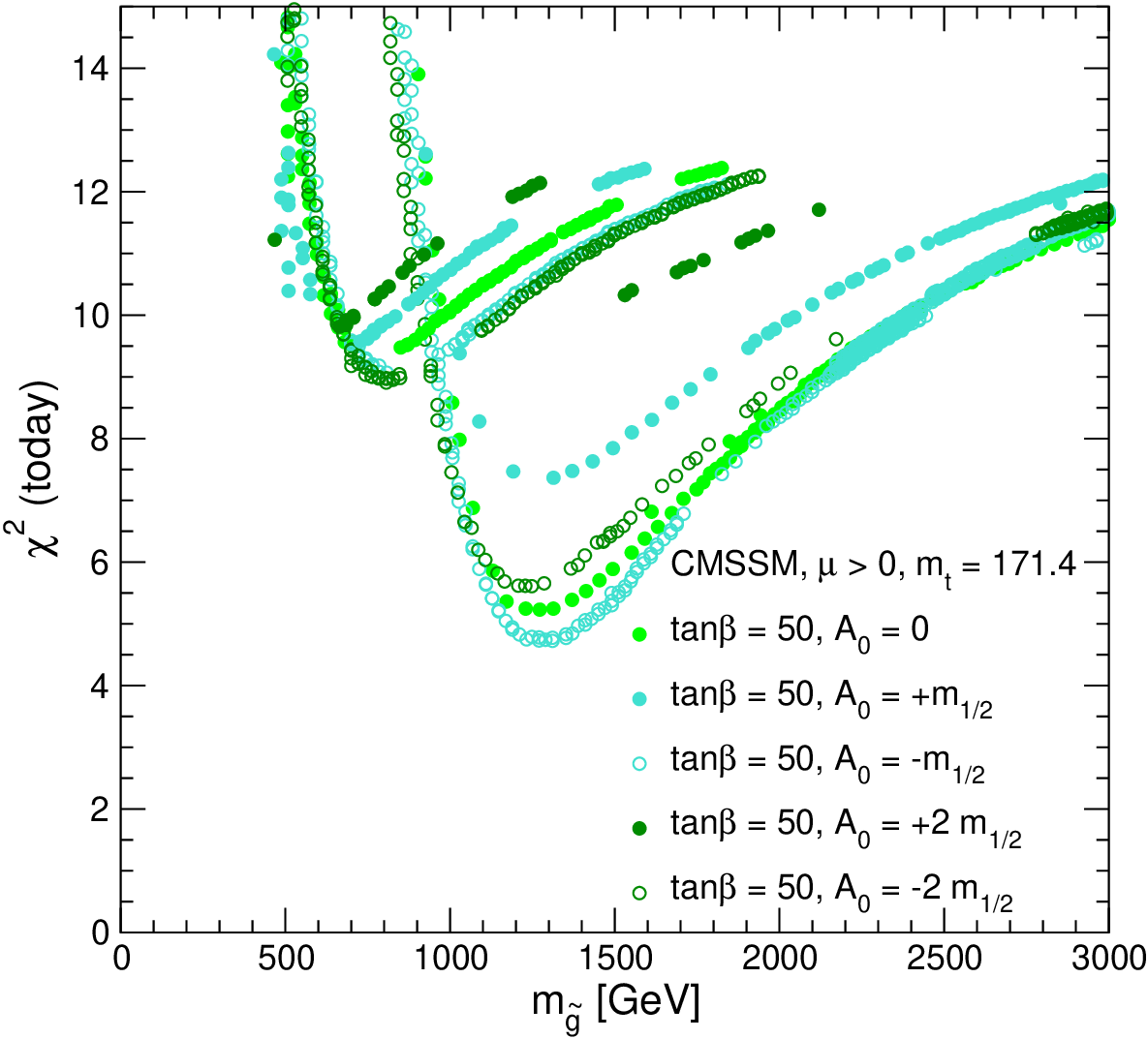}
\caption{%
Various SUSY masses are presented with their respective $\chi^2$ value in
the CMSSM for $\tb = 50$. The panels show (a)~$\mneu{1}$, (b)~$\mneu{2}$
and $\mcha{1}$ (which are very similar), (c)~$\mstaue$, (d)~$\MA$, 
(e)~$\mste$ and (f)~$\mgl$.
}
\label{fig:50masses}
\end{center}
\vspace{-1em}
\end{figure}

Analogously to the sparticle masses in Figs.~\ref{fig:10masses} and
\ref{fig:50masses}, we display in \reffi{fig:allMh} the total
$\chi^2$~functions for $\Mh$, 
as calculated in the CMSSM for $\tb = 10$ (left panel) and $\tb = 50$
(right panel). We recall that this theoretical prediction has an
intrinsic uncertainty of $\sim \pm 1.5 \gev$, which should be combined
with the experimental error in $\mt$. It is a clear prediction of 
this analysis that $\Mh$ should be very close to the LEP lower limit,
and probably $\lsim 120 \gev$, though a value as large as $\sim 123 \gev$
is possible (but is $\chi^2$ disfavoured), particularly if $\tb = 50$.

In the case of the SM, it is well known that tension between the
lower limit on $\Mh$ from the LEP direct search and the relatively low
value of $\Mh$ preferred by the EWPO has recently been
increasing~\cite{LEPEWWG,TEVEWWG}. 
This tension is strongly reduced within the CMSSM, particularly for 
$\tb = 50$.  
We display in \reffi{fig:noLEPMh} the global $\chi^2$~functions
for the EWPO and BPO, but this time {\it omitting} the
contribution for the LEP Higgs search. This corresponds to the fitted
value of $\Mh$ in the CMSSM. Comparing \reffi{fig:noLEPMh} and
\reffi{fig:allMh}, we see that all data (excluding $\Mh$) favour a
 value of $\Mh \sim 110 \gev$ if $\tb = 10$ and $\Mh \sim 115 \gev$ if
$\tb = 50$. On the other hand, 
the currently best-fit value of $\MHSM$ is $76 \gev$~\cite{LEPEWWG},
i.e.\ substantially below the SM LEP bound of $114.4 \gev$~\cite{LEPHiggsSM}.
In comparison to the favoured values {\it including} the LEP
limits we get a $\sim 5 \gev$ smaller value of $\Mh$ if $\tb = 10$,
whereas the difference is only $\sim 1 \gev$ if $\tb = 50$.%
\footnote{We also recall that the estimated theoretical uncertainty in
  $\Mh$ for fixed values of the CMSSM parameters is $\pm 3 \gev$.}%
~Correspondingly, comparing \reffi{fig:noLEPMh} and \reffi{fig:allMh},
we see that the LEP limit increases the value of $\chi^2$ by $\sim 3.5$
for the $\tb = 10$ case, but by only $\sim 1$ for the $\tb = 50$
case. However, we emphasize that for both cases there are quite
good global fits to all the EWPO and BPO with $\chi^2 \sim 4.5$.

\begin{figure}[tbh!]
\begin{center}
\includegraphics[width=.48\textwidth]{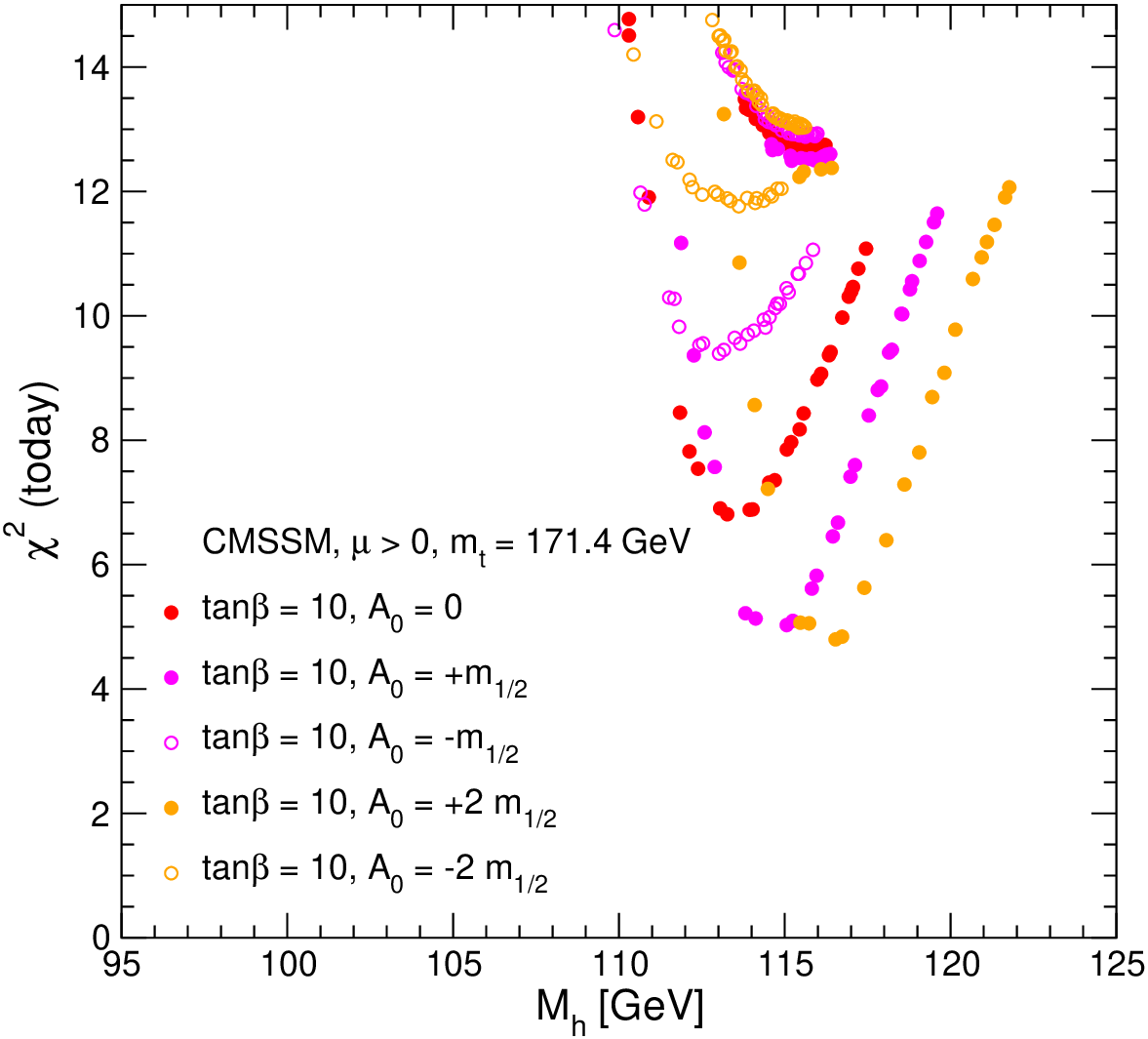}
\includegraphics[width=.48\textwidth]{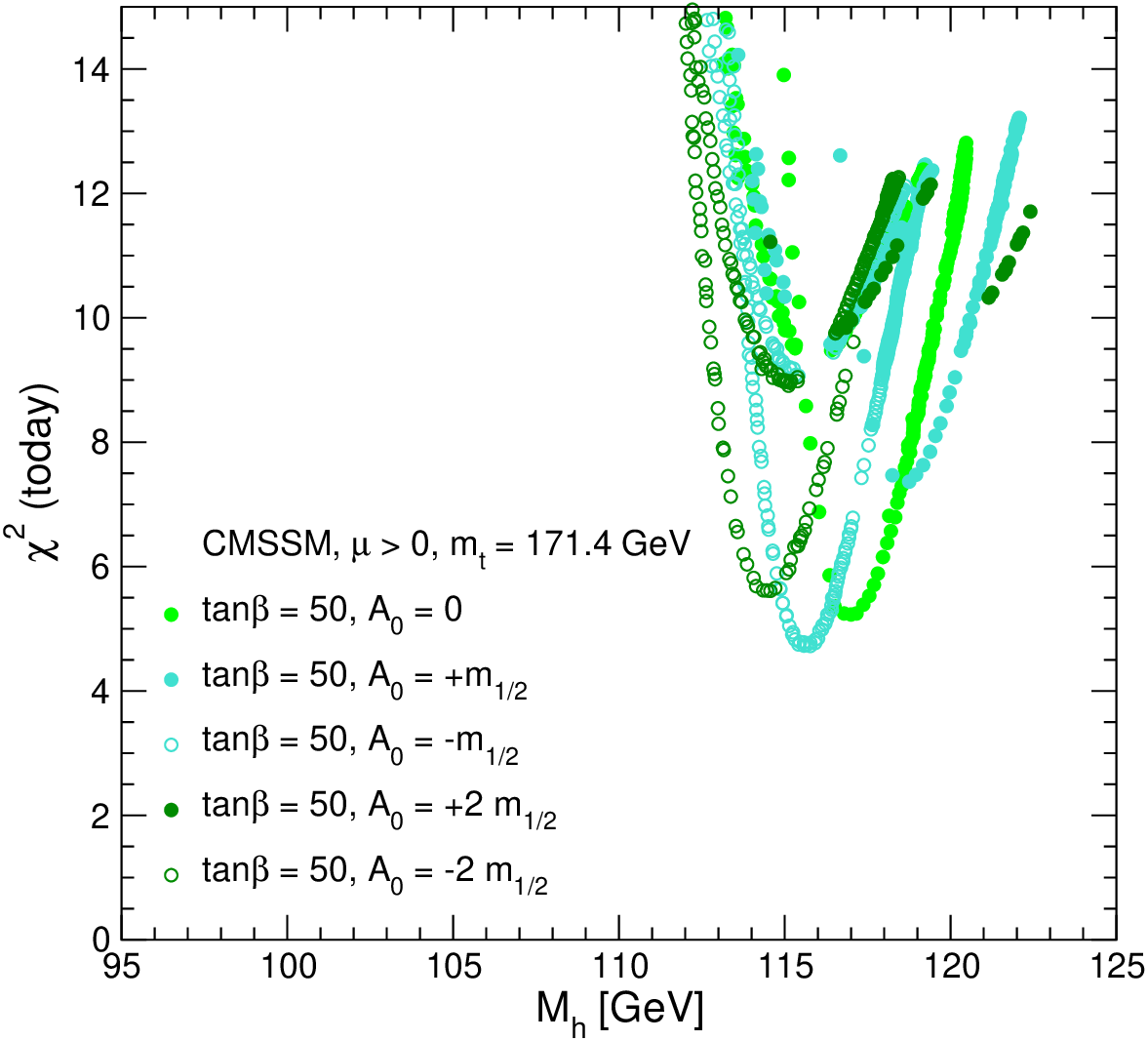}
\caption{%
The combined $\chi^2$~function
for $\Mh$, as obtained from
the combined analysis of all EWPO and BPO,
evaluated in the CMSSM for $\tb = 10$ (left) and
$\tb = 50$ (right) for various discrete values of $A_0$.
We use
$\mt = 171.4 \pm 2.1 \gev$ and $\mb(\mb) = 4.25 \pm 0.11 \gev$, and 
$m_0$ is chosen to yield the central value of the cold dark matter
density indicated by WMAP and other observations for the central values
of $\mt$ and $\mb(\mb)$.}
\label{fig:allMh}
\end{center}
\vspace{-1em}
\end{figure}

\begin{figure}[tbh!]
\begin{center}
\includegraphics[width=.48\textwidth]{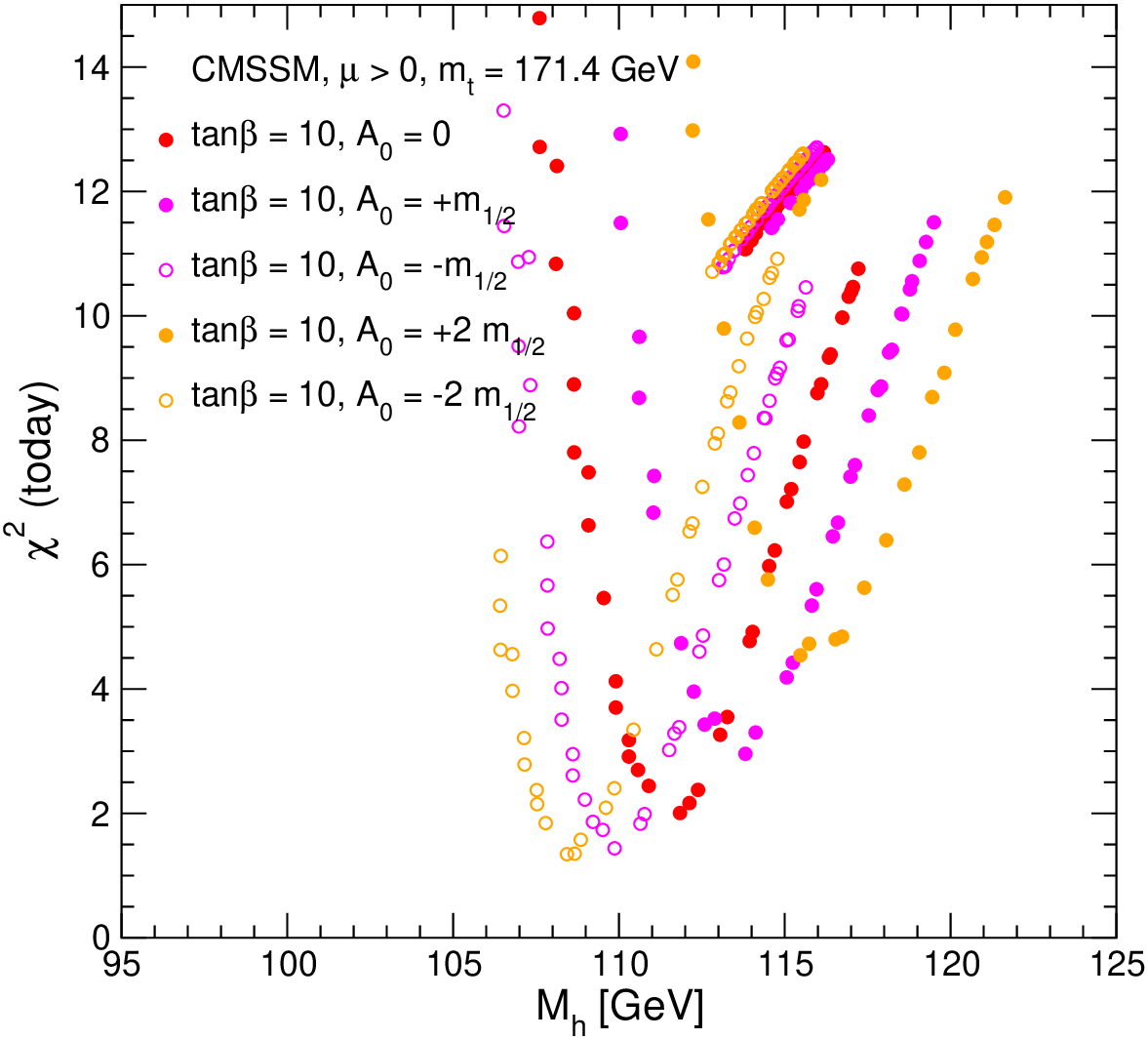}
\includegraphics[width=.48\textwidth]{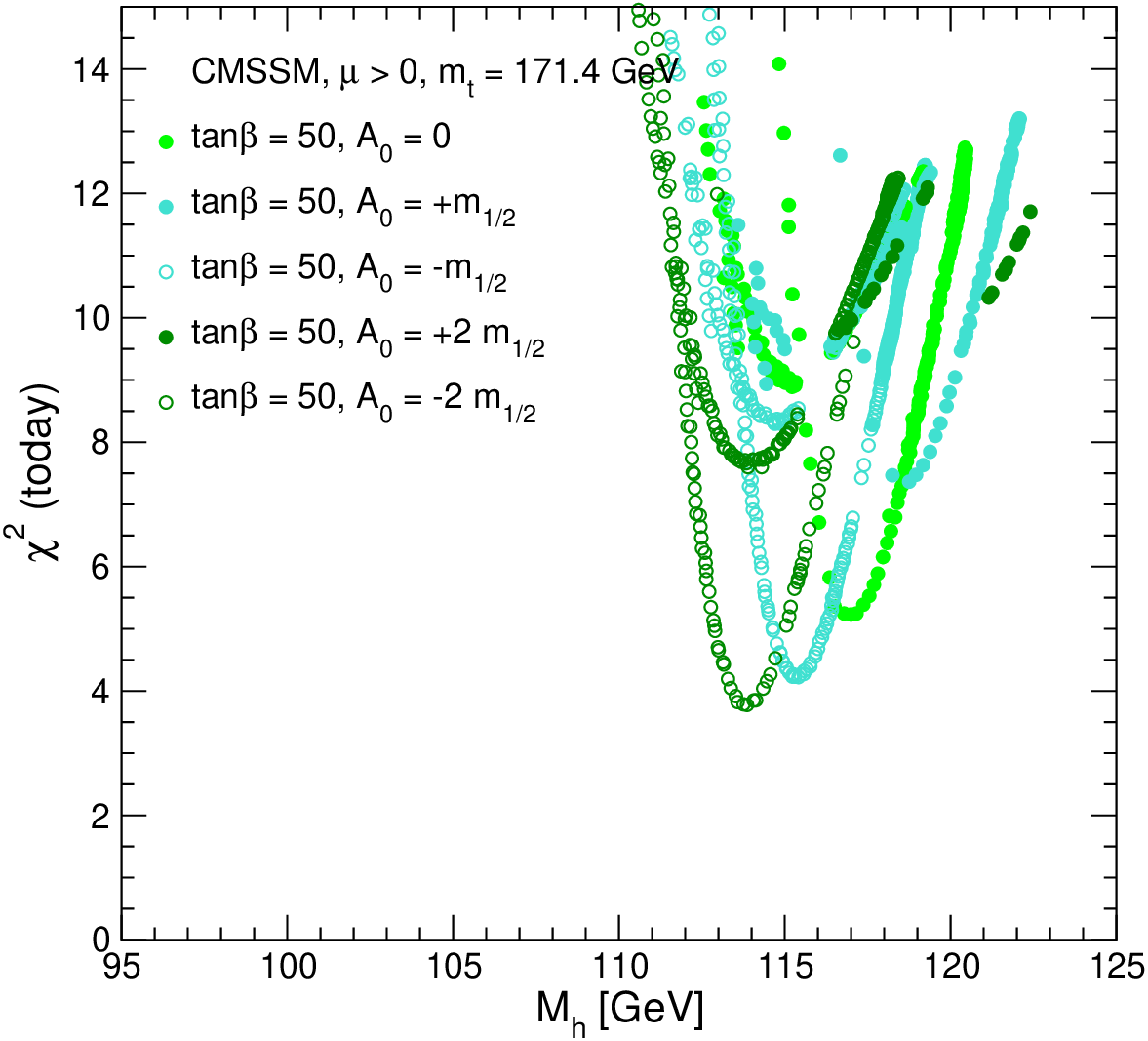}
\caption{%
The combined $\chi^2$~function for $\Mh$, as obtained from a
combined analysis of all EWPO and BPO {\it except} the LEP Higgs search,
as evaluated in the CMSSM for $\tb = 10$ (left) and
$\tb = 50$ (right) for various discrete values of $A_0$. We use
$\mt = 171.4 \pm 2.1 \gev$ and $\mb(\mb) = 4.25 \pm 0.11 \gev$, and 
$m_0$ is chosen to yield the central value of the cold dark matter
density indicated by WMAP and other observations for the central values
of $\mt$ and $\mb(\mb)$.}
\label{fig:noLEPMh}
\end{center}
\vspace{-1em}
\end{figure}


\section{NUHM Analysis Including EWPO and BPO}
\label{sec:nuhm}

The CMSSM is a very particular case of the general MSSM. It has a manageable
parameter space, but may not be able to capture all the possibilities
available in the general MSSM. Specifically, as we have seen, while
providing a better fit to the EWPO than the SM, it provides no
improvement for the BPO, and there is a slighttension between the EWPO and
the BPO within the restrictive CMSSM framework. For these and other
reasons, we now consider the NUHM. The dimensionality of the MSSM
parameter space is increased  by a manageable amount compared to the 
CMSSM, namely two extra dimensions. Moreover, there was no strong
phenomenological motivation for assuming universality for the Higgs
masses, and there is reason to hope that relaxing the Higgs universality
assumption may help reconcile the EWPO and BPO. As we have seen, the
EWPO prefer a specific range of $m_{1/2}$ ($\sim 300 \gev$ for 
$\tb = 10$ and $\sim 500 \gev$ for $\tb = 50$) in the coannihilation region,
and disfavour the focus-point region (particularly for $\tb = 10$).
Within the CMSSM, the electroweak vacuum conditions fix the
corresponding values of $|\mu|$ and $\MA$. These values are not very
crucial for the EWPO, but are potentially important for the BPO. For
example, the extra MSSM contribution to $b \to s \ga$ is small only if
the supplementary charged-Higgs and chargino diagrams cancel to some
extent, which imposes a specific condition on their masses. This may not
be satisfied within the CMSSM, but is not incompatible {\it a priori}
with the NUHM, in which $|\mu|$ and $\MA$ become (to some extent) free
parameters. 

Just as we focused attention in the previous CMSSM analysis on WMAP
lines in parameter space, where the cold dark matter density falls
within the range allowed by WMAP and other astrophysical and
cosmological observations, we also focus  on `WMAP surfaces' in in the
NUHM parameter space. Many NUHM parameter planes have been considered in
the past~\cite{NUHM1,NUHM2,ehow2} and, as in the CMSSM, generically the dark
matter constraint is satisfied only in thin strips in each NUHM
plane. Many phenomenological
studies of MSSM Higgs physics have analyzed the
possibilities in \plane{\MA}{\tb}s under different hypotheses for
other MSSM parameters. In particular, in the general MSSM framework,
\plane{\MA}{\tb}s have been suggested for phenomenological Higgs physics
analyses~\cite{benchmark,benchmark2,benchmark3}, neglecting the
constraints coming from CDM. 
In the NUHM, in order to keep the dark matter density within
the WMAP range across generic regions of such a \plane{\MA}{\tb}, one
must adjust one or more of the free parameters continuously across the
plane. We propose here two strategies for specifying suitable parameter
scans. 

We consider first a typical \plane{\MA}{\tb}\ for fixed $m_0, m_{1/2}$
and $\mu$, as shown for example in Fig.~5a of~\citere{NUHM1}, where the
choices $m_0 = 800 \gev$, $m_{1/2} = 600 \gev$ and $\mu = 1000 \gev$
were made. In this example, the relic density exceeds
the WMAP upper limit almost everywhere in the \plane{\MA}{\tb}, except along
a narrow vertical strip where $\mneone \sim \MA/2$, and rapid direct-channel
annihilation suppresses the relic density below the WMAP range. On either side
of this strip, there are thin regions where the relationship between $\MA$ and
$m_{1/2}$ is such that the relic density falls within the WMAP range whatever
the value of $\tb$. Building on this observation, we study a \plane{\MA}{\tb}
\Athree\ with the same values of $m_0 = 800 \gev$ and $\mu = 1000 \gev$,
but with $m_{1/2}$ chosen to vary across the plane so as to maintain the WMAP
relationship with $\MA$:
\begin{equation}
\frac{9}{8} \MA  - 12.5  \gev \le m_{1/2} \;  \le \frac{9}{8} \MA + 37.5\gev.
\label{A3}
\end{equation}
As we saw earlier, within the CMSSM, smaller values of $m_0$ are preferred. We
therefore consider also a \plane{\MA}{\tb}\ \Afive\ with the fixed values 
$m_0 = 300 \gev$ and $\mu = 800 \gev$, and $m_{1/2}$ again adjusted
continuously across the plane so as to maintain the WMAP relationship
with $\MA$: 
\begin{equation}
1.2 \MA  - 40 \gev \le m_{1/2} \;  \le 1.2 \MA + 40 \gev.
\label{A5}
\end{equation}
Many more examples could be chosen, but these serve as representative
examples of NUHM planes that enable us to explore the possibility of reducing
the tension between the EWPO and BPO.

We also consider two more examples, inspired by the \plane{\mu}{\MA}s also
shown in \citere{NUHM1}. There we see that, for fixed values of 
$m_{1/2}$ and $m_0$, there is a `magic' value of $\mu > 0$ which provides
a suitable value of the relic density for almost all values of $\MA$,
the exception 
being a narrow strip around the rapid-annihilation funnel where $\mneone \sim
\MA/2$. The value of $\mu$ varies with $\tb$, but a suitable scan yields a
\plane{\MA}{\tb}\ where the relic density falls within the WMAP range for all
except a sliver of $\MA$ values that broadens somewhat as $\tb$
increases. Within this sliver, the relic density falls below the WMAP
range: this region is therefore not incompatible with cosmology, but
would require a supplementary source of cold dark matter.
One example of such a plane, \Atwo, has fixed $m_{1/2} = 500 \gev$ and 
$m_0 = 1000 \gev$, with $\mu$ in the range
\begin{equation}
\mu \; =  250 - 400 \gev.
\label{A2}
\end{equation}
The other example \Afour\ has fixed $m_{1/2} = 300 \gev$ and 
$m_0 = 300 \gev$, with $\mu$ in the range
\begin{equation}
\mu \; = \; 200 - 350 \gev.
\label{A4}
\end{equation}
The four scenarios are summarized in \refta{tab:nuhmscen}. In the
analyses below, we quote the minimal values of $\chi^2$ for values of
$\mu$ within the ranges (\ref{A2}), (\ref{A4}), 
for the planes \Atwo\ and \Afour, respectively.

\begin{table}[tbh!]
\renewcommand{\arraystretch}{1.5}
\BC
\begin{tabular}{|c||c|c|c|c|}
\cline{2-5} \multicolumn{1}{c||}{}
 & $m_{1/2}$ & $m_0$ & $A_0$ & $\mu$ \\ \hline\hline
\Athree & varied & 800  & 0 & 1000   \\ \hline
\Afive  & varied & 300  & 0 & 800    \\ \hline
\Atwo   & 500    & 1000 & 0 & varied \\ \hline
\Afour  & 300    & 300  & 0 & varied \\ 

\hline\hline
\end{tabular}
\EC
\renewcommand{\arraystretch}{1}
\caption{
The four NUHM scenarios, with $\MA$ and $\tb$ kept as free parameters. All
masses are in GeV.
}
\label{tab:nuhmscen}
\end{table}

We now consider the most important contributions to the likelihood functions 
for these four \plane{\MA}{\tb}s.

The principal contributions to the 
overall $\chi^2$~value, and hence also to the
likelihood function, for the
\plane{\MA}{\tb}\ for scenario \Athree\ are shown in
\reffi{fig:A3}. We see in panel (a) that the 
$\chi^2$~%
value of $\amu$ is not very
satisfactory anywhere in the plane, but particularly not at large $\MA$
and small $\tb$. Panel (b) shows that the LEP lower limit on $\Mh$
disfavours $\MA < 300 \gev$ in this scenario. Small values of $\MA$ are
also disfavoured by $b \to s \ga$, as seen in panel (c), and large
values of $\MA$ and $\tb$ are also disfavoured, but to a lesser
extent. Panels (d) and (e) show that large values of $\tb$ and small
values of $\MA$ are disfavoured by both $B_s \to \mu^+ \mu^-$ and 
$B_u \to \tau \nu_\tau$. Finally, panel (f) shows the values of the
combined EWPO and BPO $\chi^2$~function for scenario \Athree\ throughout the
\plane{\MA}{\tb}. We see that the best-fit point has $\MA \sim 440 \gev$
and $\tb \sim 50$. This is the optimal compromise between $\amu$ and 
$b \to s \ga$ that also respects the $\Mh$ and other BPO constraints. We
note that this best-fit point has $\chi^2 = 7.1$, which is not a
significant improvement, but even slightly worse than the CMSSM fits
discussed in the previous 
Section. 
We also display in panel (f) of \reffi{fig:A3} the 
$\De \chi^2 = 2.30$ and 4.61 contours, which would correspond to the
68~\% and 95~\% C.L. contours in the \plane{\MA}{\tb} {\it if} the
overall likelihood distribution, $\cL \propto e^{-\chi^2/2}$,
was Gaussian. This is clearly only roughly the case in this
analysis, but these contours nevertheless give interesting indications
on the preferred region in the \plane{\MA}{\tb}. 

We do not show results in the upper right corners of these planes
(with high $\MA$ and high $\tb$) because there the relic density in this
region is low compared to the preferred WMAP value.  In this region, for
the choice of $m_0$ and $\mu$ and the range of $m_{1/2}$ given in
\refeq{A3}, we are sitting too close to the funnel. However, these
points could be brought into agreement with WMAP, by extending the
sampled range in $m_{1/2}$ to lower values. The lower left portions of
these planes are missing because of the finite resolution of our
scan. In these regions of fixed low values of $\MA$ and $\tb$, the relic
density is very sensitive to $m_{1/2}$, and viable points are missed
with the resolution in $m_{1/2}$ of $10 \gev$ 
that we use. One final remark on \reffi{fig:A3} concerns high values of
$\tb$. At values of $\tb > 52$, the RGE evolution may break down due to a
tendency towards a divergent bottom Yukawa coupling.

\begin{figure}[tbh!]
\begin{center}
\includegraphics[width=.40\textwidth]{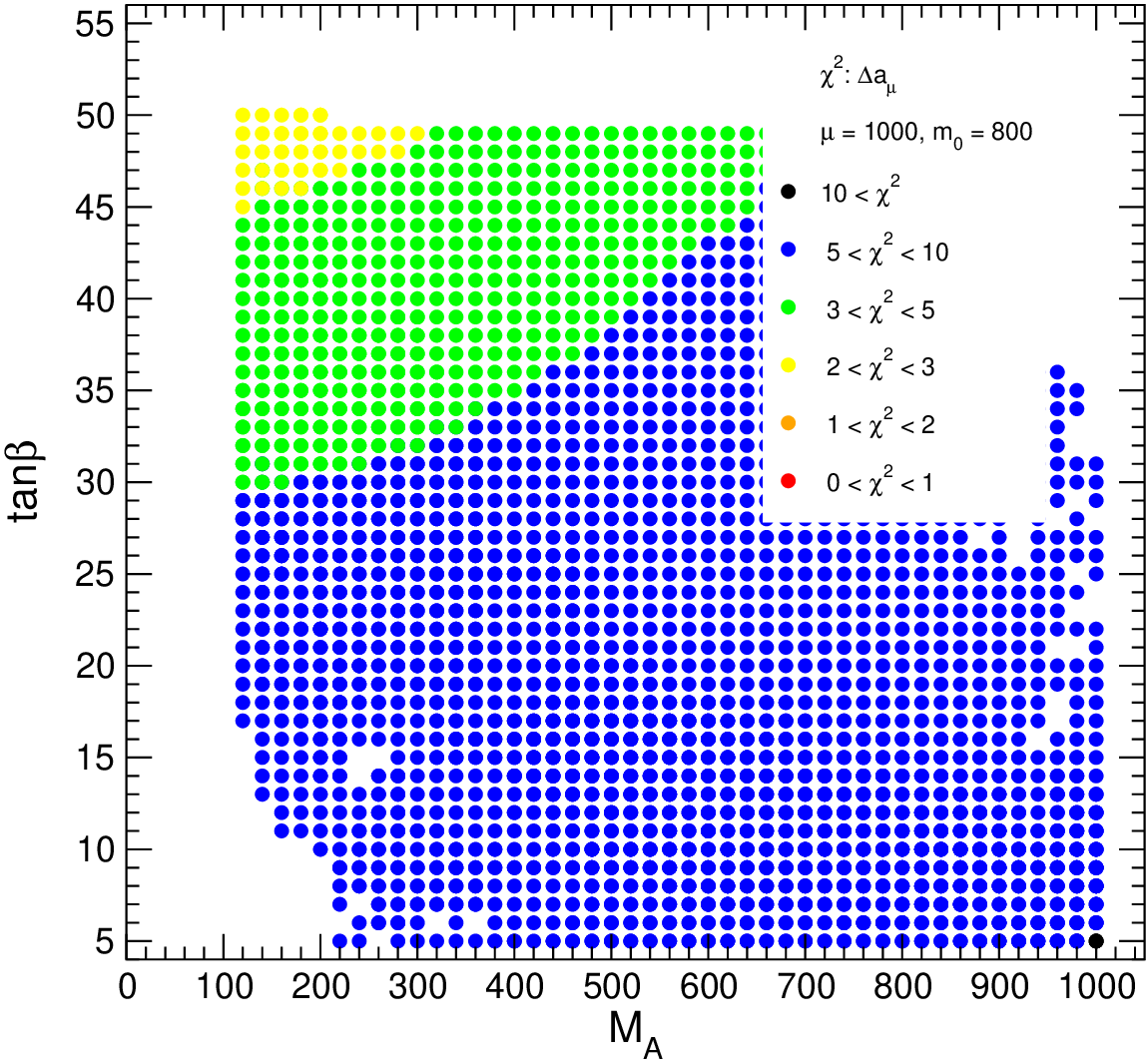}
\includegraphics[width=.40\textwidth]{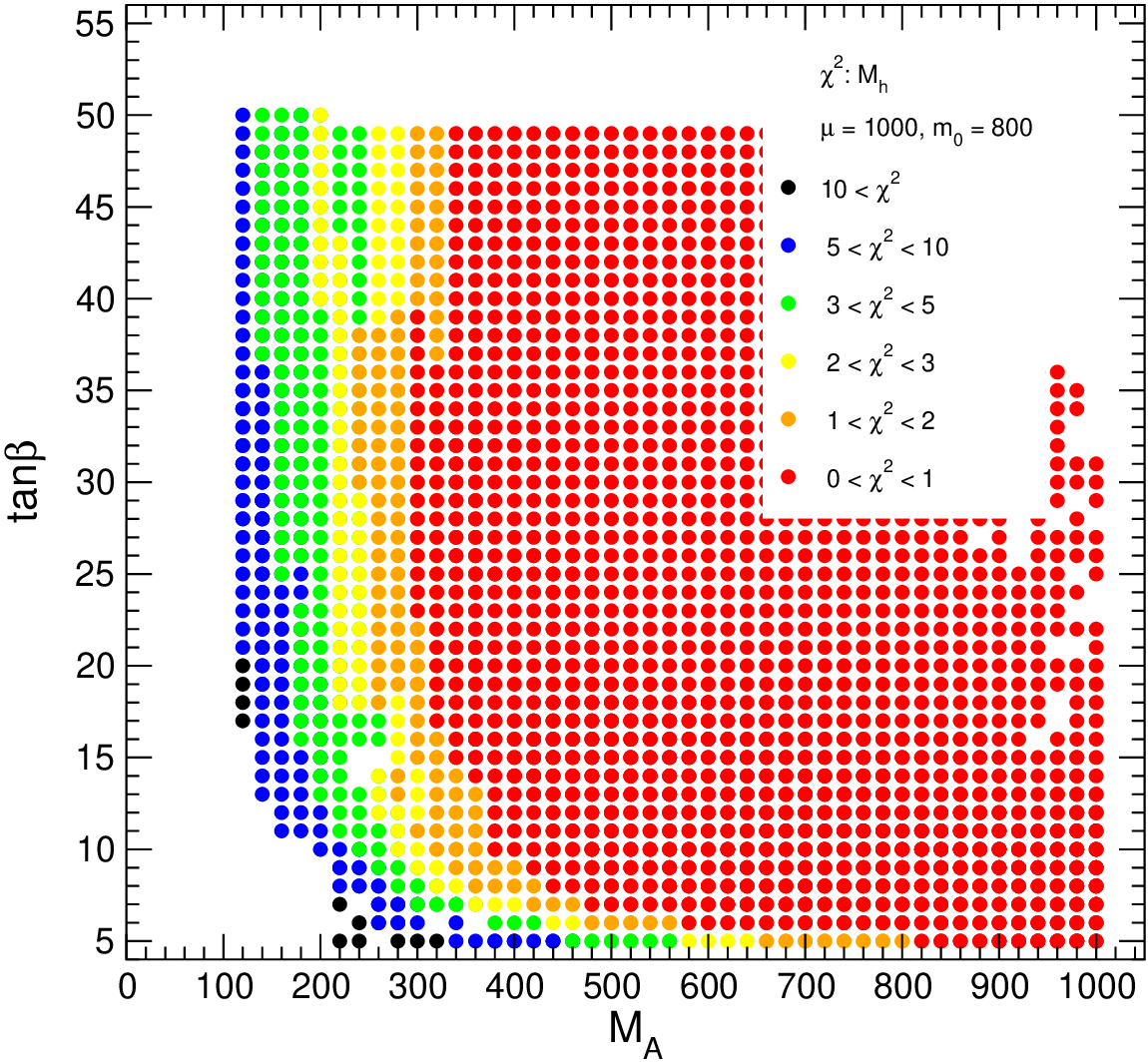}
\end{center}
\begin{center}
\includegraphics[width=.40\textwidth]{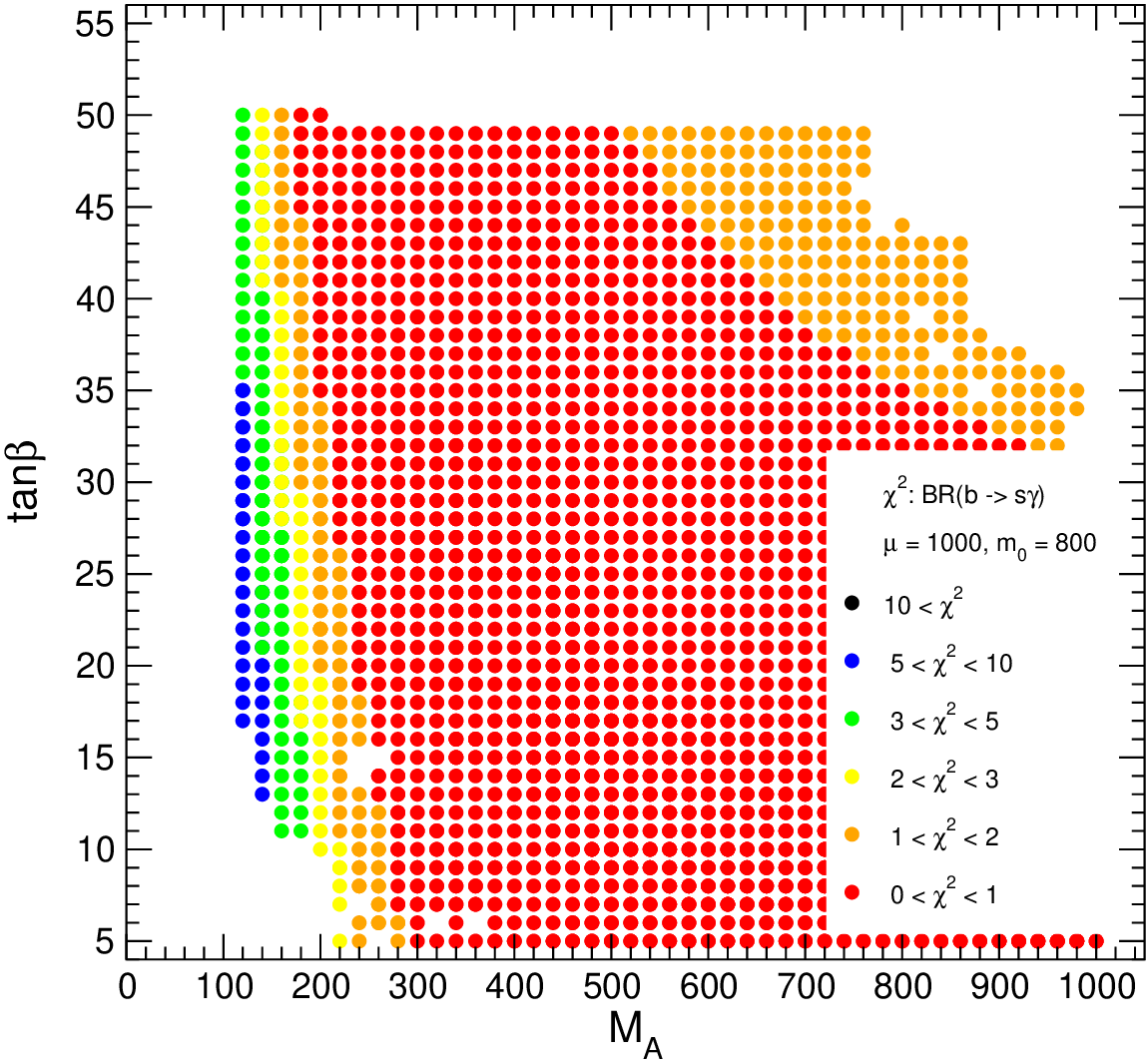}
\includegraphics[width=.40\textwidth]{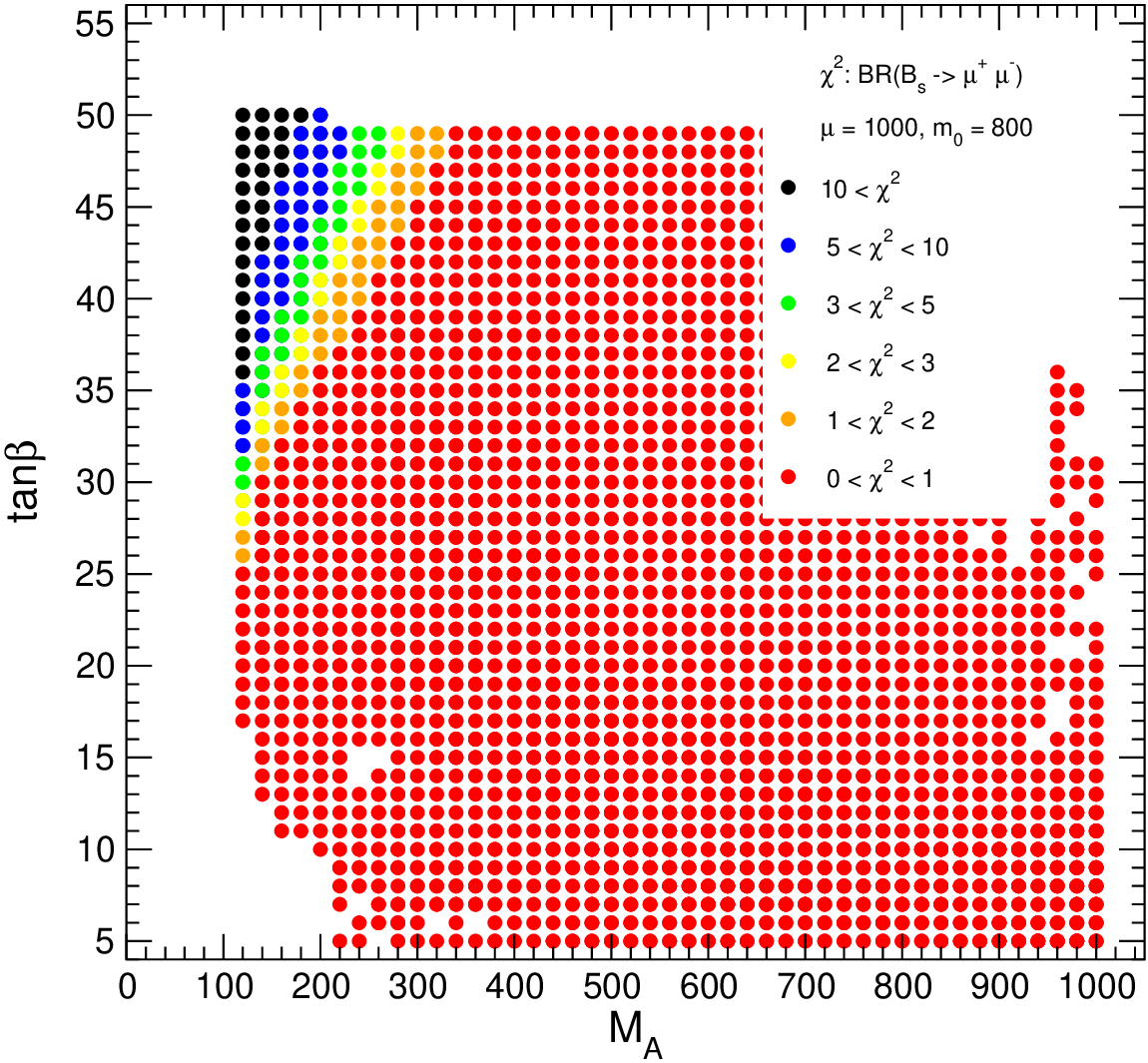}
\end{center}
\begin{center}
\includegraphics[width=.40\textwidth]{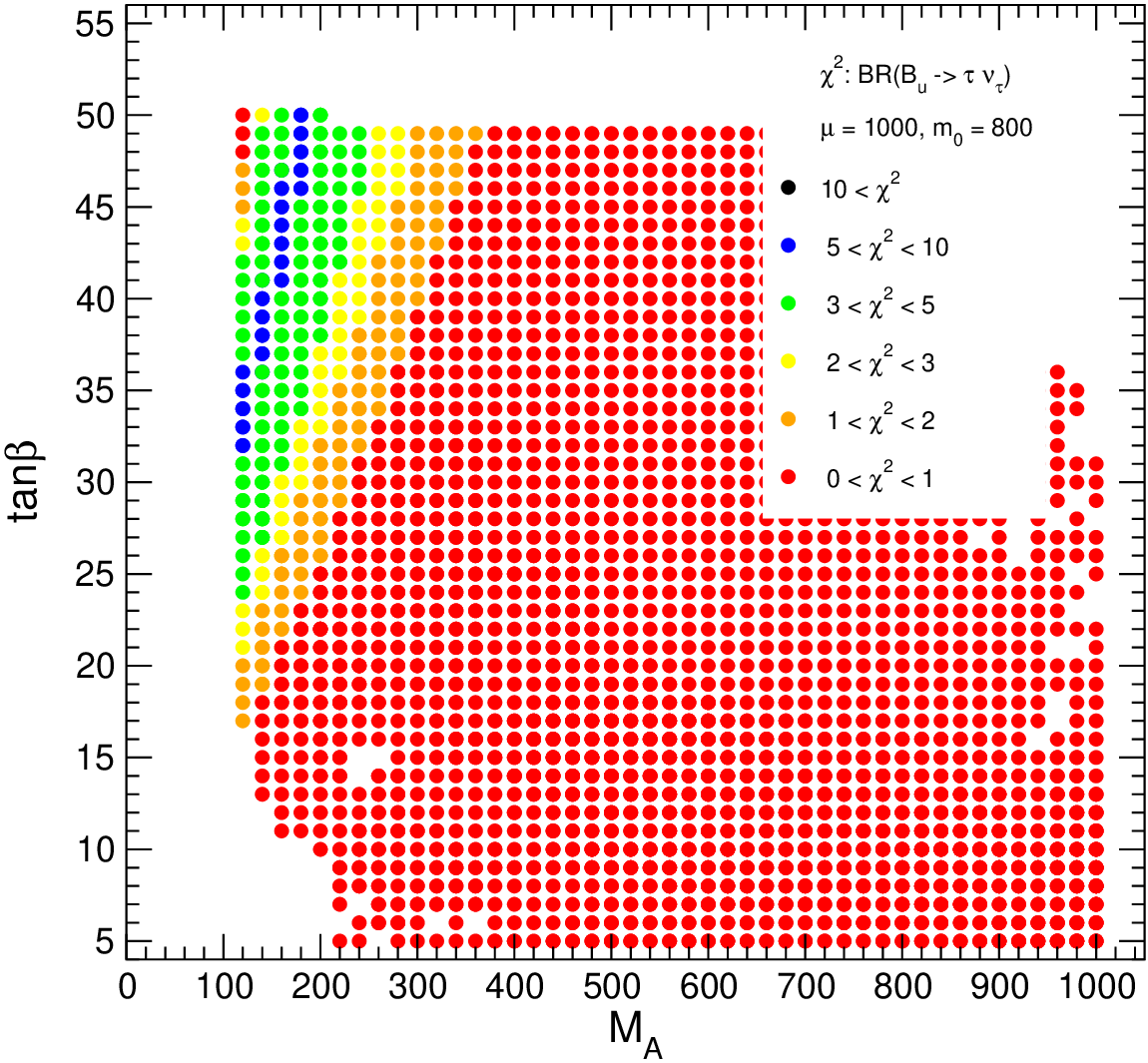}
\includegraphics[width=.40\textwidth]{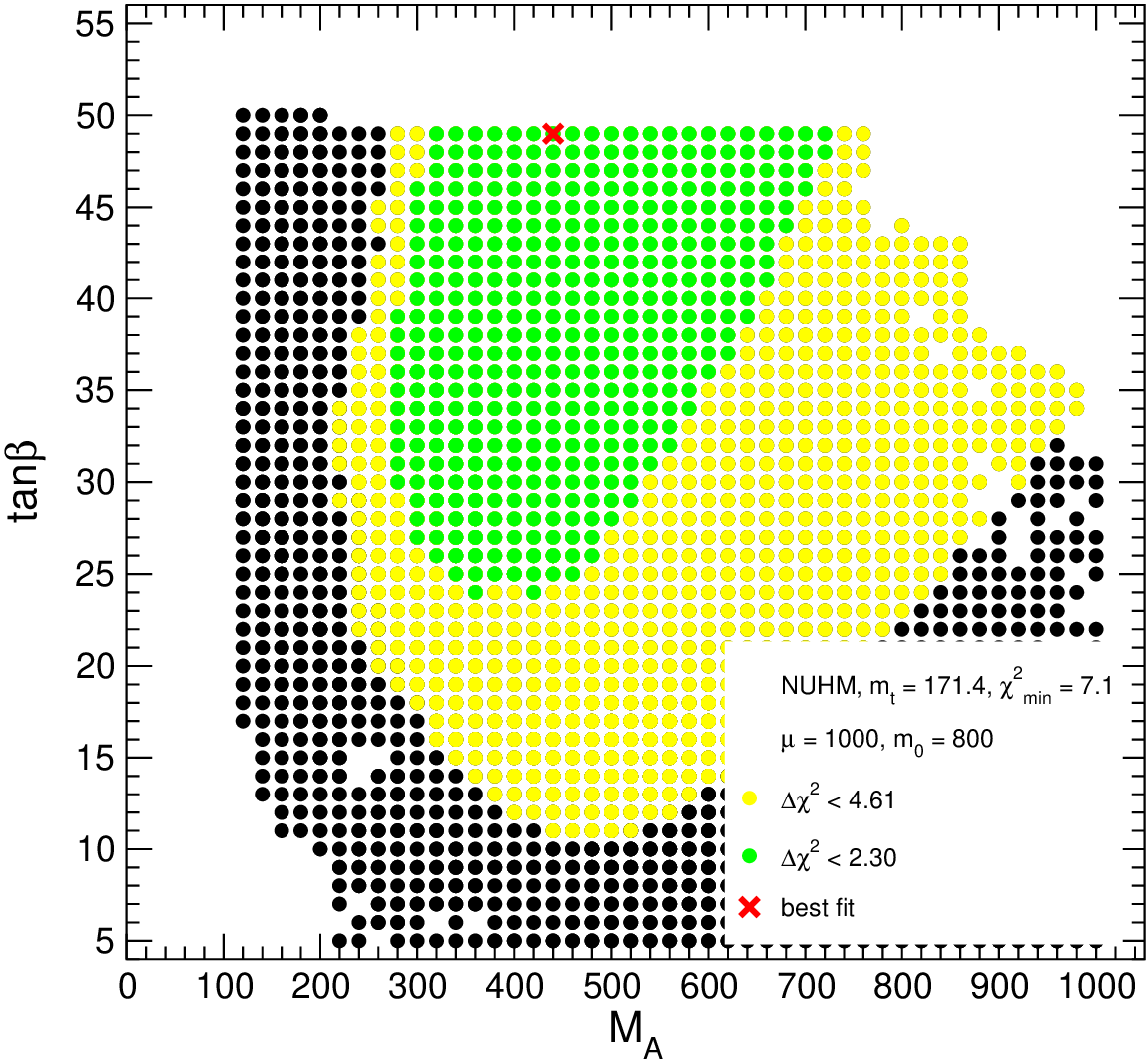}
\caption{%
The most important contributions to the total $\chi^2$~value
for the NUHM 
\plane{\MA}{\tb}\ \Athree, due to (a) $\amu$, (b) $\Mh$, (c) $b \to s
\ga$, (d) $B_s \to \mu^+ \mu^-$ and (e) $B_u \to \tau \nu_\tau$, and (f)
the combined EWPO and BPO $\chi^2$~function.
We use $\mt = 171.4 \pm
2.1 \gev$ and $\mb(\mb) = 4.25 \pm 0.11 \gev$, and $m_{1/2}$ is adjusted
continuously so as to yield the central value of the cold dark matter
density indicated by WMAP and other observations for the central values
of $\mt$ and $\mb(\mb)$.}
\label{fig:A3}
\end{center}
\vspace{-6em}
\end{figure}

The principal contributions to the total $\chi^2$~function
for the \plane{\MA}{\tb}
for scenario \Afive\ are shown in \reffi{fig:A5}. We see in panel (a)
that the value of $\amu$ is very satisfactory in a band running across
the plane from $(\MA, \tb) \sim (100 \gev$$, 15)$ to  
$\sim (400 \gev$$, 50)$. In particular, large values of $\MA$ and small
$\tb$ are disfavoured. Panel (b) shows that the LEP lower limit on $\Mh$
disfavours  $\MA < 300 \gev$ also in this scenario. Small values of
$\MA$ are also disfavoured by $b \to s \ga$, as seen in panel (c), and
large values of $\MA$ and $\tb$ are also disfavoured, but to a lesser
extent. Panels (d) and (e) show that large values of $\tb$ and small
values of $\MA$ are again disfavoured by both $B_s \to \mu^+ \mu^-$ and
$B_u \to \tau \nu_\tau$. Finally, panel (f) shows the combined EWPO and
BPO $\chi^2$~function for scenario \Afive\ throughout the
\plane{\MA}{\tb}. We see that the best-fit point has $\MA \sim 340 \gev$
and $\tb \sim 35$. This is a good fit to both $\amu$ and  $b \to s \ga$,
as well as the $\Mh$ and other BPO constraints. We note that this
best-fit point has $\chi^2 = 3.5$, which {\it is} a noticeable
improvement on the CMSSM fits discussed in the previous Section. 
We note that the $\De \chi^2 = 2.30$ and 4.61 contours are somewhat more
compact than in the case of scenario \Athree.

For the parameter choice of \Afive\, large values of $\MA$ are excluded
because the right-handed stau becomes the LSP. This could be avoided by
lowering the value of $m_{1/2}$ outside the range in \refeq{A5}, so as
to recover a neutralino LSP. However, unless we drop $m_{1/2}$
substantially below our adopted range, the relic density will be too 
small due to LSP-stau coannihilations. Finally, we note that the hole around
$(\MA, \tb) \sim (600 \gev$$, 17)$ is due to the funnel.
In this hole, the relic density is far too small to supply the preferred
amount of cold dark matter.  However, the hole could be filled if a
larger range were chosen for $m_{1/2}$.

\begin{figure}[tbh!]
\begin{center}
\includegraphics[width=.40\textwidth]{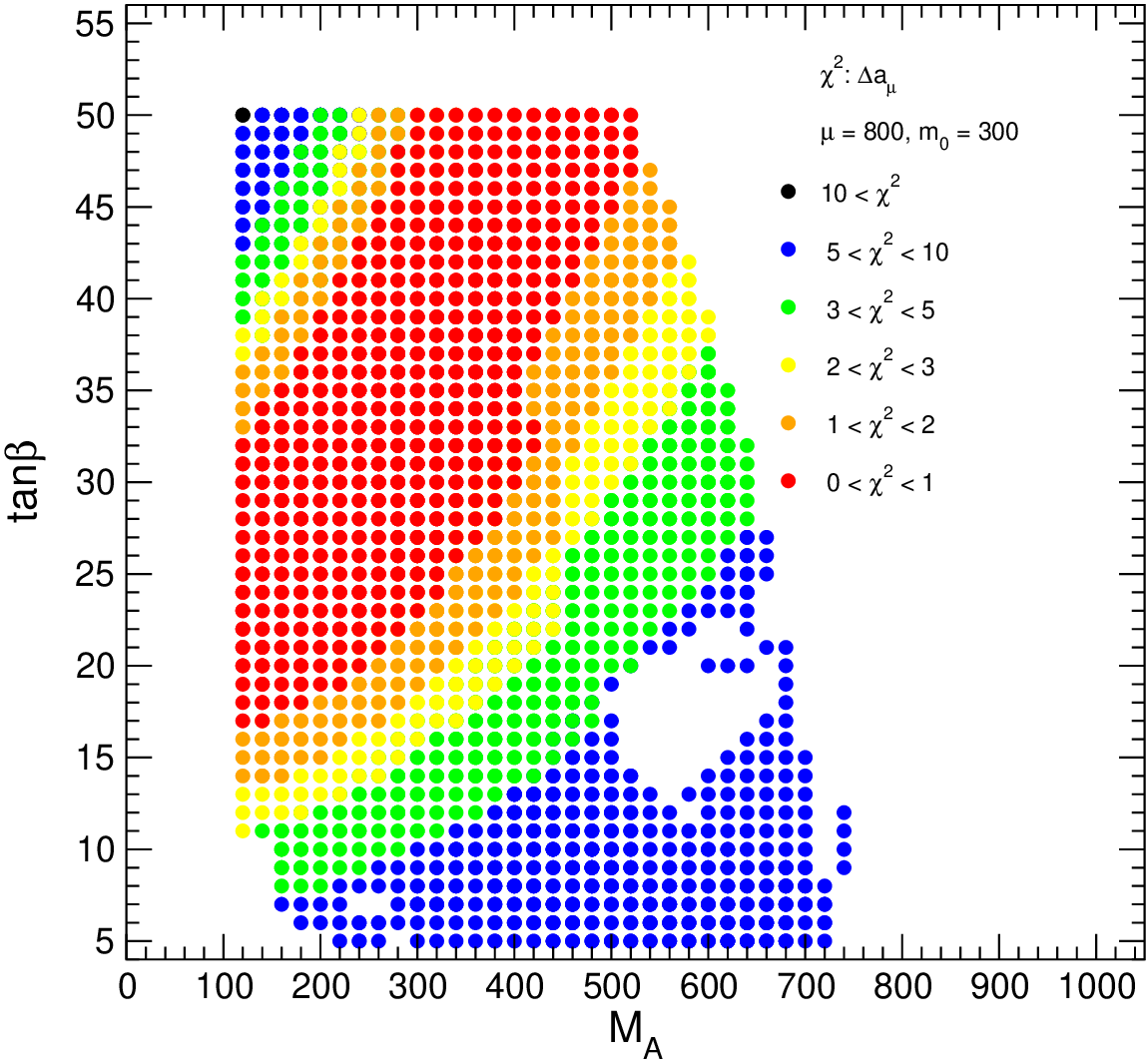}
\includegraphics[width=.40\textwidth]{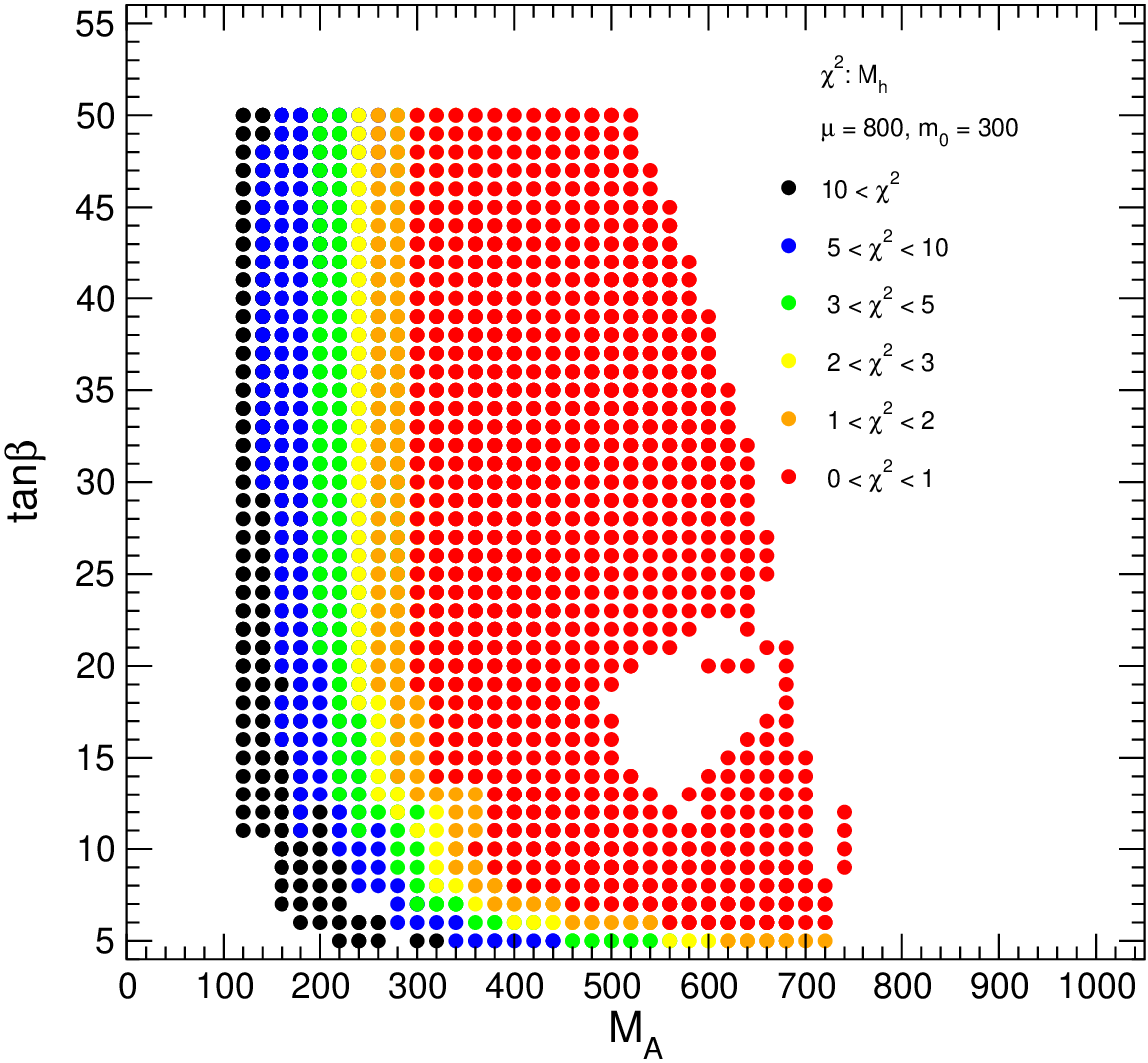}
\end{center}
\begin{center}
\includegraphics[width=.40\textwidth]{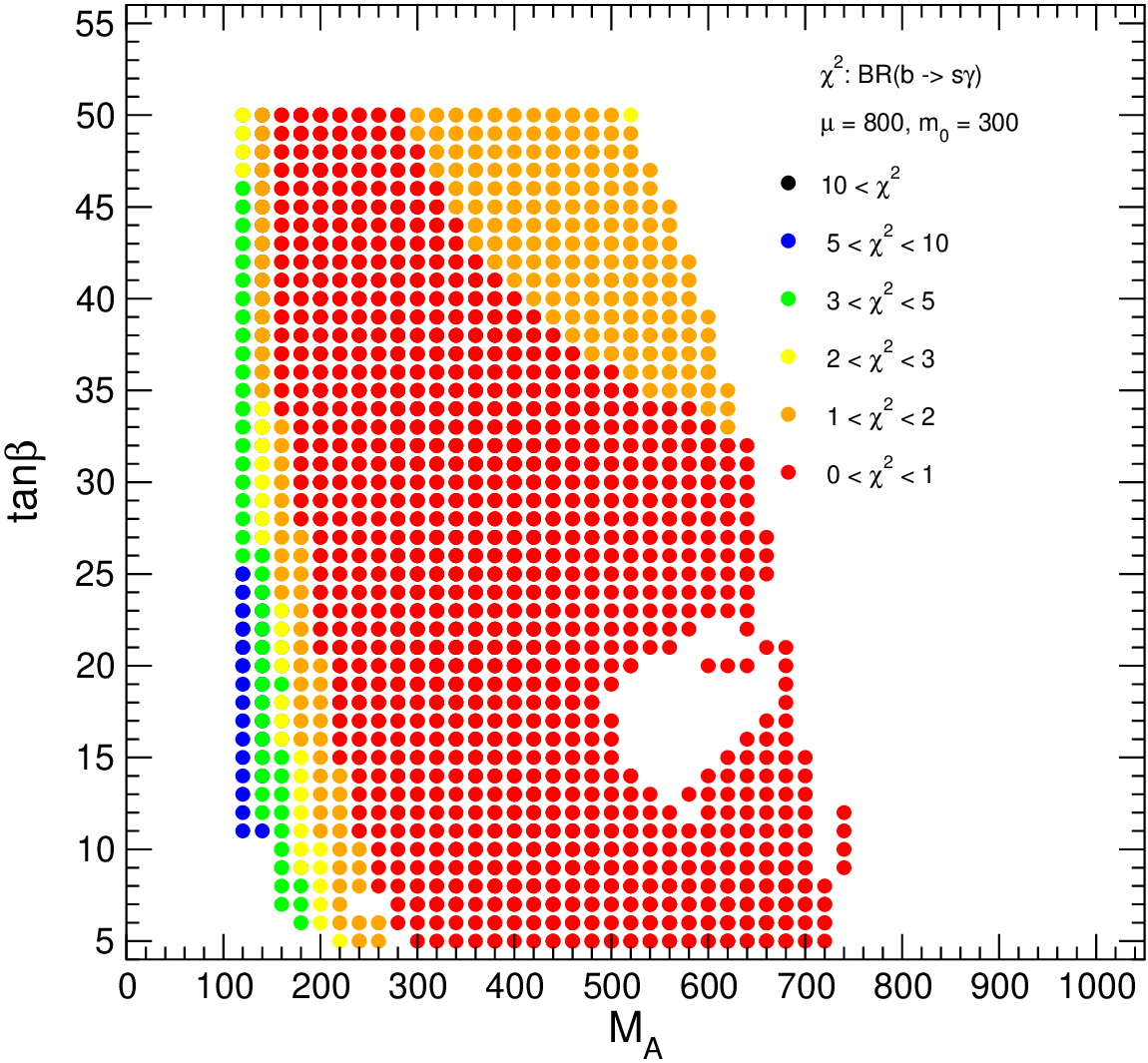}
\includegraphics[width=.40\textwidth]{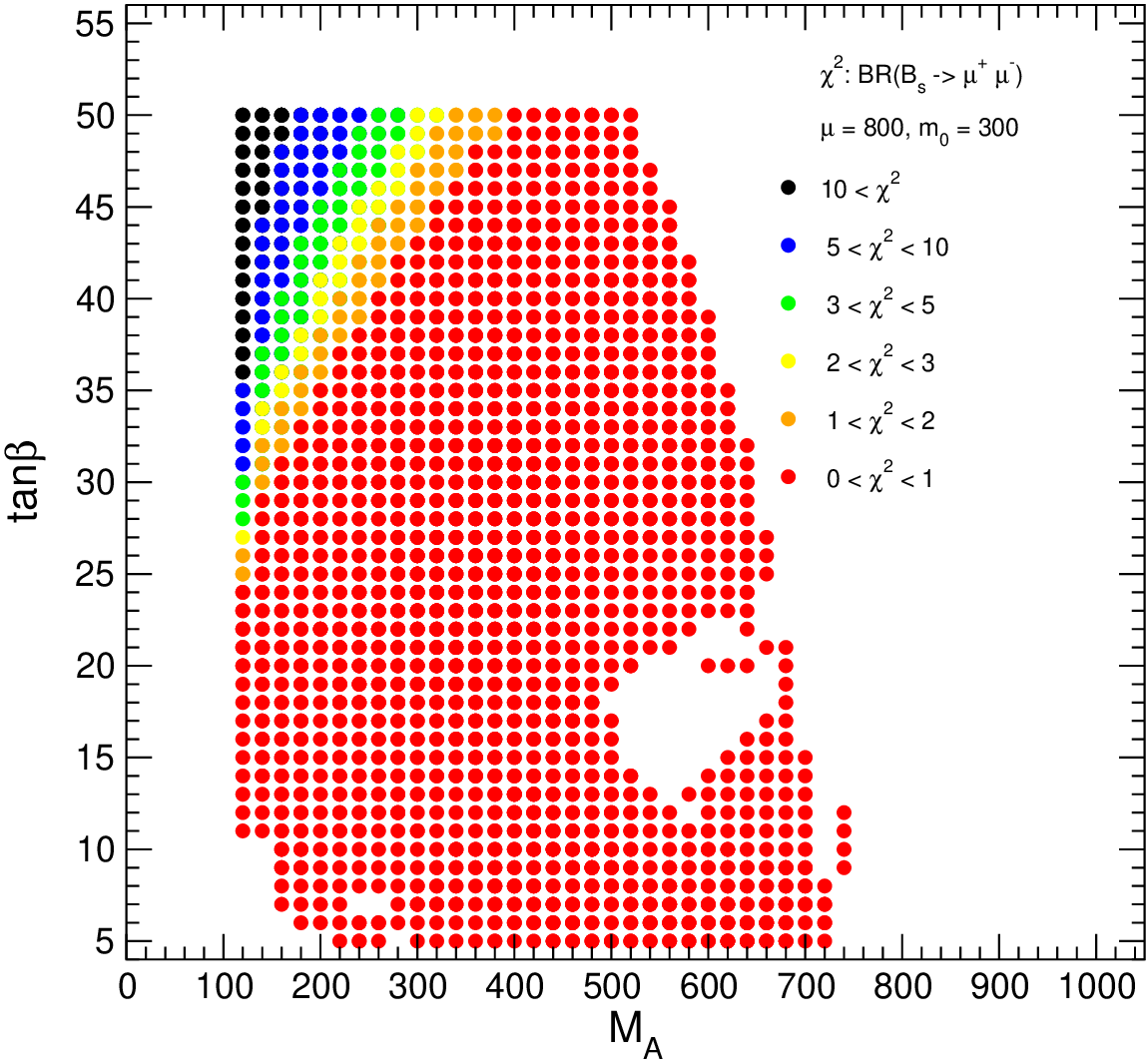}
\end{center}
\begin{center}
\includegraphics[width=.40\textwidth]{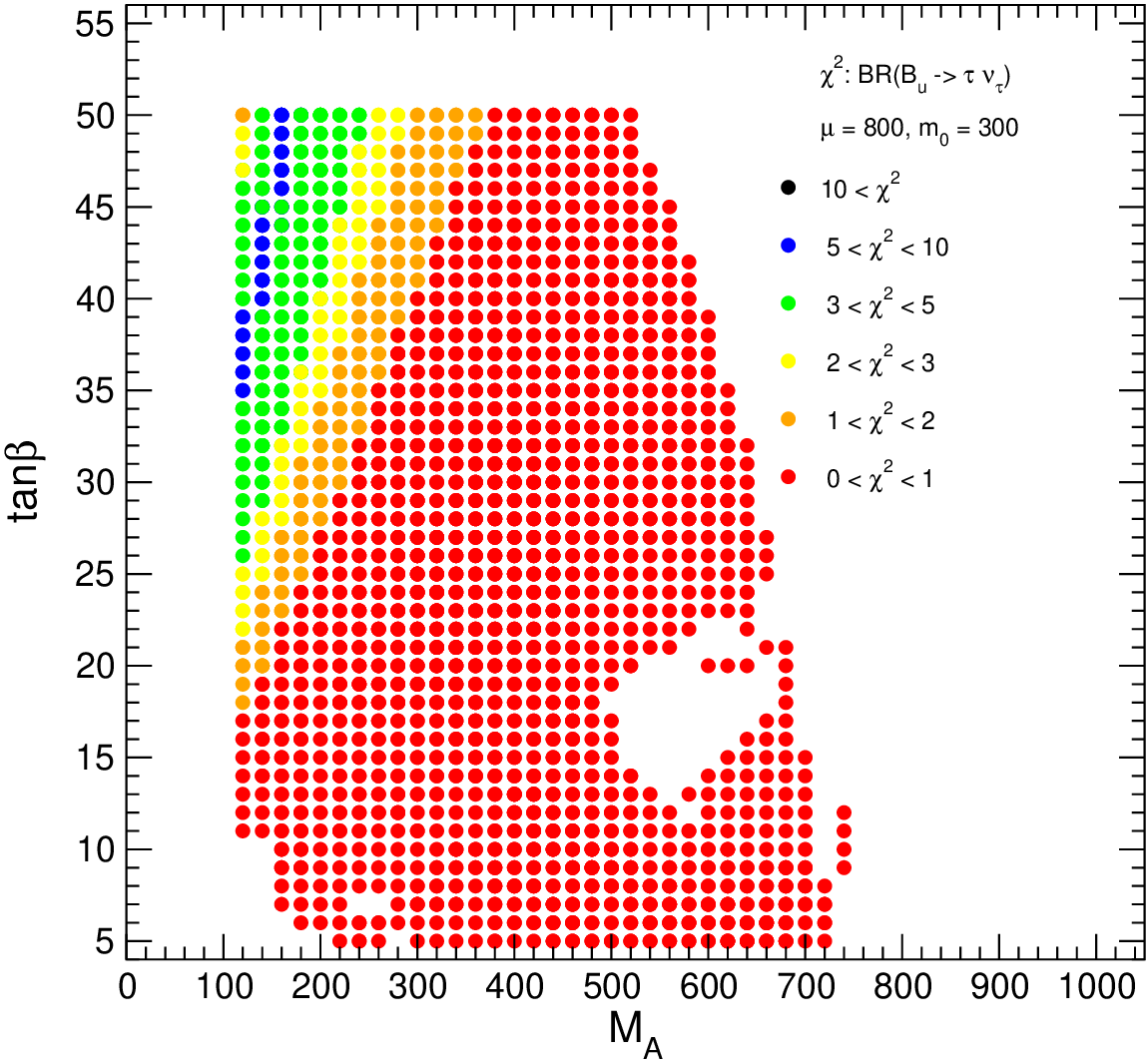}
\includegraphics[width=.40\textwidth]{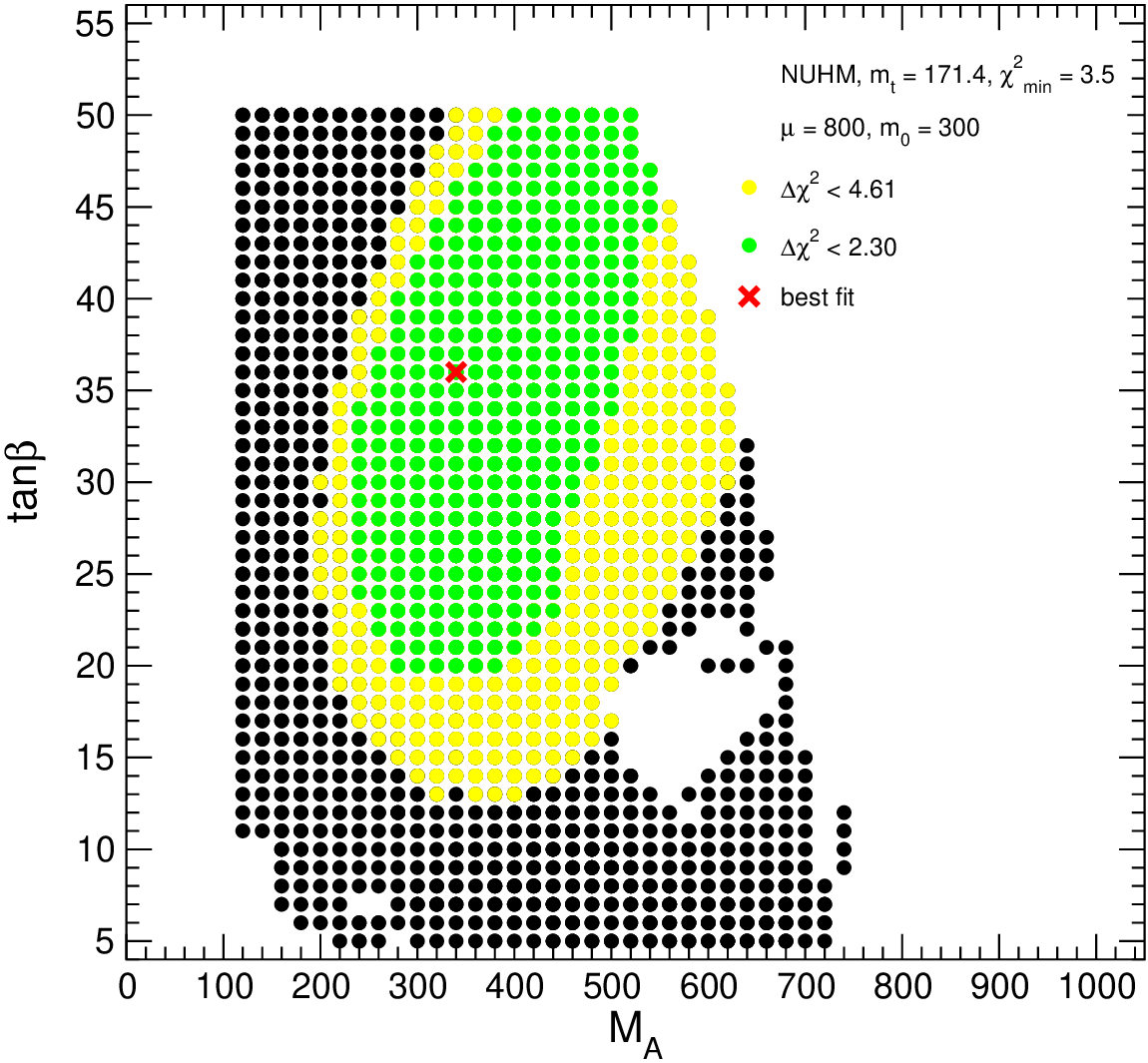}
\caption{%
The most important contributions to the total $\chi^2$~value
for the NUHM
\plane{\MA}{\tb}\ \Afive, due to (a) $\amu$, (b) $\Mh$, (c) $b \to s
\ga$, (d) $B_s \to \mu^+ \mu^-$ and (e) $B_u \to \tau \nu_\tau$, and (f)
the combined EWPO and BPO $\chi^2$~fucntion.
 We use $\mt = 171.4 \pm
2.1 \gev$ and $\mb(\mb) = 4.25 \pm 0.11 \gev$, and  $m_{1/2}$ is
adjusted continuously so as to yield the central value of the cold dark
matter density indicated by WMAP and other observations for the central
values of $\mt$ and $\mb(\mb)$.}
\label{fig:A5}
\end{center}
\vspace{-6em}
\end{figure}

The principal contributions to the total $\chi^2$~function
for the \plane{\MA}{\tb}
for scenario \Atwo\ are shown in \reffi{fig:A2}~%
\footnote{We do not display the $\chi^2$~values 
in the underdense slivers of the plane.}%
. We see in panel (a)
that the value of $\amu$ is satisfactory only for very large values of
$\tb$, almost independently of $\MA$. In particular, values of 
$\tb < 25$ are quite strongly disfavoured. Panel (b) shows that the LEP
lower limit on $\Mh$ disfavours very low values of $\tb$ and $\MA < 150
\gev$ for $\tb < 25$. Small values of $\MA$ and $\tb$ are also
disfavoured by $b \to s \ga$, as seen in panel (c), as also  are large
values of $\MA$ and $\tb$. Panels (d) and (e) show that large values of
$\tb$ and small values of $\MA$ are again disfavoured by both $B_s \to
\mu^+ \mu^-$ and $B_u \to \tau \nu_\tau$%
\footnote{We note in panel (e) the appearance of a second, narrow
favoured strip of parameter space. In this strip, the charged-Higgs
contribution to the decay amplitude is not a small perturbation, but is
$\sim - 2 \times$ the $W^\pm$ contribution!}%
.~We note that $\De M_{B_s}$ (not shown) also disfavours small $\MA$ and
large $\tb$. Finally, panel (f) shows the combined EWPO and BPO 
$\chi^2$~function for scenario \Atwo\ throughout the
\plane{\MA}{\tb}. We see that the best-fit point has $\MA \sim 300 \gev$
and $\tb \sim 35$. This is not a very good fit to $\amu$, but it is a
good fit to $b \to s \ga$, $\Mh$ and the other BPO constraints. We note
that this best-fit point has $\chi^2 = 7.4$, which is {\it not} an
improvement on the CMSSM fits discussed in the previous
Section. The $\De \chi^2 = 2.30$ and 4.61 contours are
somewhat looser than in the two previous scenarios, and extend to very
large $\MA$. 

\begin{figure}[tbh!]
\begin{center}
\includegraphics[width=.40\textwidth]{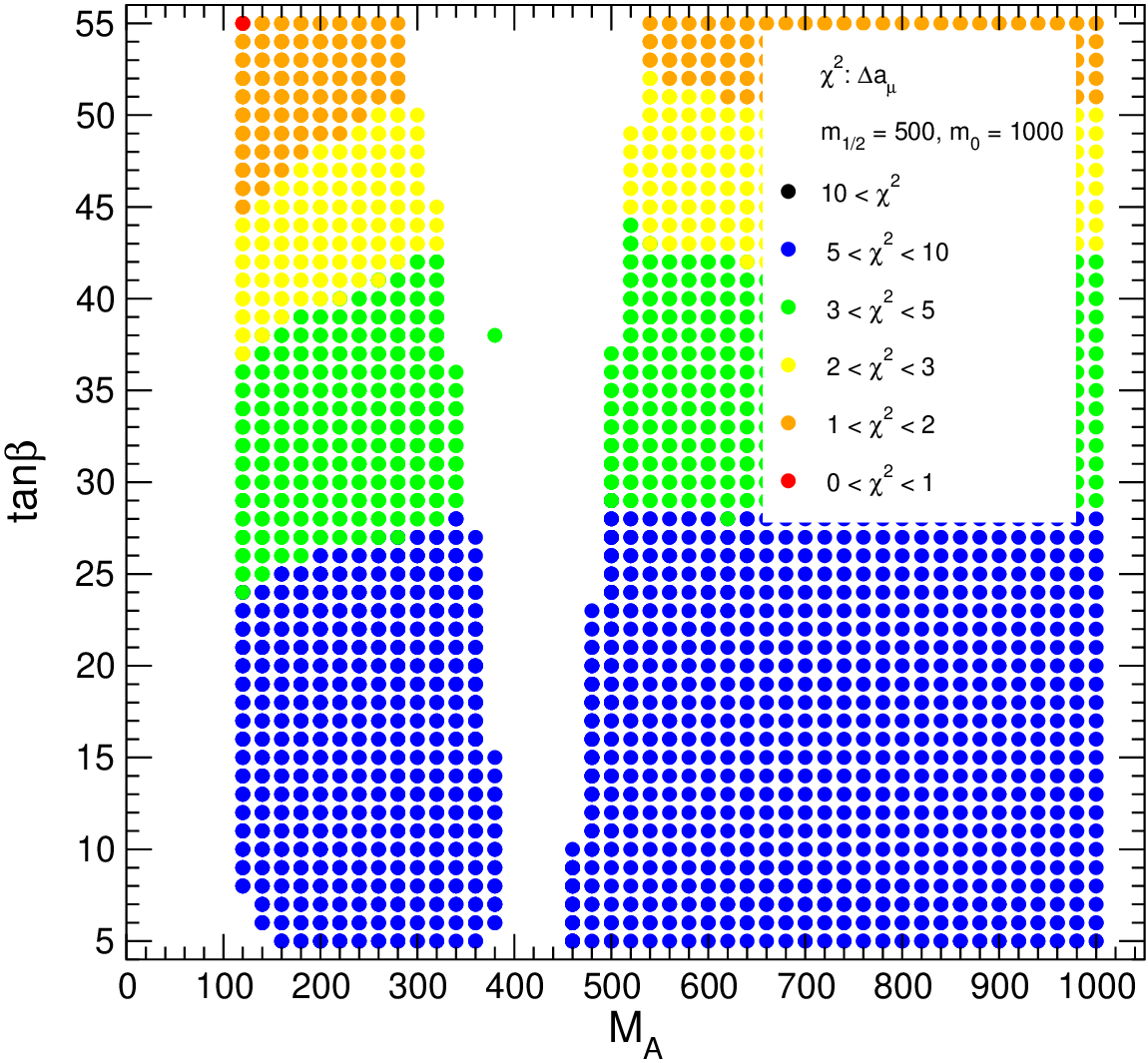}
\includegraphics[width=.40\textwidth]{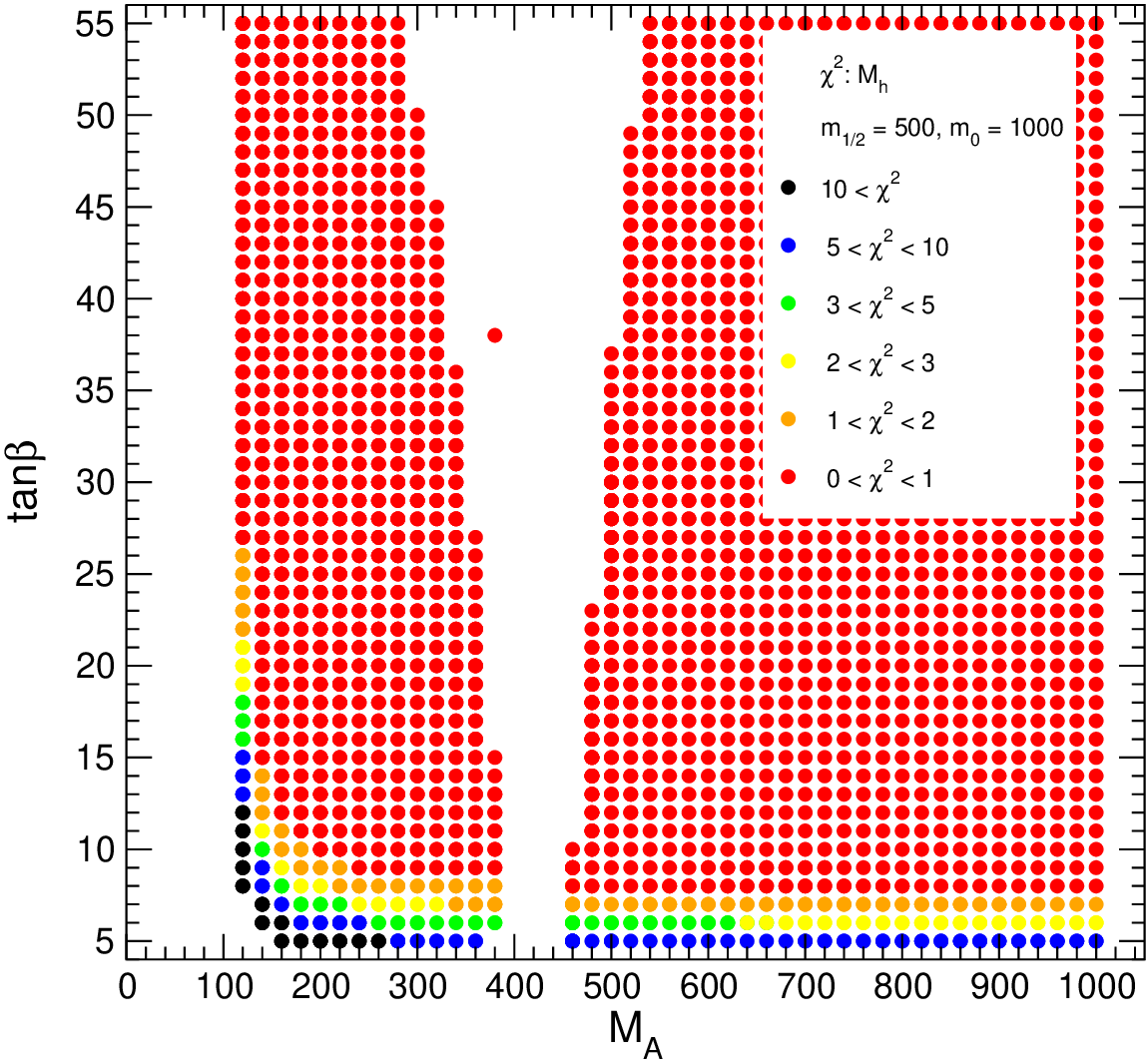}
\end{center}
\begin{center}
\includegraphics[width=.40\textwidth]{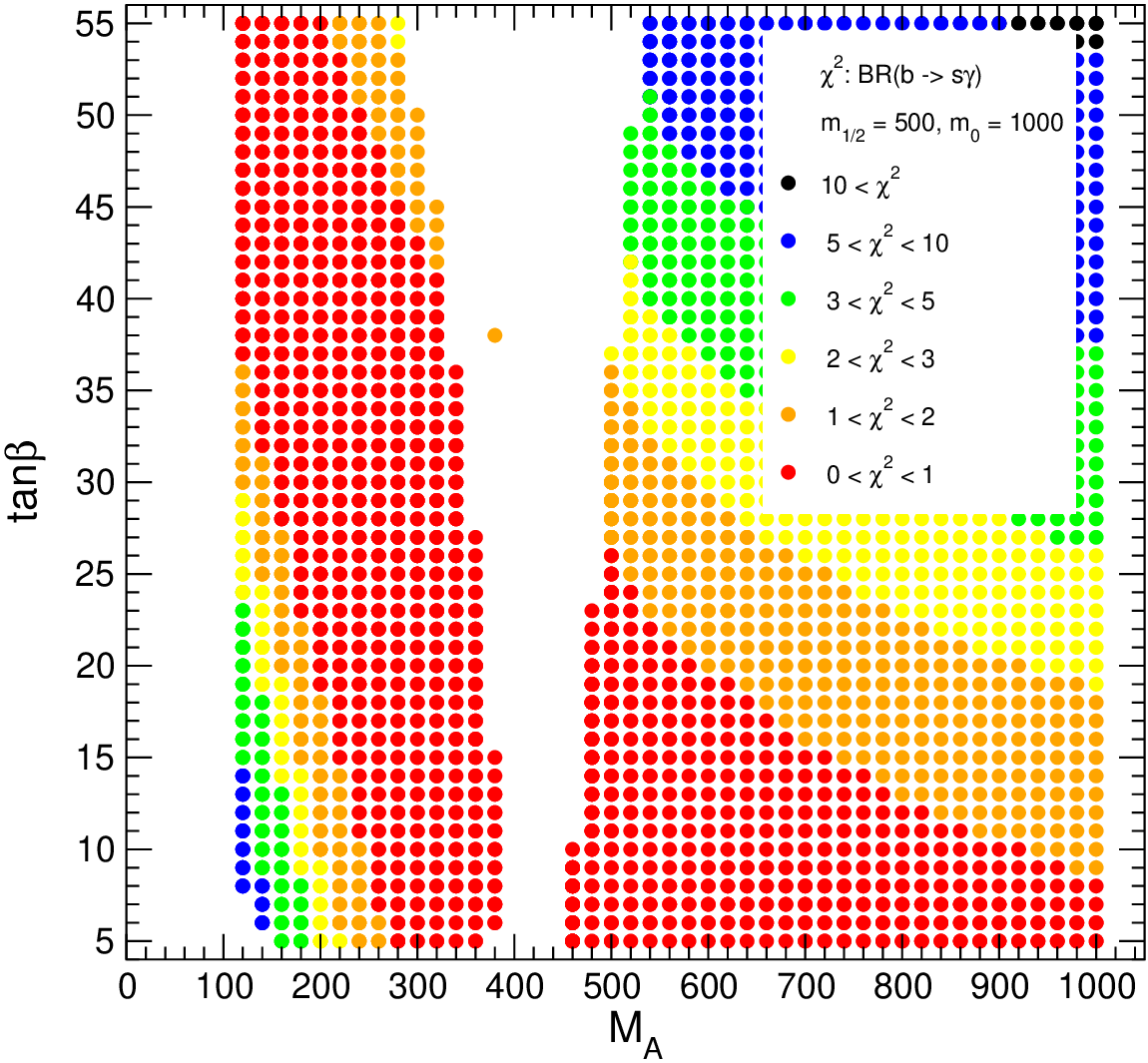}
\includegraphics[width=.40\textwidth]{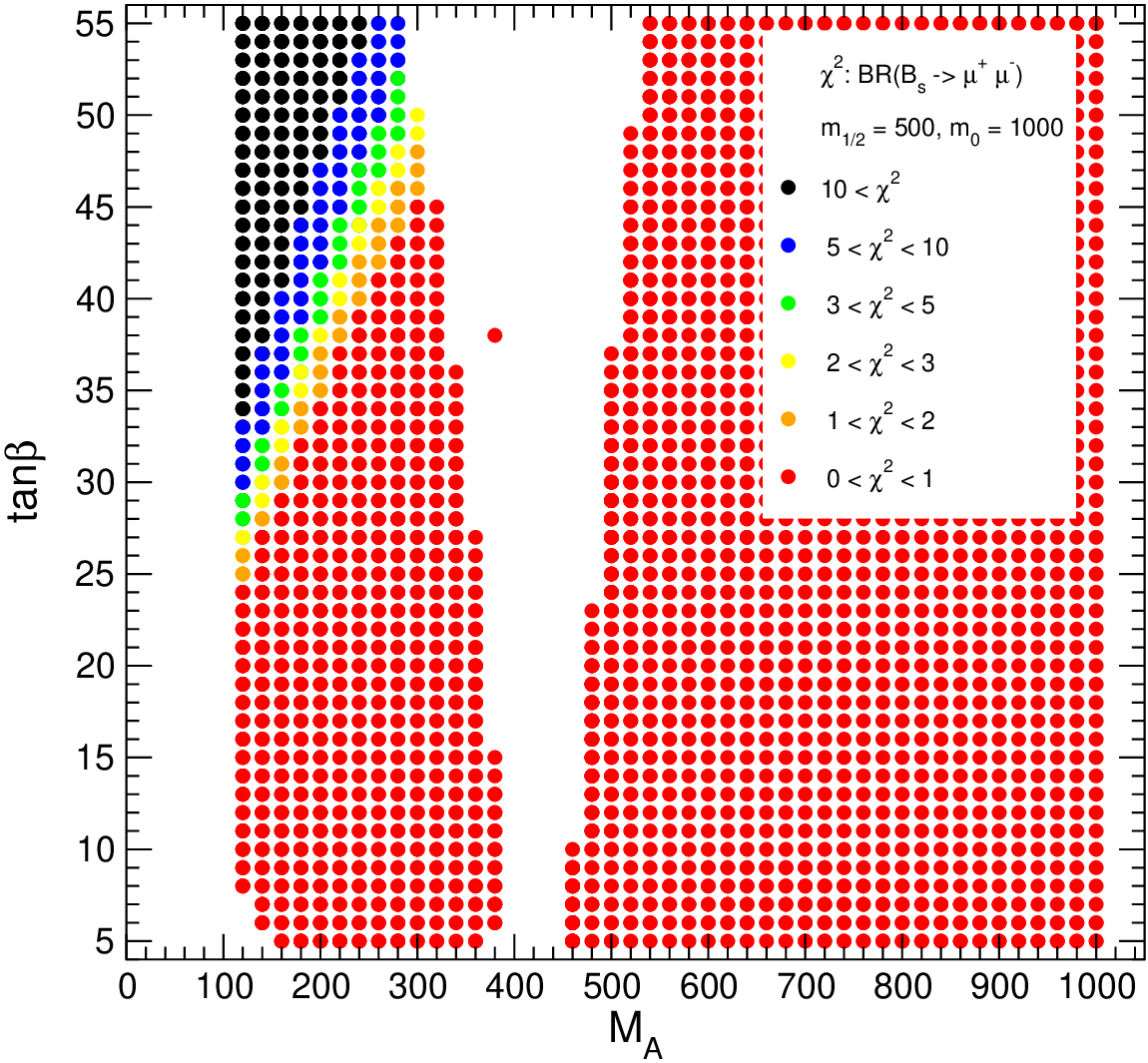}
\end{center}
\begin{center}
\includegraphics[width=.40\textwidth]{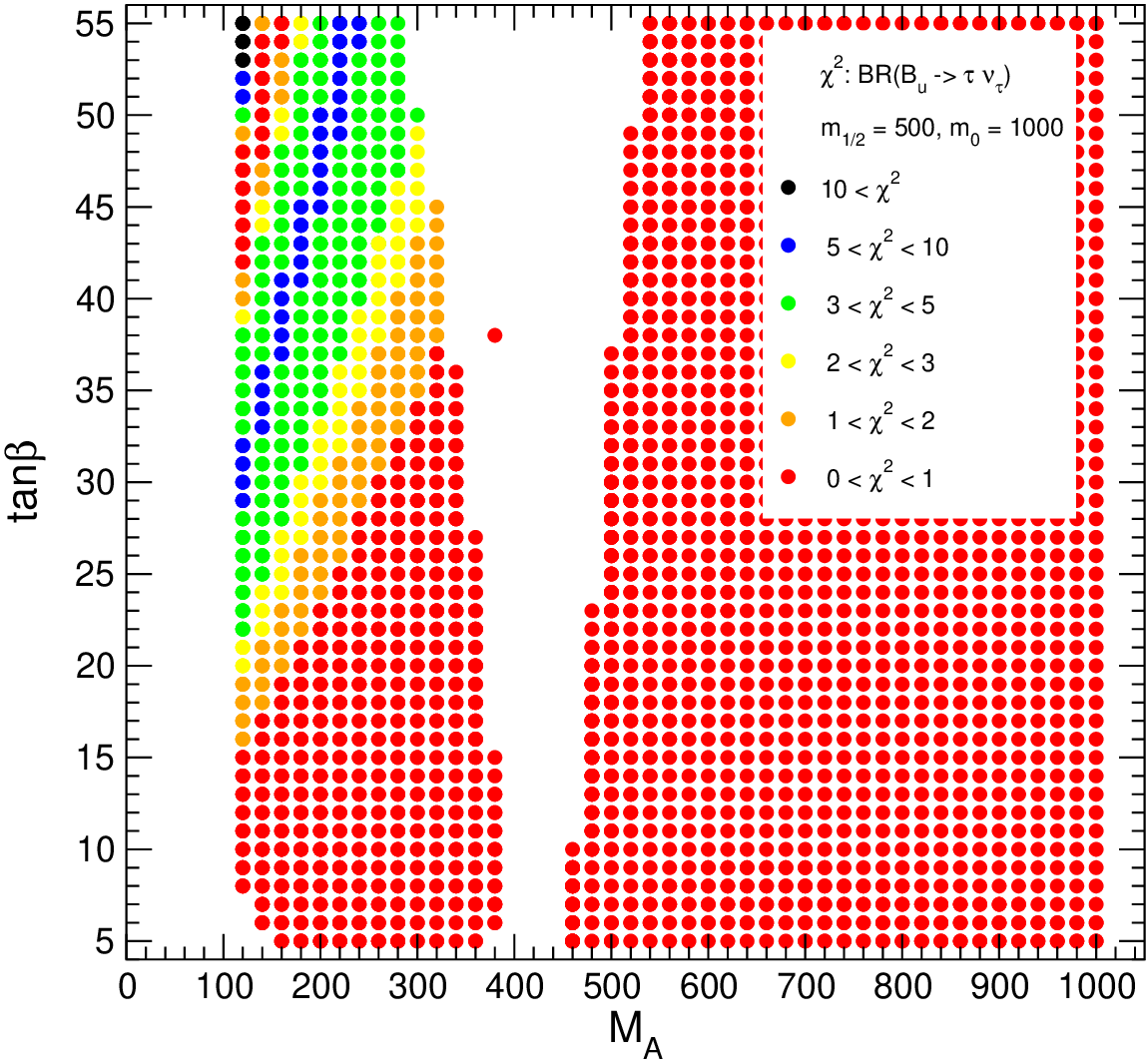}
\includegraphics[width=.40\textwidth]{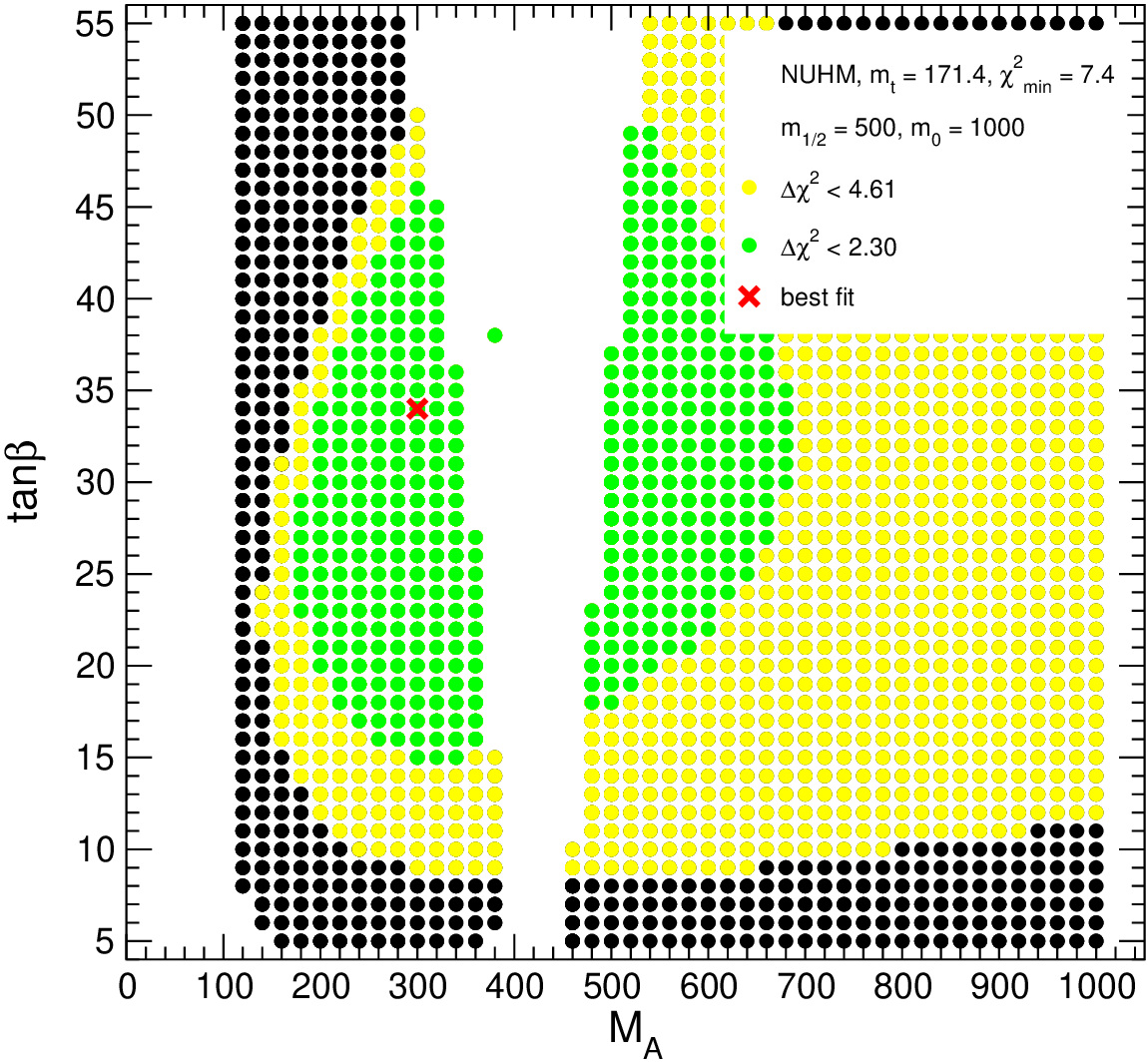}
\caption{%
The most important contributions to the total $\chi^2$~value
for the NUHM
\plane{\MA}{\tb}\ \Atwo, due to (a) $\amu$, (b) $\Mh$, (c) $b \to s
\ga$, (d) $B_s \to \mu^+ \mu^-$ and (e)  $B_u \to \tau \nu_\tau$, and
(f) the combined EWPO and BPO $\chi^2$~function.
We use $\mt = 171.4
\pm 2.1 \gev$ and $\mb(\mb) = 4.25 \pm 0.11 \gev$, and  $\mu$ is
adjusted continuously so as to yield the central value of the cold dark
matter density indicated by WMAP and other observations for the central
values of $\mt$ and $\mb(\mb)$.}
\label{fig:A2}
\end{center}
\vspace{-6em}
\end{figure}

The principal contributions to the total $\chi^2$~function
for the \plane{\MA}{\tb}\ for scenario \Afour\ are shown in
\reffi{fig:A4}~%
\footnote{Again, we do not display the $\chi^2$~values
in the underdense slivers of the plane.}%
. We see in panel (a) that $\amu$ favours a swathe with $\tb \sim 15$ to 20,
almost independently of $\MA$. In particular, values of $\tb > 25$  and
$\tb < 5$ are
quite strongly disfavoured. Panel (b) shows that the LEP lower limit on
$\Mh$ disfavours  values $\tb < 15$, the constraint becoming stronger
for $\MA < 200 \gev$. A small band of values of $\MA$ and $\tb$ are
favoured by $b \to s \ga$, as seen in panel (c), extending to $\tb > 15$
only for $\MA$ below the funnel at $\sim 250 \gev$. Panels (d) and (e)
show the familiar feature that large values of $\tb$ and small values of
$\MA$ are disfavoured by both $B_s \to \mu^+ \mu^-$ and 
$B_u \to \tau \nu_\tau$, and the same is true for $\De M_{B_s}$ (not
shown). Finally, panel (f) shows the combined EWPO and BPO $\chi^2$~function
for scenario \Afour\ throughout the \plane{\MA}{\tb}. We see
that the best-fit point has $\MA \sim 200 \gev$ and $\tb \sim 20$. This
is not a very good fit to $\Mh$, but it is a good fit to $\amu, b \to s
\ga$, and the other BPO constraints. We note that this best-fit point
has $\chi^2 = 5.6$, which is {\it similar} to the CMSSM fits discussed
in the previous Section. We note also that the $\De \chi^2 = 2.30$ and
4.61 contours are particularly tight in 
this scenario, and rule out very large values of $\MA$ and/or $\tb$.

\begin{figure}[tbh!]
\begin{center}
\includegraphics[width=.40\textwidth]{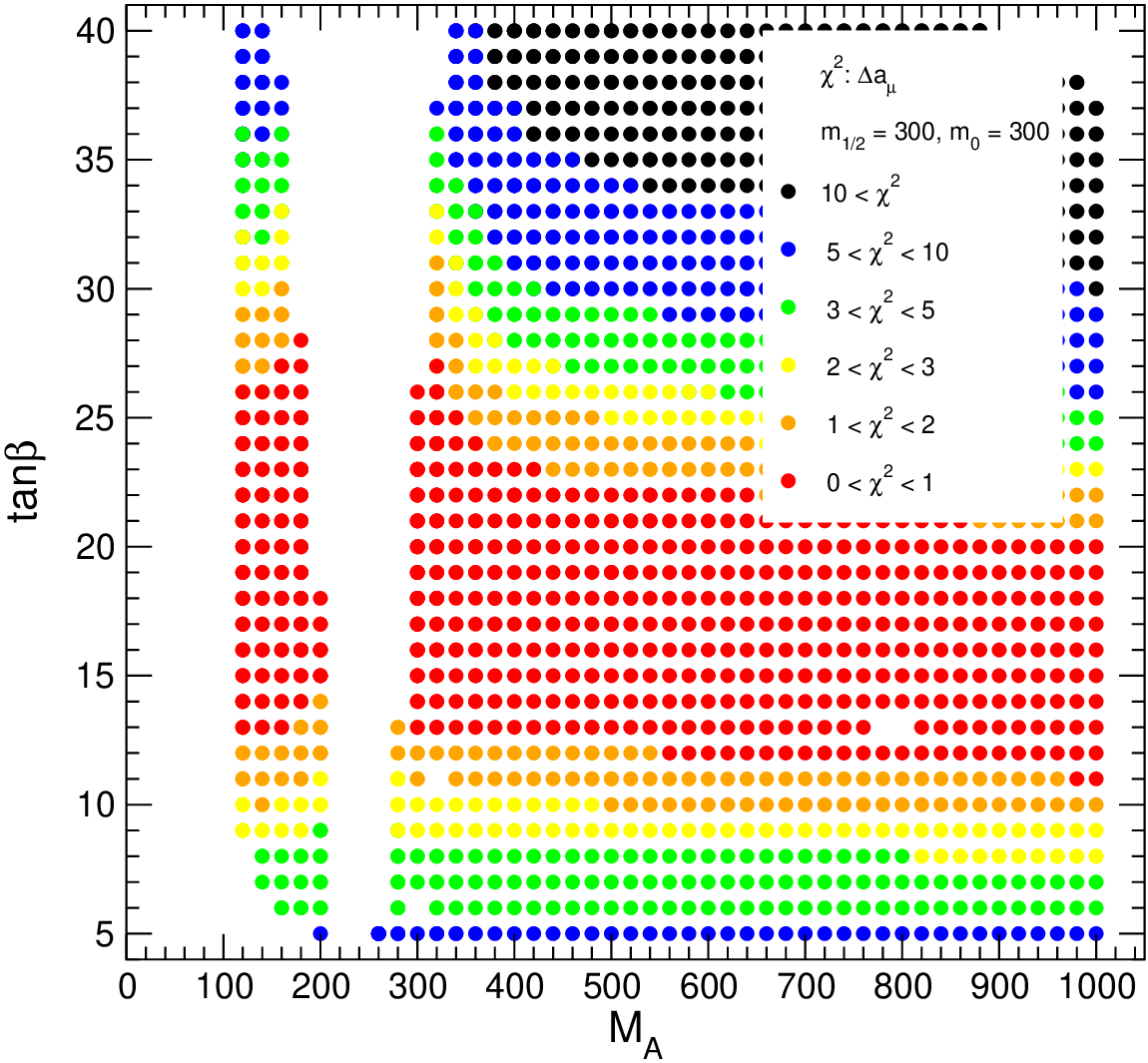}
\includegraphics[width=.40\textwidth]{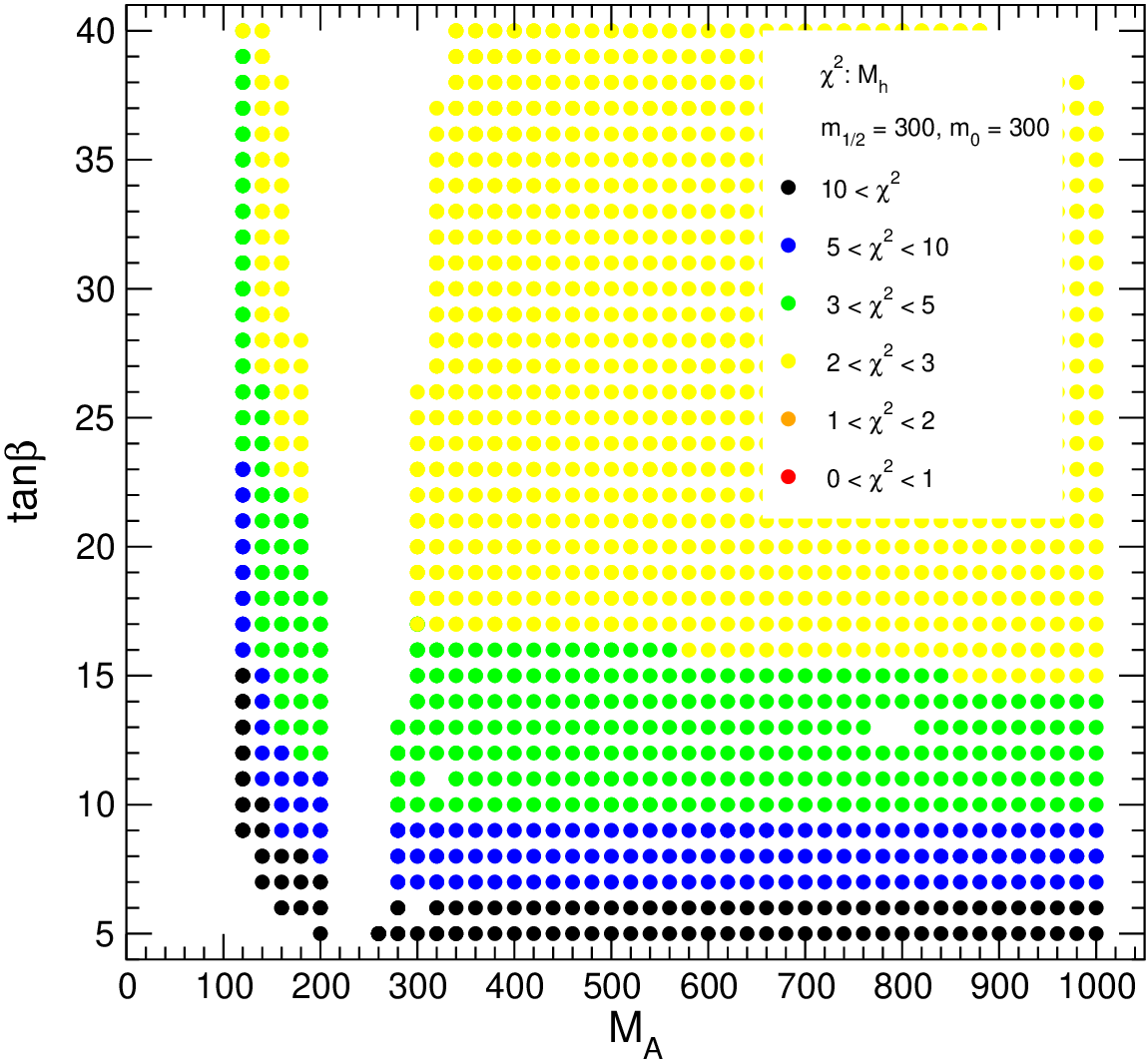}
\end{center}
\begin{center}
\includegraphics[width=.40\textwidth]{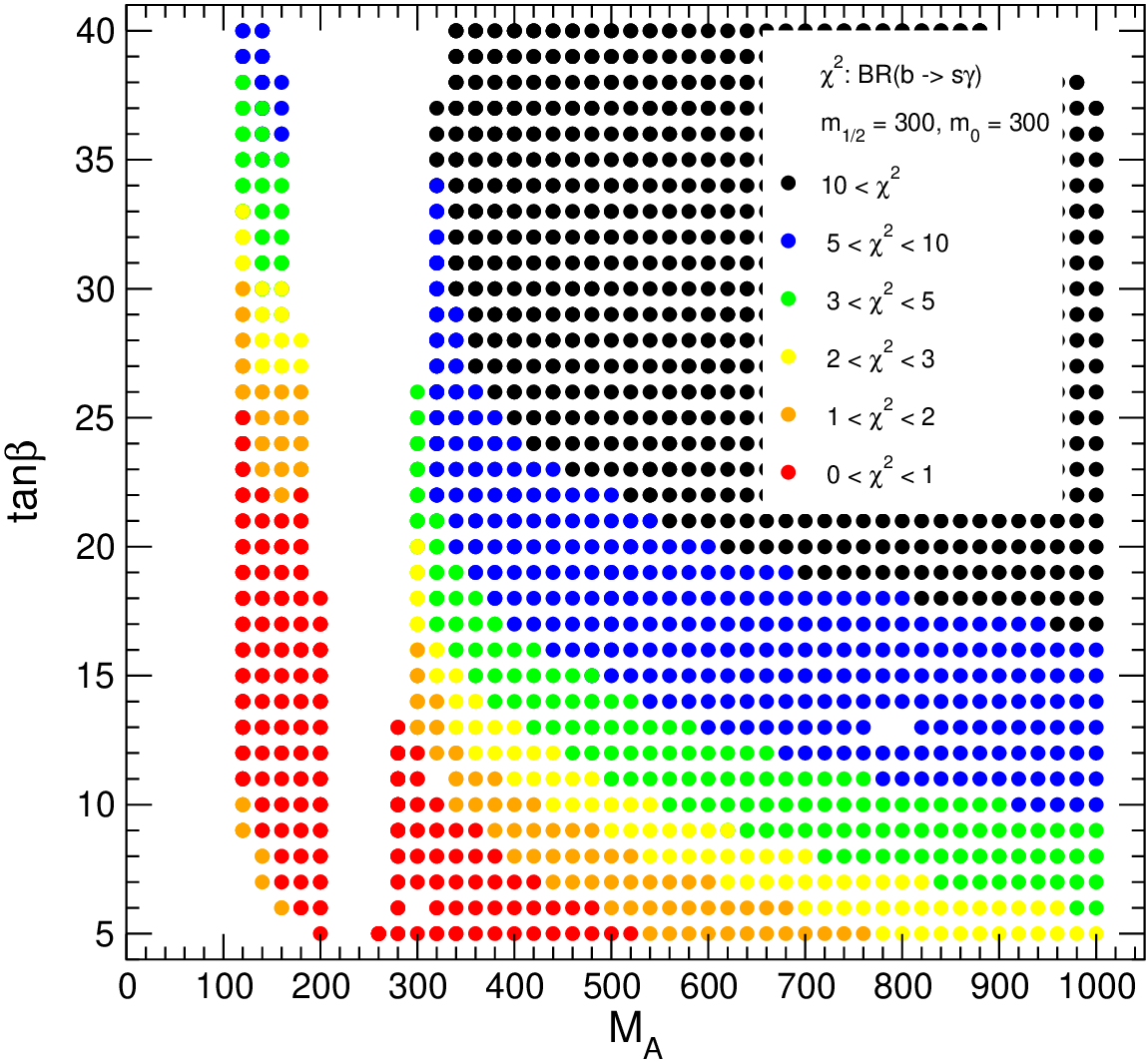}
\includegraphics[width=.40\textwidth]{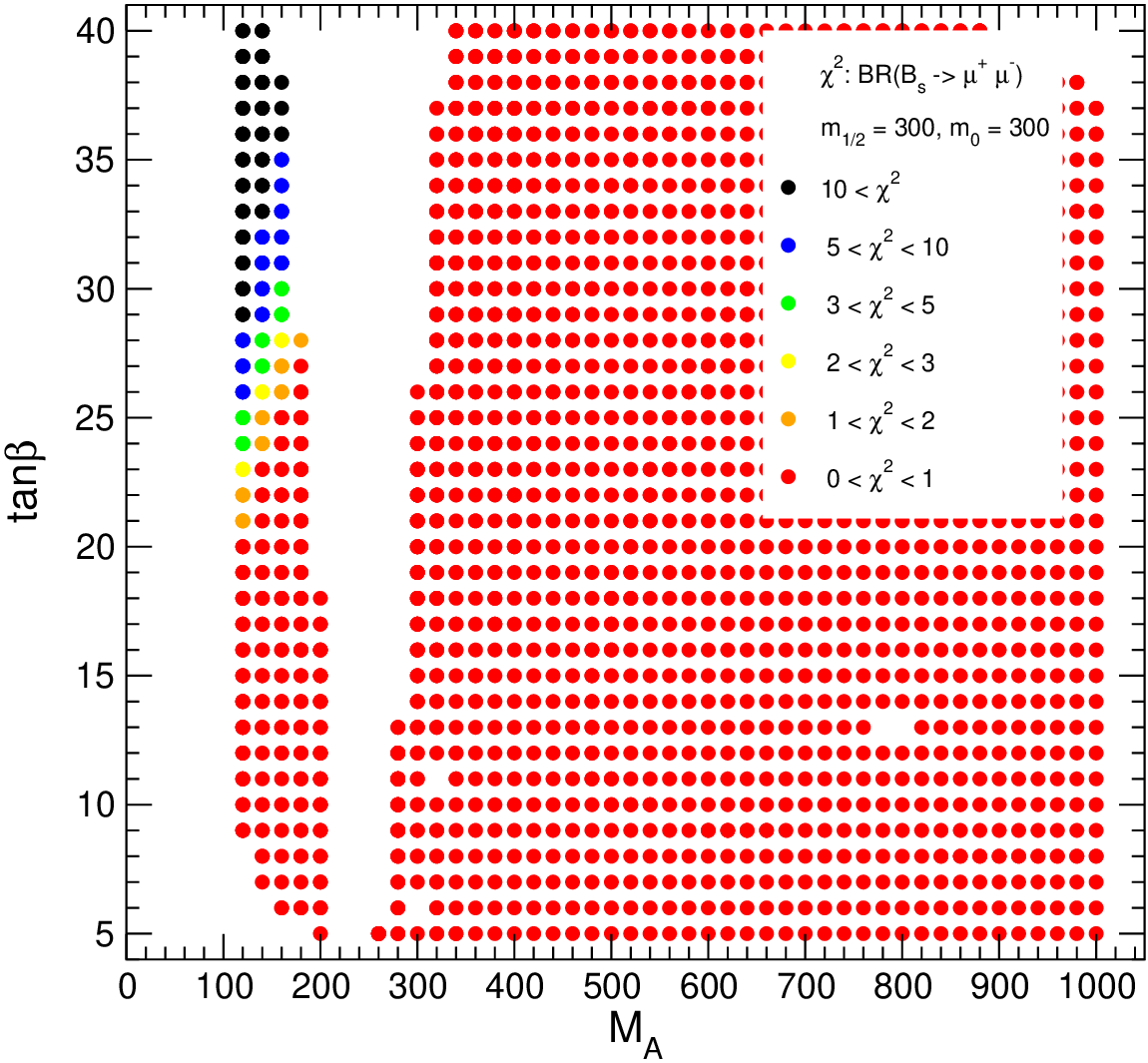}
\end{center}
\begin{center}
\includegraphics[width=.40\textwidth]{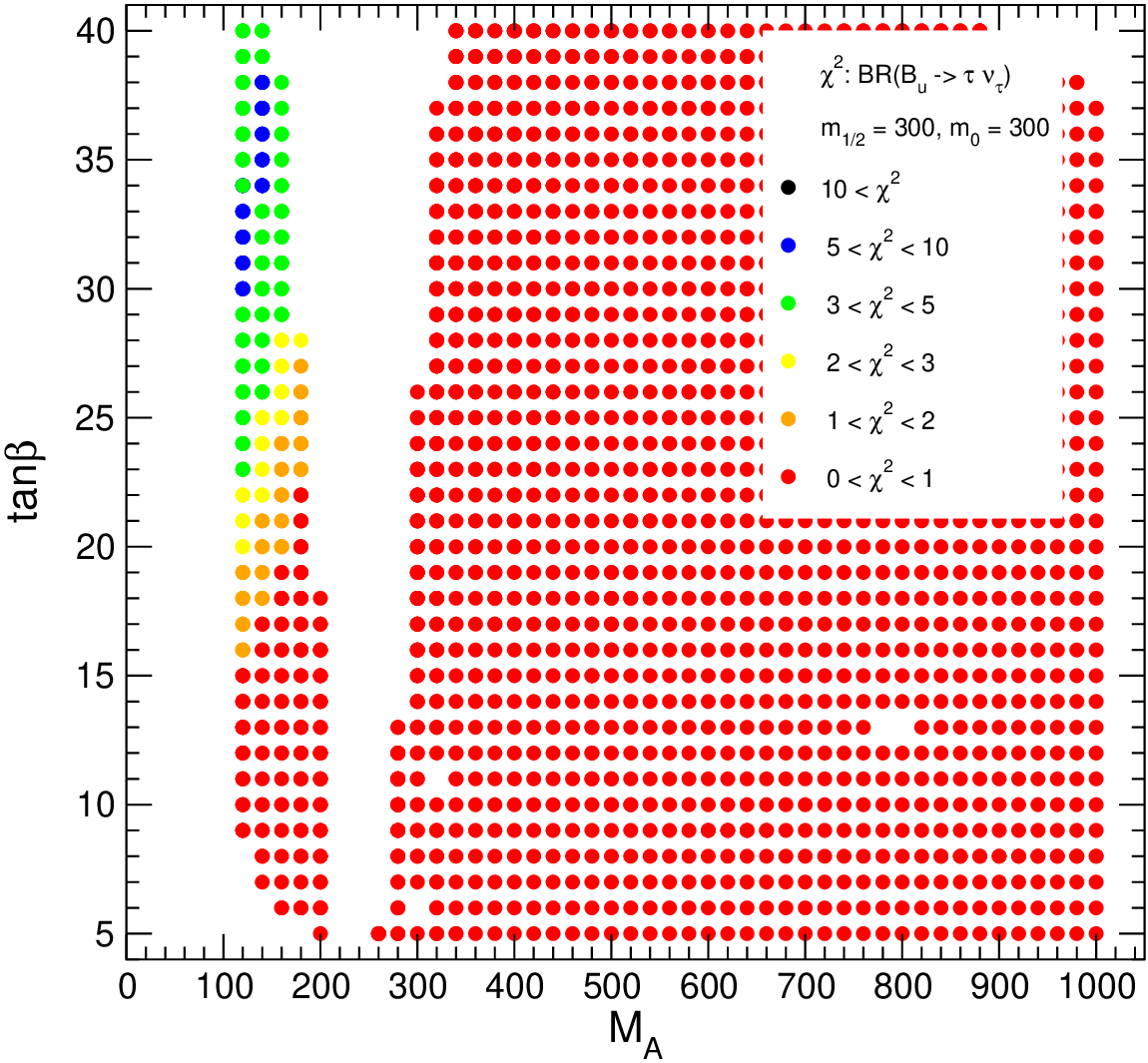}
\includegraphics[width=.40\textwidth]{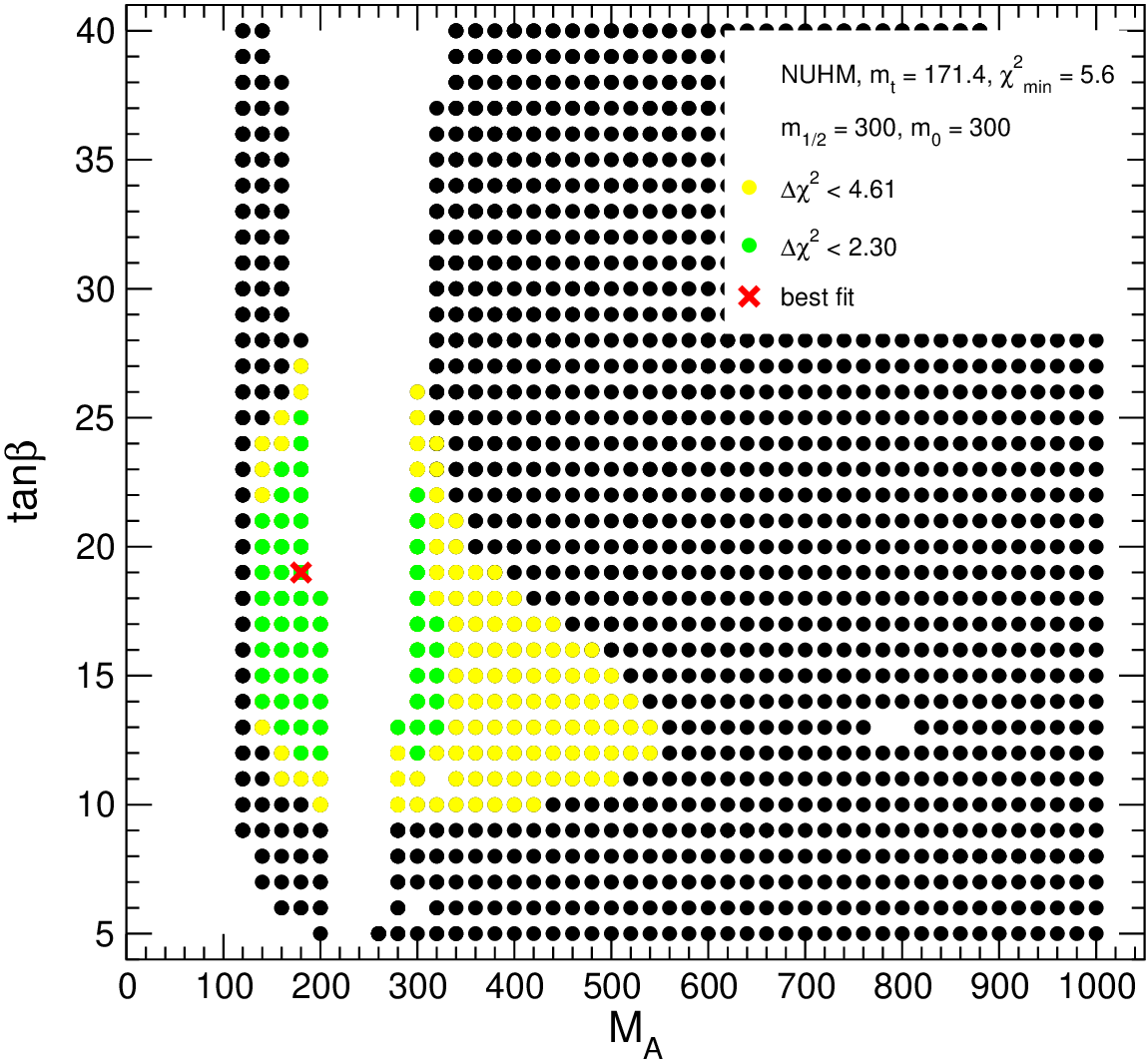}
\caption{%
The most important contributions to the total $\chi^2$~value
for the NUHM
\plane{\MA}{\tb}\ \Afour, due to (a) $\amu$, (b) $\Mh$, (c) $b \to s
\ga$, (d) $B_s \to \mu^+ \mu^-$ and (e) $B_u \to \tau \nu_\tau$, and (f)
the combined EWPO and BPO $\chi^2$~distribution.
We use $\mt = 171.4 \pm
2.1 \gev$ and $\mb(\mb) = 4.25 \pm 0.11 \gev$, and  $\mu$ is adjusted
continuously so as to yield the central value of the cold dark matter
density indicated by WMAP and other observations for the central values
of $\mt$ and $\mb(\mb)$.}
\label{fig:A4}
\end{center}
\vspace{-6em}
\end{figure}

There are some common features of these analyses for fixed $(m_{1/2},
m_0)$ and $(m_0, \mu)$. For example, we find that relatively low values
of $\MA \sim 200$ to 400~GeV are consistently favoured. This is
essentially because $\amu$ prefers moderately small values of $m_{1/2}$
which would, if left to themselves, create problems for $b \to s
\ga$. However, this tension may be mitigated if $\MA$ is correspondingly
small, providing a cancellation in the supersymmetric contributions to
the $b \to s \ga$ decay amplitude. We also note a consistent preference
for relatively large values of $\tb \sim 20$ to 50, which is
essentially due to the pressure exerted by the LEP lower limit on the
Higgs mass. 

As discussed above, the LSP would constitute (most of) the cold dark matter
across (most of) the NUHM parameter planes discussed above. Accordingly,
for completeness we discuss the prospects for direct dark matter detection in
different regions of the planes. In the cases of planes \Athree\ and
\Afive, the direct scattering rate is generally below the CDMS upper
limit~\cite{CDMS}, once one takes into account uncertainties in the
strange-quark contributions to the spin-independent scattering matrix
elements and in the local cold dark matter density. In the cases of
plane \Atwo, only in the region where $\MA$ is small and $\tb$ is high
does the dark matter scattering rate approach the CDMS upper
limit. However, there is a potential conflict with the preliminary XENON10
results~\cite{Xenon} if the strange-quark contribution and/or the local
relic density  is large. A similar situation arises 
in plane \Afour\ for small values of $\MA$, almost independent of $\tb$.
In absence of sufficient understanding of the systematic uncertainties in the
strange-quark contribution and the local cold dark matter density, we do not
attempt to include the direct dark matter searches in the overall $\chi^2$ 
function~
\footnote{For completeness, we note that along the WMAP strips
in the CMSSM the direct dark matter scattering rate is always comfortably
below the CDMS upper limit.}.

\medskip
The survey of NUHM parameter space made in this Section has not been  
exhaustive, in particular we have restricted our attention to planes
with $A_0 = 0$. Nevertheless, the values of $\chi^2$ found at the
best-fit points in the various \plane{\MA}{\tb}s are quite acceptable:
planes \Athree, \Afive, \Atwo\ and \Afour\ have 
$\chi^2 = 7.1, 3.5, 7.4$ and 5.6, respectively, in fits to 9 observables
with 2 free parameters in each case. 
It should be stressed, however, that only the \Afive\ plane has a  
minimum of $\chi^2$ noticeably lower than that for the CMSSM fits with
$\tb = 10$, which occurs when $\mu = 800 \gev$,
$m_0 = 300 \gev$, $\MA \sim 340 \gev$ and $\tb = 36$ and, moreover, at the  
point with the minimum value of $\chi^2$, the relic neutralino density
is somewhat higher than the WMAP-compatible range. One might expect a greater
reduction in $\chi^2$ in a full study  
of the NUHM, in view of its two additional parameters compared with the
CMSSM. Accordingly, we have made a preliminary study whether the quality
of the NUHM fit could be improved significantly by
varying $A_0$, assuming the same values of $m_0, \MA$ and $\tb$ as at  
the best-fit point in the \Afive\ plane, but choosing different values of
$\mu$ and $m_{1/2}$. We have investigated the possibilities 
$(\mu, m_{1/2}) = (800, 368)$ [{\bf L1}], $(800, 448)$ [{\bf L2}]  
and $(680, 448) \gev$ [{\bf L3}],
respectively, and varied $A_0$ between $\pm 1000 \gev$. \reffi{fig:L123}
shows the values of $\chi^2$ along the lines {\bf L1}, {\bf L2} and {\bf L3}.  We see that
the greatest improvement in $\chi^2$ compared to the \plane{\MA}{\tb}s
shown previously are by $\sim 0.3$ only.
Interestingly, the minimal values of $\chi^2$ are
found for small values of $A_0 \sim 0$. Undoubtedly some further  
reduction in $\chi^2$ could be found in a more complete study, but it
seems that the extra degrees of freedom in
the NUHM are not crucial for the overall quality of the fit.

\begin{figure}[tbh!]
\begin{center}
\includegraphics[width=.70\textwidth]{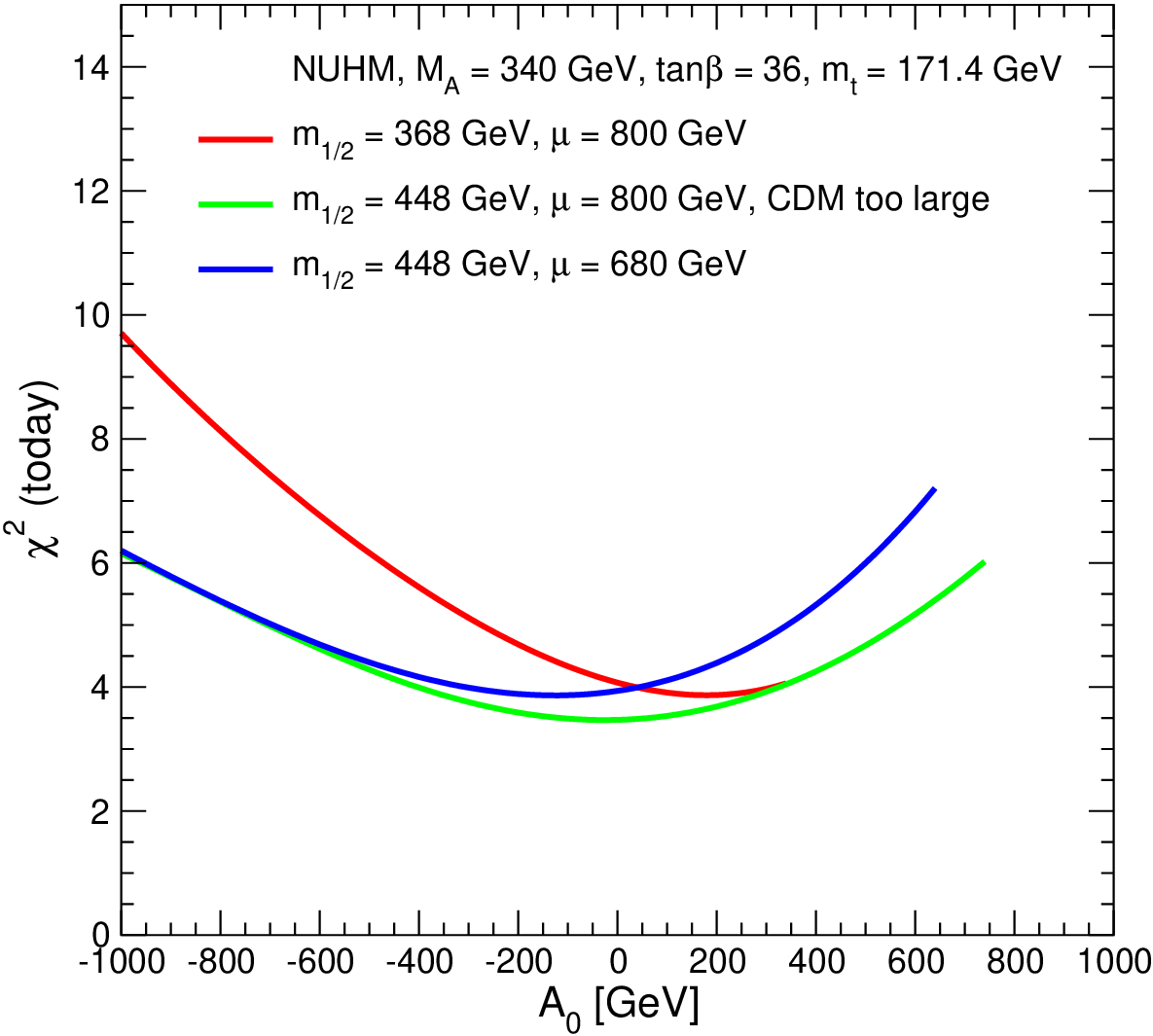}
\end{center}
\begin{center}
\caption{%
The dependence on $A_0$ of the $\chi^2$ function along lines with
$m_0 = 300 \gev$, $\MA = 340 \gev$, $\tb = 36$ and
$(\mu, m_{1/2}) = (800, 368)$ [{\bf L1}], $(800, 448)$ [{\bf L2}] and $(680, 448) \gev$ [{\bf L3}],
respectively.}
\label{fig:L123}
\end{center}
\vspace{-2em}
\end{figure}


\section{Conclusions}
\label{sec:conclusions}

We have analyzed previously the regions of the CMSSM parameter space
preferred by the EWPO~\cite{ehow3,ehow4}, and found a tendency to prefer
regions on the WMAP coannihilation strips with relatively low values of
$m_{1/2}$. These points were favoured, in particular, by the measurements of
$\amu$ and $\MW$. Both these tendencies have now been reinforced, with the
interpretation of $\amu$ based on the use of $e^+ e^-$ data to estimate the
SM contribution gaining ground, and the small decrease in $\mt$ and the
slight increase in $\MW$ tending to favour a contribution to the latter EWPO
from some physics beyond the SM. 

Previously, we incorporated just a single BPO
into our global analysis~\cite{ehow3,ehow4}, namely $b \to s \ga$. Recently,
data on $B_s \to \mu^+ \mu^-$, $B_u \to \tau \nu_\tau$ and $\De M_{B_s}$
have also become available and now impinge significantly on the CMSSM
parameter space. In this paper, for the first time, we have incorporated
all these BPO into a global analysis. 

We have found a good $\chi^2/{\rm d.o.f.}$ for this global fit. However,
it is clear that there is a slight tension between the relatively low
values of $m_{1/2}$ favoured by the EWPO and the absence of any corroborating
indication from the BPO. Nevertheless, the global $\chi^2$~analysis
favours the appearance of relatively light sparticles that should be
`easy' to see at the LHC and may offer good prospects also for the ILC.

As we also have discussed here explicitly, for the first time, the
global analysis strongly favours values of $\Mh$ only slightly above the
lower limit established by LEP. Indeed, we find that values of $\Mh <
120 \gev$ are preferred, while values above $123 \gev$ cannot be reached
in the CMSSM (for the current $\mt$ value).

Another new step in this paper, motivated by the slight tension between
the EWPO and the BPO, has been to explore the parameter space of the
NUHM. We have 
displayed for the first time \plane{\MA}{\tb}s within the NUHM over
which the WMAP constraint on the cold dark matter density is generically
respected. We have then shown the interplay of the various EWPO and BPO
in such planes. We find that, for fixed $(m_{1/2}, m_0)$ or $(m_0,
\mu)$, relatively low values of $\MA \sim 200$ to 400~GeV are favoured,
as are relatively large values of $\tb \sim 20$ to 50. It is
possible to find in this way NUHM points that have lower $\chi^2$ than
those possible in the CMSSM. 

In the future, it will be necessary to follow closely the evolutions of
both the 
EWPO and the BPO: improvements in the measurements of both $\MW$
and $\mt$ are expected, and the interpretation of $\amu$ may become
clearer. We also expect significant improvements in the measurements of
$B_s \to \mu^+ \mu^-$ and $B_u \to \tau \nu_\tau$, and possibly in the
interpretations of $b \to s \ga$ and $\De M_{B_s}$. The question will be
whether the present slight tension between the EWPO and the BPO 
within the CMSSM will strengthen or relax, and a more detailed and
systematic exploration of the NUHM parameter space will certainly be
desirable. The implications of the EWPO and the BPO for the
supersymmetric parameter space will surely be an interesting and
continuing saga.
 

\subsection*{Acknowledgements}
\vspace{-0.5em}
S.H.\ thanks C.-J.~Stephen and M.~Herndon for providing the
$\chi^2$~numbers for $\br(B_s \to \mu^+ \mu^-)$. 
S.H.\ thanks G.~Isidori and P.~Paradisi about helpful communication
about their analytical results for the BPO.
G.W.\ thanks T.~Becher and U.~Haisch for interesting discussions.
The work of K.A.O.\ was partially supported by DOE grant DE-FG02-94ER-40823. 
The work of S.H.\ was partially supported by CICYT (grant FPA2006--02315).
Work supported in part by the European Community's Marie-Curie Research
Training Network under contract MRTN-CT-2006-035505
`Tools and Precision Calculations for Physics Discoveries at Colliders'




\end{document}
